\documentclass[useAMS,usenatbib]{mnras}
\usepackage{times}
\usepackage{graphicx}
\usepackage{amsmath}
\usepackage{fmtcount}
\usepackage{amssymb}
\usepackage{gensymb}
\usepackage{natbib}
\usepackage{color}
\usepackage{booktabs}
\bibpunct{(}{)}{;}{a}{}{,}
\usepackage{psfig}
\title[The Assembly History of NGC\,1395]{Tracing the Assembly History of NGC\,1395 through its Globular Cluster System}

\author[Escudero et al.]{Carlos G. Escudero$^{1,2}$\thanks{E-mail: cgescudero@fcaglp.unlp.edu.ar}, Favio R. Faifer$^{1,2}$, Anal\'ia V. Smith Castelli$^{2,3}$, Juan C. Forte$^{3,4}$,
\newauthor 
Leandro A. Sesto$^{1,2}$, N\'elida M. Gonz\'alez$^{1,2}$, Mar\'ia C. Scalia$^{1,2}$ \\
$^{1}$Facultad de Cs. Astron\'omicas y Geof\'isicas, UNLP, Paseo del Bosque s/n, B1900FWA, La Plata, Argentina. \\
$^{2}$Instituto de Astrof\'isica de La Plata, UNLP, CONICET, Paseo del Bosque s/n, B1900FWA, La Plata, Argentina. \\
$^{3}$Consejo Nacional de Investigaciones Cient\'ificas y T\'ecnicas, Godoy Cruz 2290, C1425FQB, CABA, Argentina \\
$^{4}$Instituto Argentino de Matem\'atica Alberto Calder\'on, CONICET, Argentina. \\
}

\begin{document}

\pagerange{\pageref{firstpage}--\pageref{lastpage}} \pubyear{}
\maketitle

\label{firstpage}

\begin{abstract}
We used deep Gemini-South/GMOS $g'r'i'z'$ images to study the globular cluster (GC) system of the massive elliptical galaxy NGC\,1395, located in the Eridanus supergroup. The photometric analysis of the GC candidates reveals a clear colour bimodality distribution, indicating the presence of ``blue'' and ``red'' GC subpopulations. While a negative radial colour gradient is detected in the projected spatial distribution of the red GCs, the blue GCs display a shallow colour gradient. The blue GCs also display a remarkable shallow and extended surface density profile, suggesting a significant accretion of low-mass satellites in the outer halo of the galaxy. In addition, the slope of the projected spatial distribution of the blue GCs in the outer regions of the galaxy, is similar to that of the X-ray halo emission. Integrating up to 165\,kpc the profile of the projected spatial distribution of the GCs, we estimated a total GC population and specific frequency of 6000$\pm$1100 and $S_N$=7.4$\pm$1.4, respectively. Regarding NGC\,1395 itself, the analysis of the deep Gemini/GMOS images shows a low surface brightness umbrella-like structure indicating, at least, one recent merger event. Through relations recently published in the literature, we obtained global parameters, such as $M_\mathrm{stellar}=9.32\times10^{11}$ M$\odot$ and $M_h=6.46\times10^{13}$ M$\odot$. Using public spectroscopic data, we derive stellar population parameters of the central region of the galaxy by the full spectral fitting technique. We have found that, this region, seems to be dominated for an old stellar population, in contrast to findings of young stellar populations from the literature.    

\end{abstract}


\section{INTRODUCTION}
\label{sec:intro}
In the scenario of galaxy formation, it is clear that major fusions and tidal interactions play important roles. In the paradigm of hierarchical $\Lambda$CDM structure formation, massive early-type galaxies we see in the local Universe would have been built through a combination of two phases: {\it in situ} and {\it ex situ} formation \citep[e.g.,][]{oser10,rodriguez16}. In the first phase ($z\gtrsim2$), there is rapid {\it in situ} star formation driven by the infall of cold gas flows \citep{dekel09,naab09}, followed by a largely quiescent evolution. The second phase (i.e. {\it ex situ}) is dominated mainly by successive accretion of smaller structures \citep{zolotov10,wellons16}. These two mechanisms should leave their signatures on the dynamics, chemistry, and kinematics of the galaxies, both in their inner regions and in their halos. Recent observational studies \citep{pastorello14,duc15,carlsten17,dabrusco16} have shown the presence of gradients in the stellar properties (metallicity, surface brightness, abundance ratio) of early-type galaxies, as well as tidal structures of low-surface brightness that provide evidence and reflect, at least in part, the hierarchical history of such systems. 

Interaction processes between galaxies belonging to clusters and groups, and with their intra-cluster medium, are also very important mechanisms that determine different properties of the resulting galaxies. Several observational and simulation studies have determined that the properties of galaxies located in dense environments differ from those found in low density ones \citep{tal09,niemi10}. Although there are several mechanisms that act more efficiently in one case or another (e.g., ram pressure stripping and strangulation in clusters, and tidal interactions and mergers in groups \citep{balogh00,bekki14,mazzei14}), it is still unclear whether the different properties exhibited by galaxies located in clusters are due to the result of nature, or due to the effects of preprocessing in groups. The study of early-type galaxies located in groups and low-density environments has begun to increase in the last decade (e.g., \citealt{spitler08,cho12,lane13,salinas15,caso15,escudero15,bassino17}), since it is there that one can expect to see some transformation processes in action (e.g. accretion and/or merger processes), and not only their final product. 

In this scenario, globular clusters (GCs) have proved to be important tools for examining the early stages of galaxy formation, and also to trace major star forming episodes in a galaxy \citep{hargis12,blom14,sesto16}. The advantages of using these objects is that they are easily seen far beyond the Local Group, where, at present, it is difficul to perform direct measurements of galaxy integrated light at large galactocentric radii \citep[e.g.,][]{norris08,pota13}. In addition, the connection between GCs and the stellar populations that shape up galaxies is widely accepted. Some steps in exploring the existence of such a connection have been presented in \citet{forte07,forte09,forte12,forte14}. The basic idea of these works is that GCs form along diffuse stellar populations sharing their chemical abundances, spatial distributions and ages.

In this context, we intend to study the elliptical (E) galaxy NGC\,1395 belonging to the less dense group of the Eridanus supergroup, particularly through its GC system. The Eridanus supergroup is composed of at least three different subgroups of galaxies with different degrees of dynamic evolution: the most massive and mature subgroup associated with the E galaxy NGC\,1407, a subgroup around the lenticular galaxy NGC\,1332, and the more sparse Eridanus subgroup linked to the massive E galaxy NGC\,1395 \citep{brough06}. This supergroup together with the Fornax cluster and other neighboring groups would constitute the Fornax+Eridanus complex \citep{nasonova11,makarov11}. However, although on the scale of virial radii these groups do not overlap, they are probably bound to the main structure of the Fornax cluster which is linked to NGC\,1399 \citep{nasonova11}. 

Regarding the study of GC systems in the Eridanus supergroup, NGC\,1407 was extensively studied in the last years by different authors \citep{forbes06,trentham06,spolaor08a,spolaor08b,romanowsky09,forbes11,pota15}, and a complete analysis of its GC system is available. On the other hand, the GC system of NGC\,1332 presents a single studied performed by \citet{kundu01} using the Wide Field Planetary Camera 2 (WFPC2) snapshot survey of the {\it Hubble Space Telescope} in the $V$ and $I$ bands. 

Throughout the paper we will adopt the distance modulus of NGC\,1395 obtained from surface brightness fluctuation of \citet{tully13} of $(m-M)_0=31.88\pm0.12$ mag (Table \ref{Tab0}), which corresponds to a spatial scale of 0.11 kpc\,arcsec$^{-1}$. Some relevant parameters of NGC\,1395 are listed in Table \ref{Tab0}. 

The paper is structured as follows. In Section \ref{sec:obs} we describe the observational data, as well as the reduction and calibration processes. In Section \ref{sec:elipse} we present the isofotal analysis performed on the brightness distribution of the galaxy, highlighting the presence of a faint shell. In Section \ref{sec:GCS} we select the GC sample and describe its properties. In Section \ref{sec:correlacion} we obtain physical parameters of NGC\,1395. In Section \ref{sec:disc_concl} we present our conclusions.

\begin{table}
\centering
\caption{Relevant parameters of NGC\,1395.}
\label{Tab0}
\begin{tabular}{lccl}
\toprule
\toprule
\multicolumn{1}{c}{\textbf{Parameter}} &
\multicolumn{1}{c}{\textbf{Value}} &
\multicolumn{1}{c}{\textbf{Units}} &
\multicolumn{1}{c}{\textbf{Reference}} \\
\midrule
$\mathbf{\alpha}_{\mathbf{J2000}}$  &  03:38:29.75     & h:m:s        &  NED \\
$\mathbf{\delta}_{\mathbf{J2000}}$  &  $-$23:01:39.09  & d:m:s        &  NED \\
{\it l}                           &  216:12:42.1     & d:m:s        &  NED \\
{\it b}                           &  $-$52:07:14     & d:m:s        &  NED \\
Type                              &  E2              & --           &  RC3 \\
V${}_{\mathbf T}^{\mathbf 0}$        &  9.61            & mag          &  RC3 \\
B${}_{\mathbf T}^{\mathbf 0}$        &  10.6            & mag          &  \citet{capaccioli15} \\
(m--M)$_0$                        &  31.88$\pm$0.12  & mag          &  \citet{tully13} \\
Position angle                    &  104.5           & degree       &  \citet{capaccioli15} \\      
Helio. velocity                   &  1717$\pm$7      & km\,s$^{-1}$  &  \citet{smith00} \\
$\sigma$                          &  238$\pm$8       & km\,s$^{-1}$  &  \citet{smith00} \\
$R_\mathrm{eff}$                   &  1.10            & arcmin       &  \citet{li11}    \\
$log M_\mathrm{stellar}$           &  11.0            & $M_\odot$     &  \citet{colbert04} \\
$log L_X$                         &  40.89           & erg\,s$^{-1}$ &  \citet{osullivan01} \\
$log M_{SMBH}$                     &   8.59           & $M_\odot$     &  \citet{pellegrini10} \\
\bottomrule
\end{tabular}
\end{table}

\section{Observations and Data Reduction}
\label{sec:obs}
\begin{figure}
\resizebox{0.99\hsize}{!}{\includegraphics{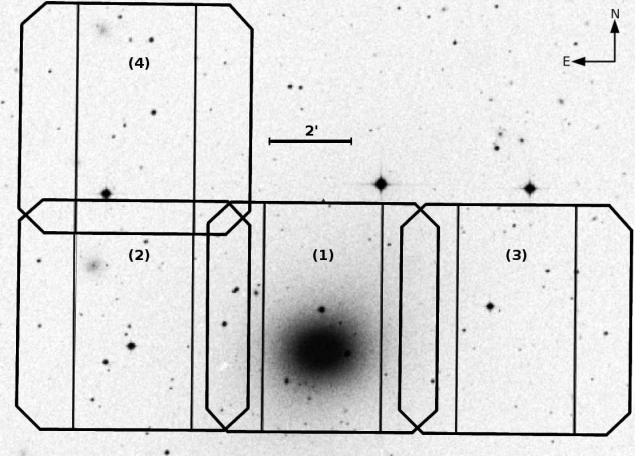}}
\caption{SDSS image of NGC\,1395 showing the four GMOS science fields. Fields 1 and 2 correspond to the observation 
programm GS-2012B-Q-44, and fields 3 and 4 to GS-2014B-Q-28.}
\label{figure1}
\end{figure}

The study of NGC\,1395 (Table \ref{Tab0}) was started using images from the camera Gemini Multi-Object Spectrograph (GMOS) mounted on the Gemini South telescope. Six fields were observed in the $g'r'i'z'$ filters \citep{fukugita96} with a binning of $2\times2$, giving a scale of 0.146 arcsec pixel$^{-1}$. Two of them correspond to comparison fields (see Section \ref{sec:comp}). Figure \ref{figure1} shows the distribution of the four observed fields around NGC\,1395. The images corresponding to fields 1 and 2 are part of Gemini programm GS-2012B-Q-44 (PI: J. C. Forte), while fields 3 and 4 correspond to the programm GS-2014B-Q-28 (PI: L. Sesto). The latter programm presents shorter exposure times compared with the first one. This difference is originated as a result of the upgrade of the GMOS detector array (E2V to Hamamatsu CCDs\footnote{http://www.gemini.edu/sciops/instruments/gmos/imaging/detector-array}) made in mid-2014. Table \ref{Tab_datos} shows a brief summary of the log of the observations used in this work. 

The reduction process of the data was performed in the usual way (see, for example, \citealp{faifer11,escudero15,sesto16}), using specific Gemini/GMOS routines within IRAF\footnote{IRAF is distributed by the National Optical Astronomical Observatories, which are operated by the Association of Universities for Research in Astronomy, Inc., under cooperative agreement with the National Science Foundation} (e.g. {\sc{gprepare}}, {\sc{gbias}}, {\sc{giflat}}, {\sc{gireduce}}, {\sc{gmosaic}}, {\sc{gifringe}}, {\sc{girmfringe}}, {\sc{imcoadd}}). Appropriate bias and flat-field frames obtained from the Gemini Observatory Archive\footnote{https://archive.gemini.edu/searchform} (GOA) as part of the standard GMOS baseline calibrations, were applied to the raw data. Since the $i'$ and $z'$ frames showed a significant fringing pattern, the correction of this effect was performed using six blank sky images close to the dates of our data. These calibration images have exposure times of 300\,s and were also downloaded from the GOA. To create and subtract the fringe pattern in our science images, we used the {\sc{gifringe}} and {\sc{girmfringe}} tasks. After correcting cosmetic effects of the data, images corresponding to the same field and filter were combined using {\sc{imcoadd}}. In this way, we obtained a final set of images to be used for the photometric analysis. The final values of FWHM are listed in Table \ref{Tab_datos}. 

In order to analyze the stellar population content of NGC\,1395, we also used public data from the final release of 6dF Galaxy Survey \citep[DR3 6dFGS;][]{jones04,jones09}. Mainly used in several studies of large-scale structures, this survey presents spectra of 136\,304 sources observed in the near IR and optical bands, using the 6dF fiber-fed multi-object spectrograph. The spectra were obtained by fibres with a beam size of 6.7 arcsec in diameter ($\sim\,R_\mathrm{eff}/10$ for NGC\,1395), and with the gratings $V$ (3900-5600\,\AA) and $R$ (5400-7500\,\AA). The resolution obtained in the final spectra is 5.8\,\AA\,with an average S/N$\geqslant$12\,\AA$^{-1}$. The spectrum of NGC\,1395 used in this work was classified as high-quality redshift ($Q=4$, \citealt{jones04}).

\begin{table*}
\centering
\caption{Set of observations of NGC\,1395. The values of airmass, exposure times and FWHM correspond to the final co-added images.} 
\label{Tab_datos}
\begin{tabular}{@{}lc@{}c@{}ccc@{}c@{}c@{}cccccc@{}c@{}c@{}c@{}}
\toprule
\toprule
\multicolumn{1}{@{}c}{\textbf{Science Data}} &
\multicolumn{1}{c}{\textbf{Gemini ID}} &
\multicolumn{1}{c}{\textbf{Field}} &
\multicolumn{1}{c}{\textbf{RA}} &
\multicolumn{1}{c}{\textbf{DEC}} &
\multicolumn{4}{c}{\textbf{Airmass}} &
\multicolumn{4}{c}{\textbf{$t_\mathrm{exp}$(s)}} &
\multicolumn{4}{c}{\textbf{FWHM ($''$)}} \\
\multicolumn{3}{c}{} &
\multicolumn{1}{c}{(h:m:s)} &
\multicolumn{1}{c}{(d:m:s)} &
\multicolumn{1}{c}{$g'$} &
\multicolumn{1}{c}{$r'$} &
\multicolumn{1}{c}{$i'$} &
\multicolumn{1}{c}{$z'$} &
\multicolumn{1}{c}{$g'$} &
\multicolumn{1}{c}{$r'$} &
\multicolumn{1}{c}{$i'$} &
\multicolumn{1}{c}{$z'$} &
\multicolumn{1}{c}{$g'$} &
\multicolumn{1}{c}{$r'$} &
\multicolumn{1}{c}{$i'$} &
\multicolumn{1}{c}{$z'$} \\
\multicolumn{1}{c}{(1)} &
\multicolumn{1}{c}{(2)} &
\multicolumn{1}{c}{(3)} &
\multicolumn{1}{c}{(4)} &
\multicolumn{1}{c}{(5)} &
\multicolumn{4}{c}{(6)} & 
\multicolumn{4}{c}{(7)} &
\multicolumn{4}{c}{(8)} \\
\midrule
NGC\,1395 & GS-2012B-Q-44  & 1  & 03:38:29.75 & -23:00:53 & 1.40 & 1.33 & 1.28 & 1.22  &  4$\times$180 & 4$\times$120 & 4$\times$150 & 6$\times$400  & 0.9 & 0.8 & 0.7 & 0.8 \\
           & GS-2012B-Q-44  & 2  & 03:38:49.53 & -23:00:53 & 1.16 & 1.12 & 1.10 & 1.07  &  4$\times$180 & 4$\times$120 & 4$\times$150 & 6$\times$400  & 1.1 & 1.1 & 1.0 & 1.0 \\
	   & GS-2014B-Q-28  & 3  & 03:38:09.17 & -23:00:54 & 1.10 & 1.13 & 1.15 & 1.18  &  6$\times$130 & 4$\times$100 & 4$\times$100 & 4$\times$400  & 0.9 & 0.8 & 0.8 & 0.7 \\
	   & GS-2014B-Q-28  & 4  & 03:38:48.09 & -22:56:00 & 1.19 & 1.23 & 1.26 & 1.31  &  4$\times$130 & 4$\times$100 & 4$\times$100 & 4$\times$400  & 0.6 & 0.6 & 0.5 & 0.6 \\
Comp. Field 1 & GS-2012B-Q-44  & -  & 03:42:18.18 & -23:02:58 & 1.14 & 1.10 & 1.08 & 1.06  &  4$\times$180 & 4$\times$120 & 4$\times$150 & 4$\times$400  & 1.1 & 0.8 & 0.8 & 0.7 \\
Comp. Field 2 & GS-2014B-Q-28  & -  & 03:31:40.30 & -22:21:31 & 1.01 & 1.09 & 1.11 & 1.15  &  7$\times$130 & 4$\times$100 & 4$\times$100 & 4$\times$400  & 0.9 & 0.8 & 0.7 & 0.8 \\
\bottomrule
\end{tabular}
\end{table*}


\section{Photometric Data Handling}
\label{sec:phot}
\subsection{Photometry}
\label{photometry}
According to the distance assumed for NGC\,1395 (Table \ref{Tab0}), it is expected that GCs are observed as unresolved objects. In order to detect as much sources as possible in our Gemini/GMOS images, we used a script that combines the detection capabilities of the SE{\sc{xtractor}} software \citep{bertin96} together with {\sc{iraf}} median filters. This allows us to model and subtract the sky background and the galaxy light, providing a better detection and subsequent estimation of the magnitudes of sources towards the inner region of the galaxy. From a list of initial reference objects, the script also provides a final catalogue for each field with all the common objects detected in the four filters. In our case, we used as a reference the object list obtained from the $r'$ images, since they present the highest number of detections in comparison with the $g', i'$ and $z'$ frames. 

The {\sc{daophot}} package (Stetson 1987) within {\sc{iraf}} was used to model the point spread function (PSF) on each science field. Each PSF model was built through the selection of $\sim30$ bright unresolved sources uniformly distributed over the GMOS field of view. The separation between resolved and unresolved sources was carried out using the parameter {\sc{class star}} of SE{\sc{xtractor}}. The classification is performed by a neural network assigning values between 0 (resolved objects) and 1 (unresolved objects). In this work we have considered 0.5 as the reference value to distinguish between resolved and unresolved sources. In addition, we used the PSF Extractor (PSFEx; \citealt{bertin11}) software in order to evaluate {\sc{daophot}} performance in the PSF modeling, and we also test a new galaxy/star classifier called {\sc{spread model}} (see Appendix \ref{DAO_PSF}). The value adopted in this case to separate resolved from unresolved sources was $0.0035$.

The instrumental psf magnitudes for each object were obtained with the {\sc{allstar}} task using the psf models previously built. In addition, we conducted an aperture correction to these magnitudes using the {\sc{mkapfile}} task. This procedure was applied to all filters. As a final step, we built a master photometric catalogue with all the detected unresolved objects in each field.

In order to transform the instrumental magnitudes to the standard system, we used standard star fields observed during the same nights as Field 1 (E2V CCDs) and Field 4 (Hamamatsu CCDs). These images were taken as part of the observing program mentioned in Section \ref{sec:obs}. The expression used to calibrate our data was:
\begin{eqnarray}
\label{cero}
 m_{std} = m_{zero} + m_{inst} - K_{CP} (X-1)  + CT (m_1-m_2)_{std},
\end{eqnarray}
\noindent where $m_{std}$ are the standard magnitudes, $m_\mathrm{zero}$ is the photometric zero point, $m_{inst}$ are instrumental magnitudes, $K_\mathrm{CP}$ is the mean atmospheric extinction at Cerro Pach\'on, $X$ is the airmass, CT is the coefficient of the colour term and $(m_1-m_2)$ are the colours indicated in the fifth column of Table\,\ref{Tab_cero}. Owing to the small number of standard stars present in our standard fields, it was not possible to obtain reliable values for the colour terms. Therefore, we have adopted the values published in \citet{escudero15}, since they were obtained with the same instrument and the same filters (Table \ref{Tab_cero}). The values obtained by these authors are similar to those given by the Gemini Observatory. 

The photometric zero-points obtained for fields 1 and 4 are listed in Table \ref{Tab_cero}. The data corresponding to fields 2 and 3 were calibrated using common sources between these fields and Field 1. Subsequently, the zero-points differences between Field 2 and Field 4 were obtained, resulting in values lower than 0.07 mag in all filters. These small differences were applied to Field 4 to homogenize our photometry to the field containing the galaxy. Finally, we applied to our catalogue the Galactic extinction coefficients given by \citet{schlafly11} (sixth column in Table \ref{Tab_cero}).

\begin{table}
\centering
\caption{Coefficients used in the transformation to the standard system. m$_\mathrm{zero}^{*}$: final values of zero-point for the E2V and Hamamatsu detectors. CT: colour term obtained by \citet{escudero15}. A$_{\lambda}$: Galactic extinction coefficients in each filter given by \citet{schlafly11}.}
\label{Tab_cero}
\begin{tabular}{@{}cccccc@{}}
\toprule
\toprule
\multicolumn{1}{l}{\textbf{Filter}} &
\multicolumn{1}{c}{\textbf{m$_\mathrm{zero}^{*}$(E2V)}} &
\multicolumn{1}{c}{\textbf{m$_\mathrm{zero}^{*}$(Ham)}} &
\multicolumn{1}{c}{\textbf{CT}} &
\multicolumn{1}{c}{\textbf{(m$_1$-m$_2$)}} &
\multicolumn{1}{c}{\textbf{A$_{\lambda}$}} \\
\midrule
$g'$ &  28.25$\pm$0.02 & 28.09$\pm$0.03 & ~0.08$\pm$0.02  & $(g'-r')$ & 0.076 \\
$r'$ &  28.35$\pm$0.03 & 28.31$\pm$0.03 & ~0.03$\pm$0.01  & $(g'-r')$ & 0.053 \\
$i'$ &  27.95$\pm$0.03 & 28.29$\pm$0.03 & -0.02$\pm$0.05  & $(r'-i')$ & 0.039 \\
$z'$ &  26.84$\pm$0.02 & 28.13$\pm$0.02 & ~0.0            & $(i'-z')$ & 0.029 \\
\bottomrule
\end{tabular}
\end{table}


\subsection{Completeness test}
\label{sec:test}
We performed completeness tests in each science field, in order to determine reliable limits in magnitude in our photometric catalogue. This was carried out adding 200 artificial sources in intervals of 0.2 mag in the magnitude range $21<r'_0<29$ mag. A total number of 8000 sources were added to the images using the IRAF tasks {\sc{starlist}} and {\sc{addstar}}. To simulate the spatial distribution of GCs around NGC\,1395, a power law function (i.e. $r^{-1}$) was considered in the field that contains the galaxy (Field 1 in Figure \ref{figure1}), and a uniform spatial distribution in the remaining fields. Subsequently, the recovery of these artificial sources was done using the identification script mentioned in Section\,\ref{photometry}. 

\begin{figure}
\resizebox{0.9\hsize}{!}{\includegraphics{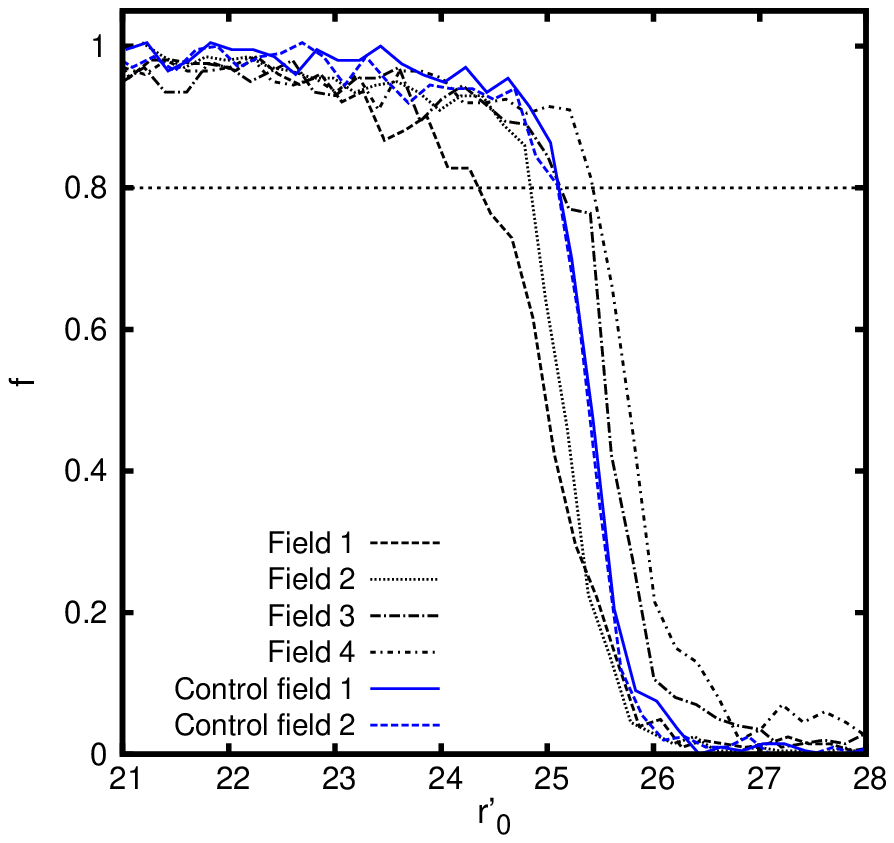}}
\resizebox{0.9\hsize}{!}{\includegraphics{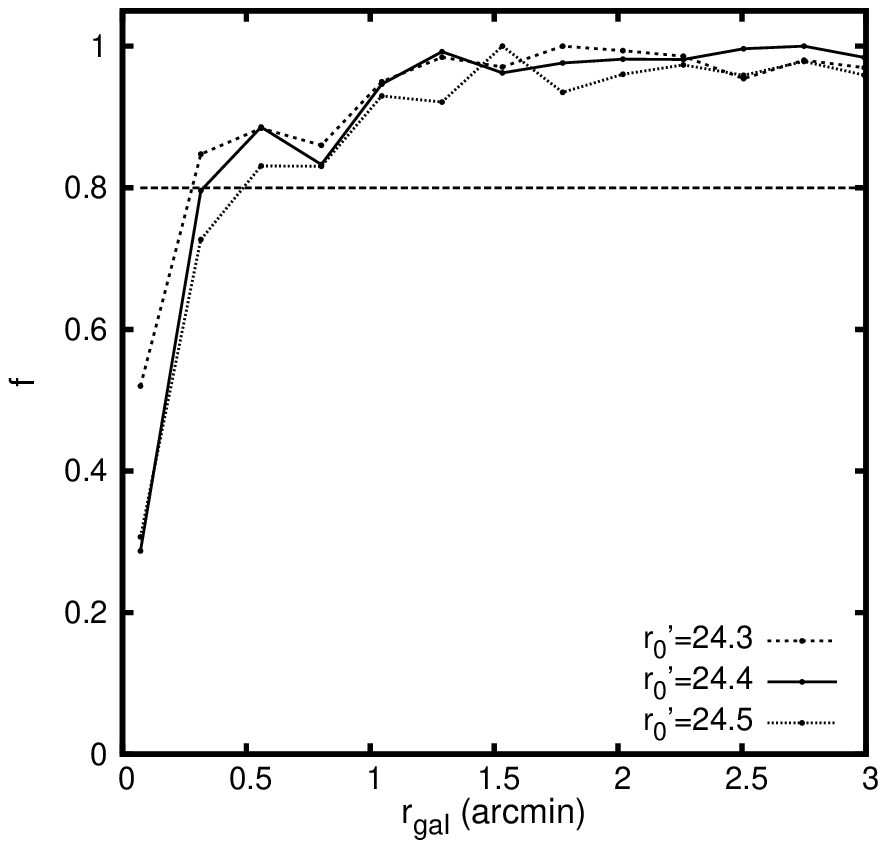}}
\caption{{\it Top panel:} completeness fraction as a function of $r'_0$ magnitude. The solid and dashed blue lines show the completeness tests performed on the comparison fields used in this work. {\it Bottom panel:} completeness fraction as a function of galactocentric radius for field 1. The horizontal dashed line shows the 80 per cent completeness level.}
\label{figure2}
\end{figure}

Top panel in Figure\,\ref{figure2} shows the fraction of recovered objects ({\it f}) as a function of the reference magnitude. This figure shows that our GMOS mosaic has a completeness level greater than 80 per cent at $r'_0\sim25$ mag, decreasing to $r'_0\sim24.4$ mag for the case of Field 1 (the one containing NGC\,1395). On the other hand, this figure also shows that the fields that were observed with the Hamamatsu detectors (Field 3 and Field 4), are slightly deeper than those observed with the E2V CCDs. This difference is due to the new CCDs which have a better quantum efficiency and an improved red sensitivity.

Also, we perform completeness tests as a function of galactocentric radius ($r_\mathrm{gal}$) for different limiting magnitudes. As we have found in previous works \citep[e.g.,][]{escudero15}, there is a strong spatial dependence in the completeness, mainly when we move towards the inner region of the galaxy. In order to improve the detection of objects in that region, we initially used the {\sc{ellipse}} model obtained in Section\,\ref{sec:elipse} and, subsequently, the detection script mentioned in Section\,\ref{photometry}. This procedure allowed us to detect a significant number of objects in the central region of the galaxy up to 0.33 arcmin (bottom panel in Figure \ref{figure2}). At $r'_0\sim24.4$ mag, our photometry is 80 per cent complete, only losing objects within the radius mentioned above.

\begin{figure}
\centering
\resizebox{0.95\hsize}{!}{\includegraphics{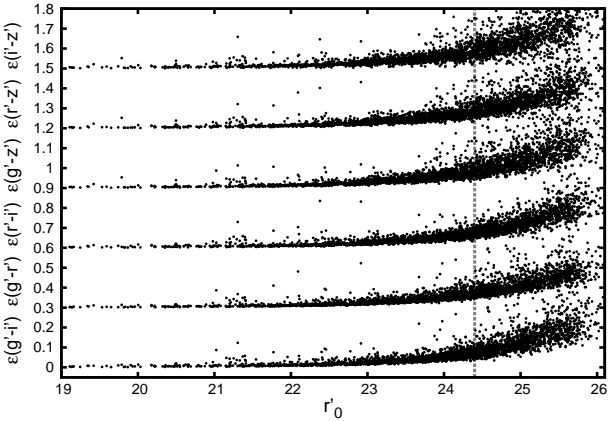}}
\caption{Photometric errors for different colour indices as a function of the $r'$ magnitude (arbitrarily shifted upwards). Vertical dashed line indicates the limiting magnitude $r'_0=24.4$ mag adopted in this work.}
\label{figure3}
\end{figure}

Observing the photometric errors of the colour indices as a function of the magnitude $r'$ for all the unresolved sources (Figure\,\ref{figure3}), we obtain that at $r'_0=24.4$ mag the median colour errors are $\lesssim0.1$ mag. Therefore, we have adopted this value as a lower limit in magnitude for the subsequent analysis.

\subsection{Comparison Fields}
\label{sec:comp}
The estimation of the contamination level in our data was carried out using two comparison fields (Table\,\ref{Tab_datos}), corresponding to the programms GS-2012B-Q-44 and GS-2014B-Q-28, respectively. They were observed in the same filters and with the same exposure times as fields 1-2 and 3-4, respectively. 

The reduction process of the images and the photometry performed over the unresolved sources ({\sc{class star}}$>$0.5) detected in these fields, were carried out following the guidelines presented in Sections\,\ref{sec:obs} and \ref{sec:phot}. Although these fields were observed in different nights, with different integration times and with different CCD cameras, they turned out to be photometrically similar (see the top panel of Figure\,\ref{figure3}).

\begin{figure}
\resizebox{0.99\hsize}{!}{\includegraphics{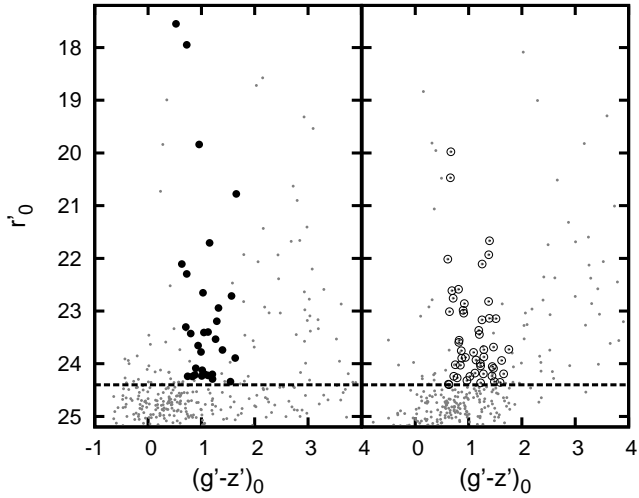}}
\caption{Colour-magnitude diagrams with all unresolved sources (grey dots) detected in the two comparison fields (programms GS-2012B-Q-44 and GS-2014B-Q-28, respectively). Closed and open black circles show the objects with colours and magnitudes inside the ranges adopted for the GC candidates in NGC\,1395. The horizontal dashed line depicts $r'_0=24.4$ mag which corresponds to a 80 per cent completeness level in our science fields.}
\label{figure4}
\end{figure}

In Figure\,\ref{figure4} we show the colour-magnitude diagram $r'_0$ versus $(g'-z')_0$ with all the unresolved sources detected in both fields. This figure shows a low number of unresolved sources with magnitudes $r'_0<24.4$ mag and with colour ranges similar to those adopted for GC candidates (see Section\,\ref{sec:color}). Most of these objects correspond to medium-high redshift galaxies and foreground stars \citep{fukugita95}. From the comparison fields, we estimate a low contamination in our GC sample of $1.36$ objects\,arcmin$^{-2}$ for $r'_0<24.4$ mag.

\section{NGC\,1395}
\label{sec:elipse}
\subsection{Surface brightness profile}
\label{sec:brightness_prof}
We obtained the surface brightness profiles of NGC\,1395 in the filters $g'$, $r'$, $i'$ and $z'$, in order to compare it with the spatial distribution of its associated GCs. To do this, we built a mosaic in each filter with the images of Field 1 and Field 2 corresponding to the E2V detectors. Using the objects in the overlapping area of these fiels, we equalized the signal and the sky level before combining them with SWarp \citep{bertin02}. 

We modelled the galaxy light using the IRAF tasks {\sc{ellipse}} and {\sc{bmodel}} on the $g'$ filter allowing the free variation of the center, ellipticity ($e$) and position angle ($PA$; measured from the north counterclockwise) of the isophotes during the fit. Subsequently, the obtained model was applied on the remaining filters. To avoid the light contribution from bright objects in the field, they were masked before performing the fit. Towards the outer regions of the galaxy ($\sim3.3$ arcmin) the isofotal parameters were fixed during the modelling, because the fit becomes unstable given the low surface brightness of the galaxy in these zones. 

\begin{figure}
\resizebox{0.99\hsize}{!}{\includegraphics{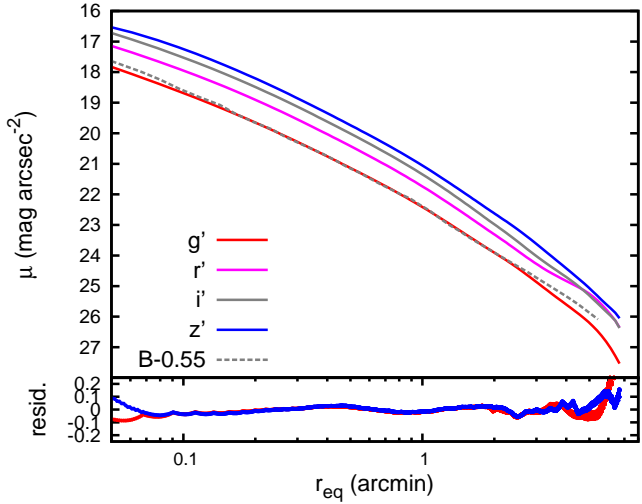}}
\caption{{\it Top panel:} surface brightness profiles of NGC\,1395 in the filters $g'$,$r'$,$i'$ and $z'$ (solid lines). As a comparison, we also included the $B$ profile from the CGS (dashed line), which was shifted vertically by adding a constant value. {\it Bottom panel:} residuals obtained from the fit of the S\'ersic function on the $g'$ and $z'$ profiles (red and blue lines).}
\label{figure5}
\end{figure}

At this point, an important factor to consider in the built of the surface brightness profiles is the determination and subtraction of the sky level. An underestimation or overestimation of such values can introduce spurious structures in the obtained profiles \citep{erwin08}. In our case, we considered the sky level using the counts of the outermost isophote of the model (SMA$\sim$7.7 arcmin).
Figure\,\ref{figure5} show the radial surface brightness profiles of NGC\,1395 as a function of the equivalent radius\footnote{$r_\mathrm{eq}=\sqrt{ab}=a\sqrt{1-e}$, with $a$ and $b$, semi-major and semi-minor axes of the ellipses} ($r_\mathrm{eq}$) in the different filters. 

In order to test the reliability of our GMOS profiles, we downloaded the brightness profiles of the galaxy in the $BVRI$ filters from the Carnegie-Irvine Galaxy Survey \citep[CGS;][]{li11}. Subsequently, they were transformed to the Sloan photometric system using the expression given by \citet{fukugita95} for E galaxies (their table 3). Both sets of profiles show a good agreement up to $r_\mathrm{eq}\sim4$ arcmin, after which a small difference is observed, possibly as a result of the sky values adopted in each case. As an example, Figure\,\ref{figure5} shows the agreement between the $B$ and $g'$ profiles.

With the aim at obtaining structural parameters of NGC\,1395, we fit to its brightness profiles the analytical expression of the S\'ersic function \citep{sersic68}, expressed in surface-brightness units (mag\,arcsec$^{-2}$):
\begin{equation}\label{eq:sersic}
\mu(r)= \mu_\mathrm{eff} + \left(\frac{2.5\,b_n}{\ln 10}\right)\left[\left(\frac{r}{R_\mathrm{eff}}\right)^{(1/n)}-1\right].
\end{equation}
Here, $R_\mathrm{eff}$ is the effective radius of the galaxy, $\mu_\mathrm{eff}$ is the surface brightness at that radius, $n$, the S\'ersic index, and $b_n=1.9992n-0.3271$. 

The resulting parameters in the different fits are listed in Table\,\ref{sersic_ajuste}. From these parameters, we also calculated the total magnitudes of NGC\,1395 in the different bands. The lower panel of Figure\,\ref{figure5} shows the residual of the fits in the $g'$ and $z'$ filters, exhibiting an excellent agreement between the model and the observed profile ($\Delta\,\mathrm{mag}<0.05$ up to $r_\mathrm{eq}\sim4.5$ arcmin). From our fits, we obtained for the effective radius of NGC\,1395, a mean value of $\langle R_\mathrm{eff} \rangle=1.07\pm0.03$ arcmin ($\sim$$7.1\pm0.2$ kpc). This value is similar to that obtained by \citet{li11} of 1.10 arcmin. However, it should be noted that the value obtained by these authors was estimated in different photometric bands.
For the subsequent analysis of the GC system of NGC\,1395, we will take our mean value as representative of the half-light radius of the galaxy.

\begin{table}
\centering
\caption{Values of $\mu_\mathrm{eff}$, $R_\mathrm{eff}$ and $n$ obtained by fitting S\'ersic profiles in the filters $g'$,$r'$,$i'$ and $z'$ to the surface brightness profiles of NGC\,1395. Last column shows the total magnitude ($m_\mathrm{tot}$) of the fitted profile.}
\label{sersic_ajuste}
\begin{tabular}{cllll}
\toprule
\toprule
\multicolumn{1}{c}{\textbf{Filter}} &
\multicolumn{1}{c}{\textbf{$\mu_\mathrm{eff}$}} &
\multicolumn{1}{c}{\textbf{$R_\mathrm{eff}$}} &
\multicolumn{1}{c}{\textbf{$n$}} &
\multicolumn{1}{c}{\textbf{$m_\mathrm{tot}$}}\\
\multicolumn{1}{c}{\textbf{}} &
\multicolumn{1}{c}{\text{(mag\,arcsec$^{-2}$)}} &
\multicolumn{1}{c}{\text{(arcmin)}} &
\multicolumn{1}{c}{\textbf{}} &
\multicolumn{1}{c}{\textbf{}} \\
\midrule
$g'$ &  22.90$\pm$0.02  & 1.26$\pm$0.02  & 5.12$\pm$0.08 & 9.98$\pm$0.02 \\
$r'$ &  21.86$\pm$0.03  & 1.04$\pm$0.01  & 4.97$\pm$0.08 & 9.35$\pm$0.05 \\
$i'$ &  21.27$\pm$0.03  & 0.97$\pm$0.02  & 5.11$\pm$0.12 & 9.00$\pm$0.05 \\
$z'$ &  21.15$\pm$0.02  & 1.03$\pm$0.01  & 4.99$\pm$0.07 & 8.68$\pm$0.04 \\
\bottomrule
\end{tabular}
\end{table}

\subsection{Isophotal analysis and tidal features}
\label{sec:tidal}
\begin{figure}
\resizebox{0.99\hsize}{!}{\includegraphics{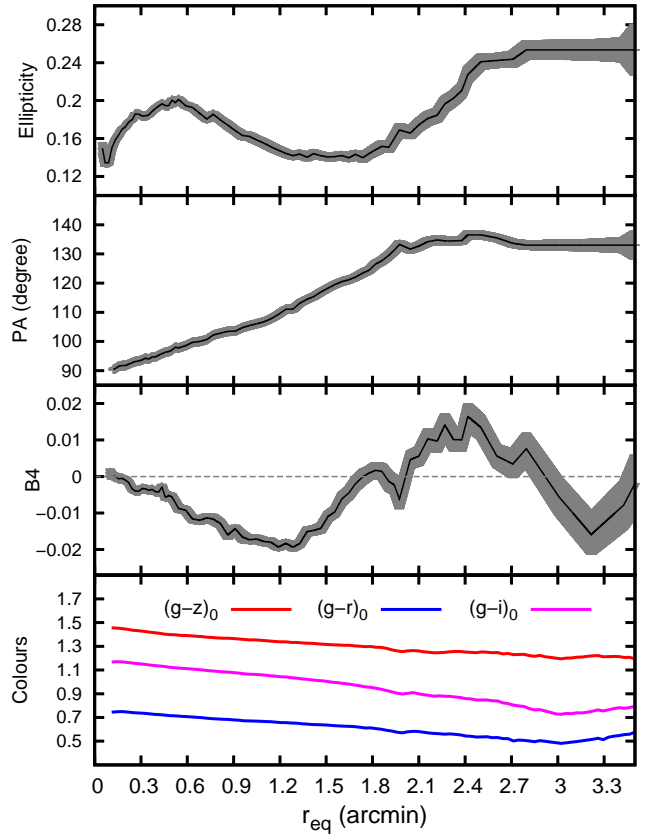}}
\caption{Variations of the isophotal parameters $e$, $PA$ and $B_4$ with its uncertainties (black line and shaded regions) in the filter $g'$ versus equivalent galactocentric radius. Last panel shows the colour indices $(g'-z')_0$, $(g'-r')_0$ and $(g'-i')_0$ as a function of the equivalent galactocentric radius (red, blue and magenta lines, respectively).}
\label{figure6}
\end{figure}

The analysis of the variations of the isophotal parameters of NGC\,1395 can provide important clues about possible merger and/or interaction events with objects of lower mass \citep{lane13,escudero15}. In Figure\,\ref{figure6} we show the variation of the ellipticity ($e$), position angle ($PA$), the Fourier coefficient $B_4$ and the colours $(g'-z')_0$, $(g'-i')_0$ and $(g'-r')_0$ as a function of $r_\mathrm{eq}$. The figure shows a significant variation mainly in $PA$ changing approximately 45$^\circ$ within 2.5 arcmin, while $e$ ranges from 0.14 to 0.24 in this region, after which both parameters remain constant. On the other hand, the coefficient $B_4$ indicates that the galaxy posses boxy isophotes up to $\sim2$ arcmin ($B_4<0$ and $B_4>0$ describe boxy and discy isophotes, respectively) and after $\sim$3 arcmin. The presence of boxy isophotos are generally related to recent events of interactions and/or mergers \citep{kormendy96}. The results obtained here are in good agreement with those of \citet{li11}. However, it is necessary to mention that our observations are photometrically deeper than that work, which allows us to obtain greater detail in the distribution of the surface brightness of NGC\,1395.

\citet{malin83} and \citet{tal09} mention that NGC\,1395 presents low contrast shells within its bright envelope, in a northwesterly direction. However, both papers do not present a characterization or description of them. Therefore, given the excellent quality of the GMOS data, we performed an analysis on our images with the aim at detecting some of these substructures. 

At first glance, it is possible to identify a faint shell in our original images without further processing, at a galactocentric radius of approximately 3 arcmin ($\sim$20 kpc) in the northwest direction. To highlight the shell due to its low brightness, and any other possible substructure present in the images, we subtracted the underlying galaxy light using the model in the $g'$ filter obtained with {\sc{ellipse}}. Then, we used SE{\sc{xtractor}} to detect and remove all the sources from the image, considering a high value of the parameter {\sc{back\_size}} ({\sc{back\_size}}=256), in order to avoid including small-scale structures in the background model. Subsequently, we applied the unsharp masking technique in the resulting image with a Gaussian kernel of $\sigma=80$ pixels ($\sim$0.2 arcmin) to highlight any possible remaining low surface brightness structure. 

\begin{figure}
\resizebox{0.95\hsize}{!}{\includegraphics{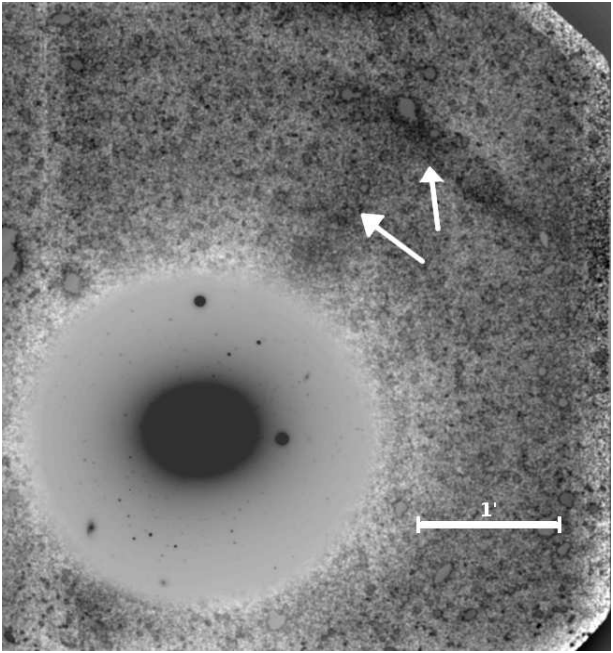}}
\caption{Image obtained after the subtraction of the sources present in the field and the subsequent smoothing by the unsharp masking technique. We have superimposed the $g'$ image of the galaxy for a better visualization of the different structures. White arrows indicate the location of a shell in NGC\,1395 at a galactocentric radius of $\sim3$ arcmin ($\sim$20 kpc), as well as a radial feature. North is up and East to the left.}
\label{figure7}
\end{figure}

Figure\,\ref{figure7} shows the location of the shell in the $g'$ image, which can also be seen in the $r'$ and $i'$ frames. In addition, a radial feature perpendicular to the shell is observed in the final image, forming an umbrella-like structure. This type of configuration likely created during the accretion of low-mass and gas-free galaxies on a nearly radial orbit \citep{tal09,sanderson13}, is frequently observed in early-type galaxies with stellar masses $>10^{10.5}$ $M\odot$ \citep{malin83,tal09,atkinson13,duc15,bilek16}. Since the dynamic age of these morphological disturbances seems to be $\sim0.5-3$ Gyr \citep{nulsen89,pop17}, the presence of these substructures in NGC\,1395 would indicate that the galaxy has recently experienced, at least, one minor merger. 

\section{GLOBULAR CLUSTER SYSTEM} 
\label{sec:GCS}
\subsection{Globular cluster colours}
\label{sec:color}
\begin{figure}
\centering
\resizebox{0.98\hsize}{!}{\includegraphics{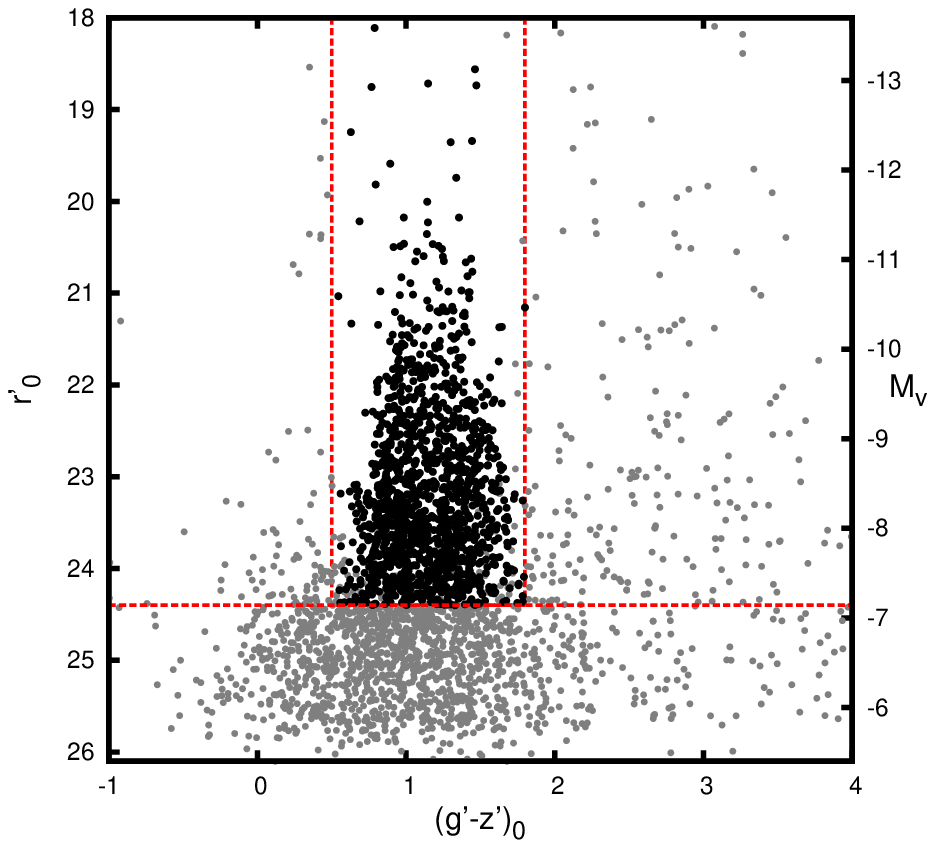}}
\resizebox{0.9\hsize}{!}{\includegraphics{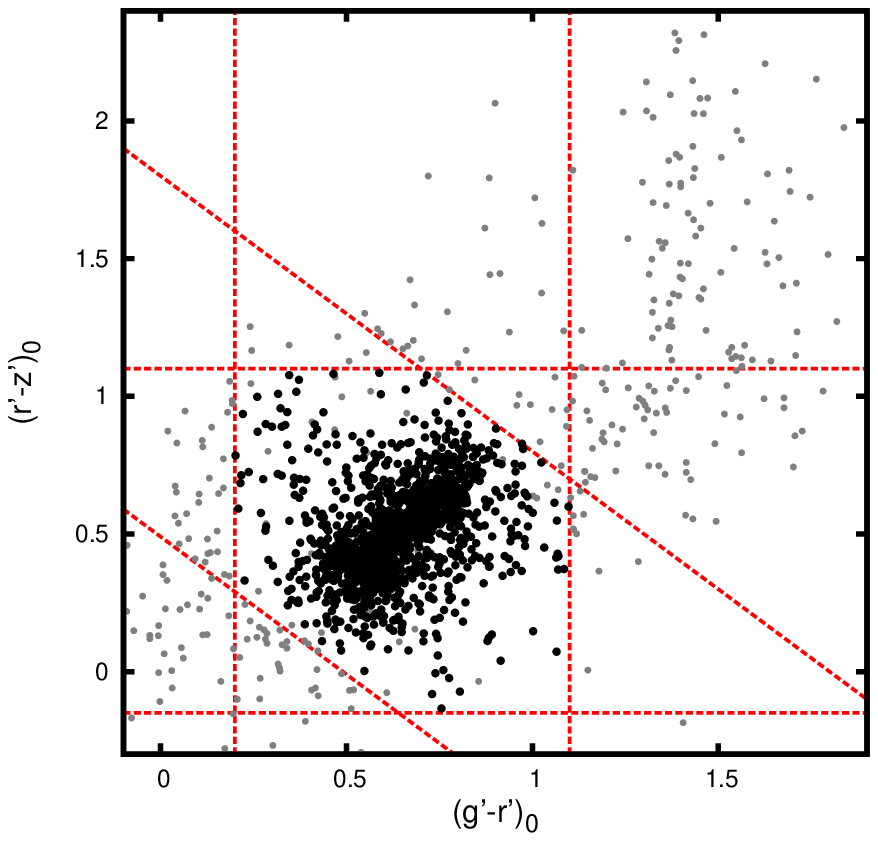}}
\caption{Colour-magnitude {\it (top panel)} and colour-colour {\it (bottom panel)} diagrams with the unresolved sources detected in the mosaic with magnitude $r'_0<24.4$ mag (grey and black points). Black points indicate the final sample of GC candidates selected from the different colour cuts. The dashed red lines show some of the colour ranges and magnitude used for the selection of GC candidates.}
\label{figure8}
\end{figure}

Figure\,\ref{figure8} shows the $r'_0$ versus $(g'-z')_0$ and the $(r'-z')_0$ versus $(g'-r')_0$ diagrams of the all unresolved sources detected in our GMOS mosaic. Since GCs are grouped around specific colours ranges in this type of diagrams, the selection of the GC candidates of NGC\,1395 was performed using the following colour cuts: \\
\noindent$0.2<(g'-r')_0<1.1$ mag; $0.4<(g'-i')_0<1.5$ mag; $0.5<(g'-z')_0<1.8$ mag; $-0.1<(r'-i')_0<0.7$ mag; $-0.15<(r'-z')_0<1.1$ mag; $-0.3<(i'-z')_0<0.75$ mag.
\noindent Some of these limits are plotted in Figure\,\ref{figure8} with vertical, horizontal and diagonal red dashed lines. The colour ranges chosen in this work result similar to that used by \citet{peng06}, \citet{faifer11} and \citet{escudero15} for early type galaxies, and they were adopted in order to obtain a sample of GC candidates as clean as possible. 
In addition, we chose those objects brighter than $r'=24.4$ mag, corresponding to the completeness analysis performed in Section\,\ref{sec:test}. On the other hand, we do not consider a magnitude cut toward the bright end, because we have not identified a significant number of Milky Way stars, massive clusters and/or ultra compact dwarf candidates in the colour-magnitude diagram.
The final number of GC candidates in our sample is 1328, and they are displayed with black points in both panels of Figure\,\ref{figure8}. The same colour and magnitude cuts were applied in the comparison fields, estimating $\sim8$ per cent of contamination due to background objects in our final sample.

To investigate the different GC subpopulations present in NGC\,1395, we used the Gaussian Mixture Modeling (GMM) code of \citet{muratov10}, on the different GC colour distributions. GMM is a statistical algorithm that allows us to calculate if the data are better represented by a unimodal or multimodal distribution, using three different statistics: (1) parametric bootstrap method (low values of $P(\chi^2)$ indicate a multi-modal distribution), (2) separation of the peaks between Gaussians ($D>2$ implies a multimodal distribution) and (3) kurtosis of the input distribution ($k<0$ condition necessary but not sufficient for bimodality). 

We run GMM on the different colour distributions considering the heteroscedastic (different dispersion) and homoscedastic (same dispersion) modes. Table\,\ref{gmm_2comp} shows the values of mode peaks ($\mu_n$), the dispersions ($\sigma_n$), the fraction of objects assigned to the blue GC subpopulation ($p_n$), and the statistical values $P(\chi^2)$, $D$ and $k$ obtained by GMM. As shown in this table, although both modes show a similar level of significance ($>$99 per cent), the homoscedastic fit may cause the position of the peaks to move away from their true value \citep{peng06,harris16}. 

\begin{figure}
\centering
\resizebox{0.49\hsize}{!}{\includegraphics{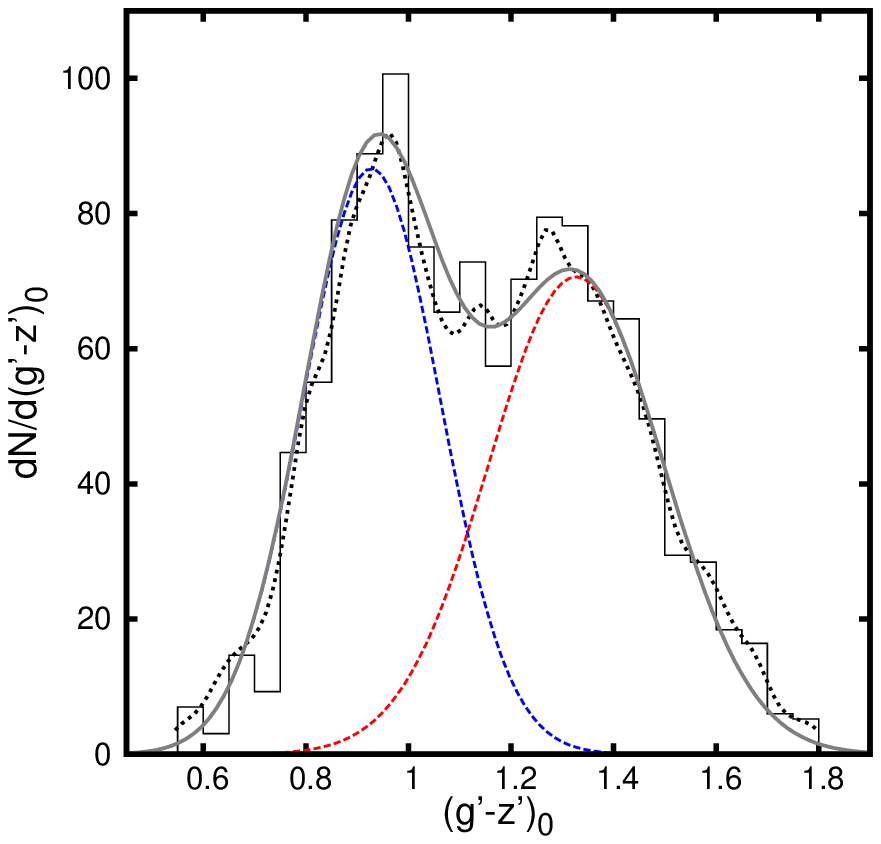}}
\resizebox{0.49\hsize}{!}{\includegraphics{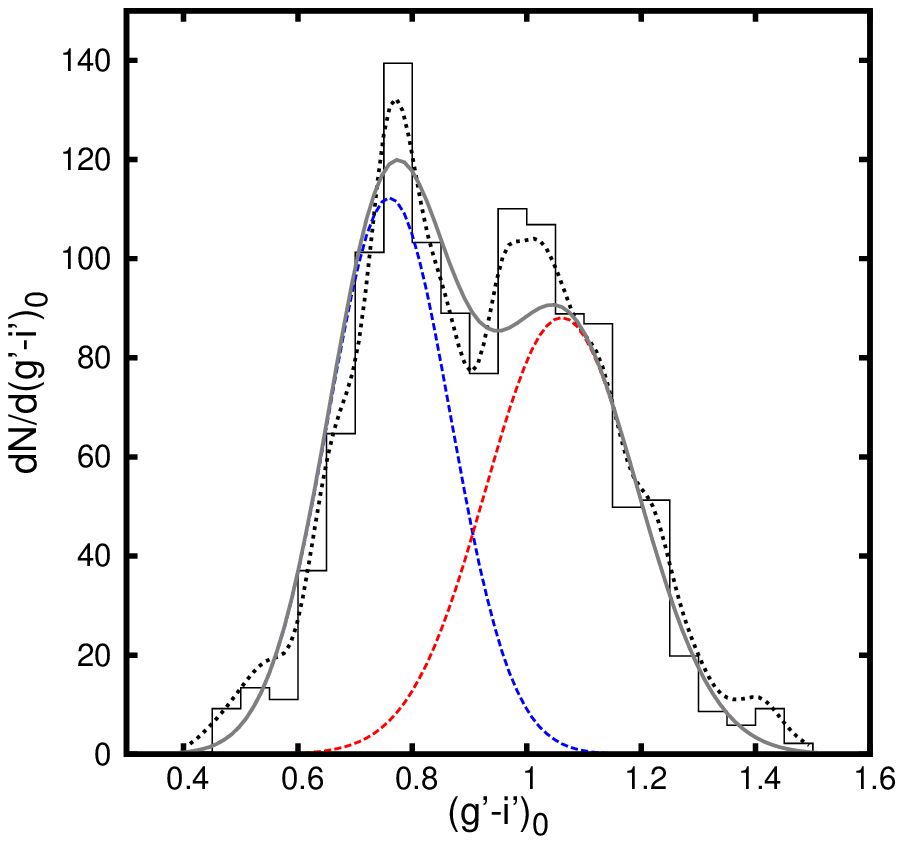}}
\resizebox{0.49\hsize}{!}{\includegraphics{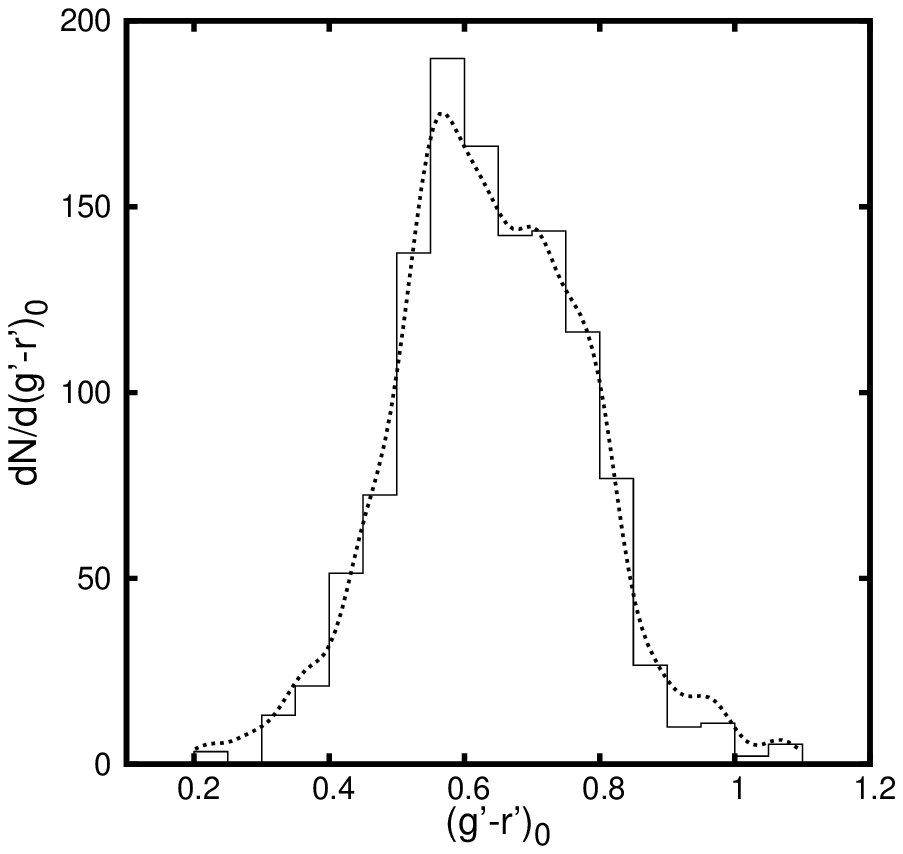}}
\resizebox{0.49\hsize}{!}{\includegraphics{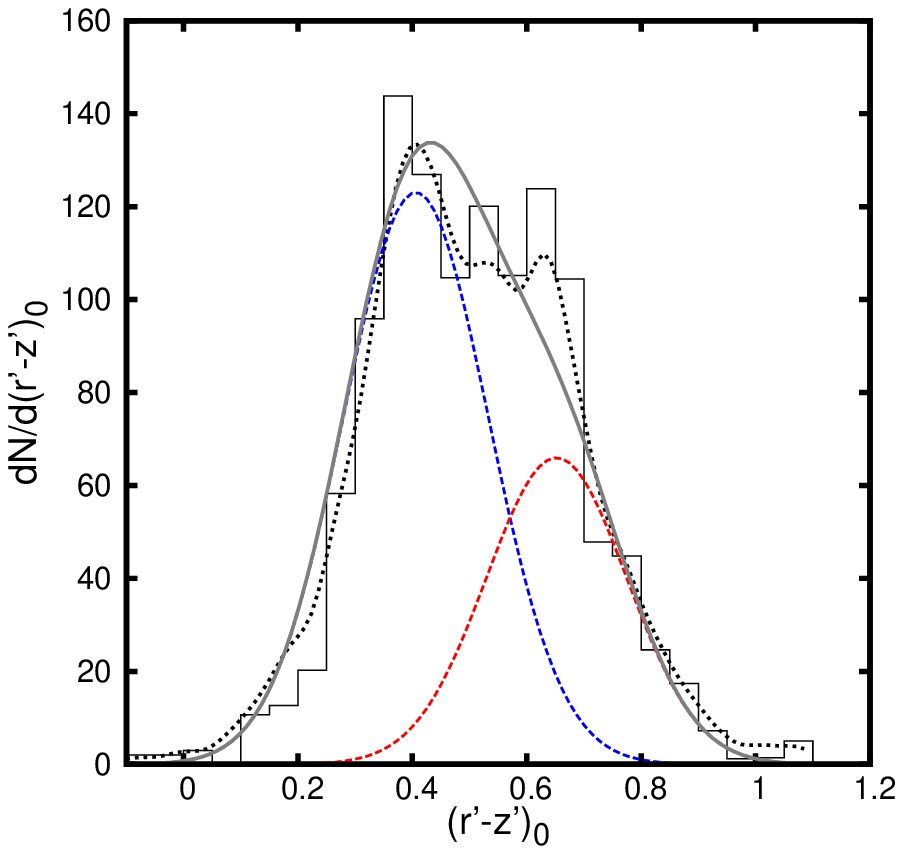}}
\resizebox{0.49\hsize}{!}{\includegraphics{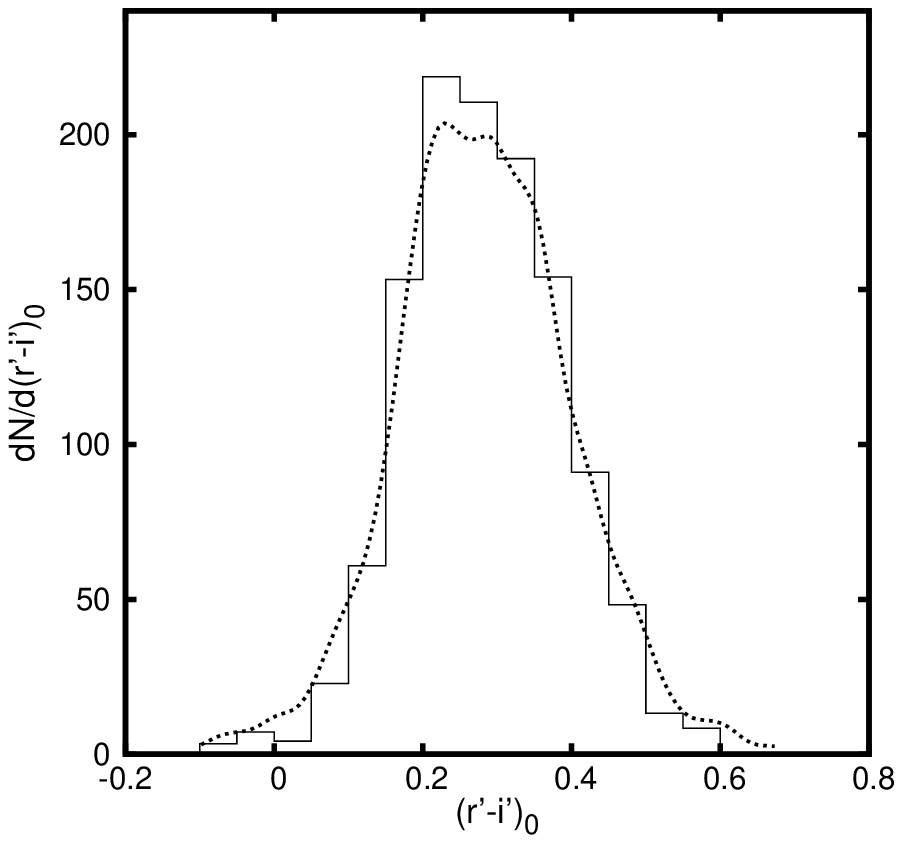}}
\resizebox{0.49\hsize}{!}{\includegraphics{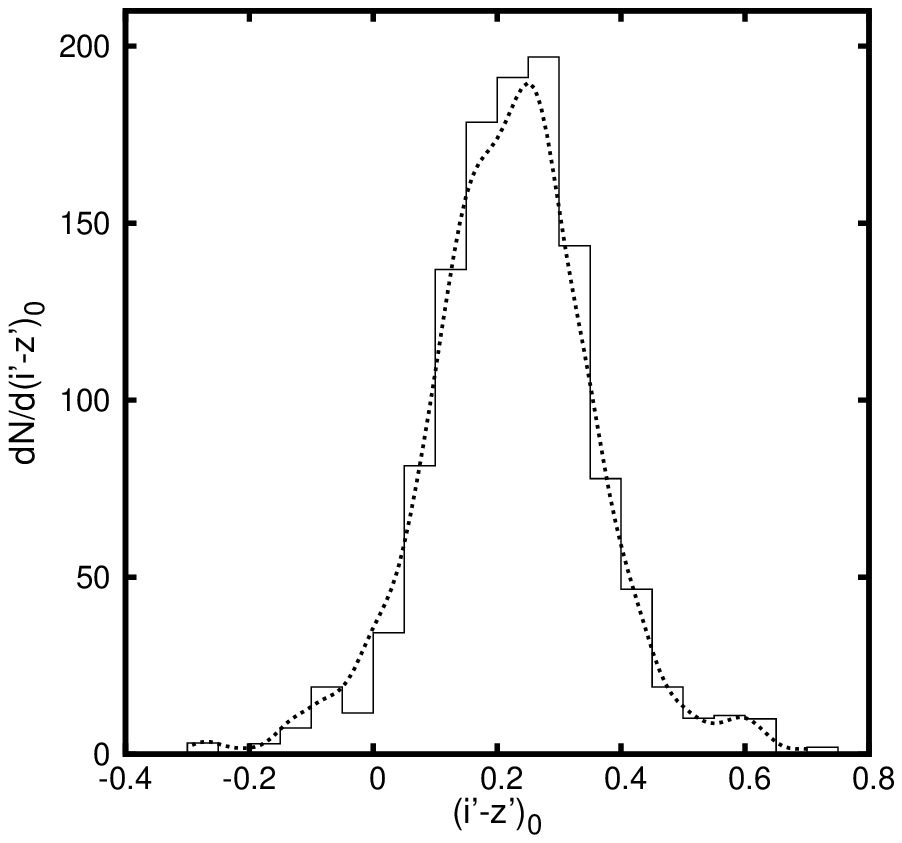}}
\caption{From left to right and from top to bottom, we show the colour histograms $(g'-z')_0$, $(g'-i')_0$, $(g'-r')_0$, $(r'-z')_0$, $(r'-i')_0$ and $(i'-z')_0$. The dotted black line represents the smoothed colour distribution. In the histograms that show evidence of bimodality, blue and red dashed lines show the best fit Gaussian components obtained by GMM for blue and red GC subpopulations. Their sum is depicted by the grey solid line.}
\label{figure9}
\end{figure}

In Figure\,\ref{figure9}, we show background-corrected colour histograms for different combinations of photometric bands. Each histogram was constructed adopting a bin size comparable with the maximum photometric errors in the considered colour. According to the values obtained in Table\,\ref{gmm_2comp}, the colours $(g'-z')_0$, $(g'-i')_0$ and $(r'-z')_0$ display clear evidence of a bimodal distribution, indicating the presence of at least two GC subpopulations (``blue'' or metal-poor and ``red'' or metal-rich) in NGC\,1395. In Figure\,\ref{figure9}, we show explicitly this both gaussian components only in the cases in which the bimodality is detected.

\begin{table*}
\centering
\caption{Values obtained with GMM for the different colour distributions considering heterocesdastic and homoscedastic modes (first and second line). The values of the mode peaks ($\mu_n$), dispersions ($\sigma_n$) and the fraction of objects assigned to the blue GC subpopulation ($p_n$) are listed along with the statistical values $P(\chi^2)$, $D$ and $k$.}
\label{gmm_2comp}
\begin{tabular}{ccccccccc}
\toprule
\toprule
\multicolumn{1}{c}{\textbf{Colour}} &
\multicolumn{1}{c}{\textbf{$\mu_b$}} &
\multicolumn{1}{c}{\textbf{$\sigma_b$}} &
\multicolumn{1}{c}{\textbf{$\mu_r$}} &
\multicolumn{1}{c}{\textbf{$\sigma_r$}} &
\multicolumn{1}{c}{\textbf{$f_b$}}  &
\multicolumn{1}{c}{\textbf{$P(\chi^2)$}} &
\multicolumn{1}{c}{\textbf{$D$}} &
\multicolumn{1}{c}{\textbf{$k$}} \\
\multicolumn{1}{c}{(mag)} &
\multicolumn{1}{c}{(mag)} &
\multicolumn{1}{c}{(mag)} &
\multicolumn{1}{c}{(mag)} &
\multicolumn{1}{c}{(mag)} &
\multicolumn{4}{c}{} \\
\midrule
$(g'-z')$   &  0.93$\pm$0.01  &  0.13$\pm$0.02  &  1.33$\pm$0.02  &  0.17$\pm$0.01  &  0.46$\pm$0.03  &  0.010  &  2.60$\pm$0.10  &  -0.83  \\
            &  0.96$\pm$0.01  &  0.15$\pm$0.01  &  1.36$\pm$0.01  &  0.15$\pm$0.01  &  0.54$\pm$0.02  &  0.010  &  2.60$\pm$0.09  &  -0.83  \\
$(g'-i')$   &  0.76$\pm$0.01  &  0.11$\pm$0.01  &  1.06$\pm$0.01  &  0.13$\pm$0.01  &  0.47$\pm$0.04  &  0.010  &  2.49$\pm$0.13  &  -0.73  \\
            &  0.78$\pm$0.01  &  0.12$\pm$0.01  &  1.08$\pm$0.01  &  0.12$\pm$0.01  &  0.55$\pm$0.02  &  0.010  &  2.49$\pm$0.12  &  -0.73  \\
$(g'-r')$   &  0.56$\pm$0.02  &  0.10$\pm$0.01  &  0.72$\pm$0.04  &  0.11$\pm$0.01  &  0.51$\pm$0.15  &  0.020  &  1.56$\pm$0.33  &  -0.27  \\
            &  0.57$\pm$0.01  &  0.10$\pm$0.01  &  0.74$\pm$0.01  &  0.10$\pm$0.01  &  0.61$\pm$0.06  &  0.010  &  1.56$\pm$0.18  &  -0.27  \\
$(r'-i')$   &  0.27$\pm$0.02  &  0.10$\pm$0.03  &  0.48$\pm$0.09  &  0.07$\pm$0.03  &  0.95$\pm$0.33  &  0.070  &  2.33$\pm$0.99  &   0.03  \\
            &  0.26$\pm$0.01  &  0.10$\pm$0.01  &  0.40$\pm$0.04  &  0.10$\pm$0.01  &  0.82$\pm$0.08  &  0.010  &  2.33$\pm$0.59  &   0.03  \\
$(r'-z')$   &  0.41$\pm$0.02  &  0.13$\pm$0.01  &  0.65$\pm$0.03  &  0.12$\pm$0.01  &  0.61$\pm$0.09  &  0.010  &  2.00$\pm$0.17  &  -0.40  \\
            &  0.40$\pm$0.01  &  0.12$\pm$0.01  &  0.65$\pm$0.01  &  0.12$\pm$0.01  &  0.58$\pm$0.03  &  0.010  &  2.00$\pm$0.16  &  -0.40  \\
$(i'-z')$   &  0.22$\pm$0.01  &  0.13$\pm$0.01  &  ---            &  ---             &  ---            &  ---   &  ---            &   0.50  \\
            &  0.21$\pm$0.01  &  0.13$\pm$0.01  &  0.24$\pm$0.01  &  0.13$\pm$0.01  &  0.65$\pm$0.06  &  0.010  &  0.03$\pm$0.07  &   0.50  \\
\bottomrule
\end{tabular}
\end{table*}

In order to examine in greater detail the GC colour distribution of NGC\,1395, and subsequently determine appropriate colour boundaries to separate both GC families, we considered subsamples of clusters with the same number of objects (332 candidates) at different galactocentric radii ($r_\mathrm{gal}<$80, 80-152, 152-296 and $>$296 arcsec). We again use GMM in the different radial bins considering the heteroscesdastic fit. Table\,\ref{gmm_2comp_rgal} lists the values obtained for each galactocentric range. 

\begin{figure}
\centering
\resizebox{0.49\hsize}{!}{\includegraphics{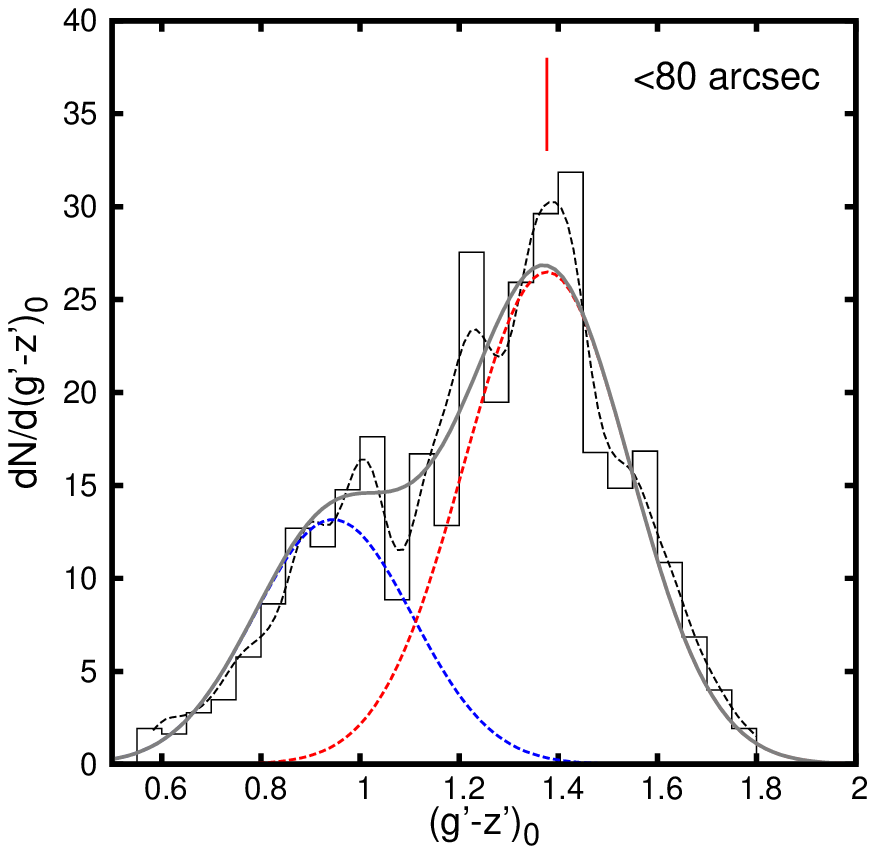}}
\resizebox{0.49\hsize}{!}{\includegraphics{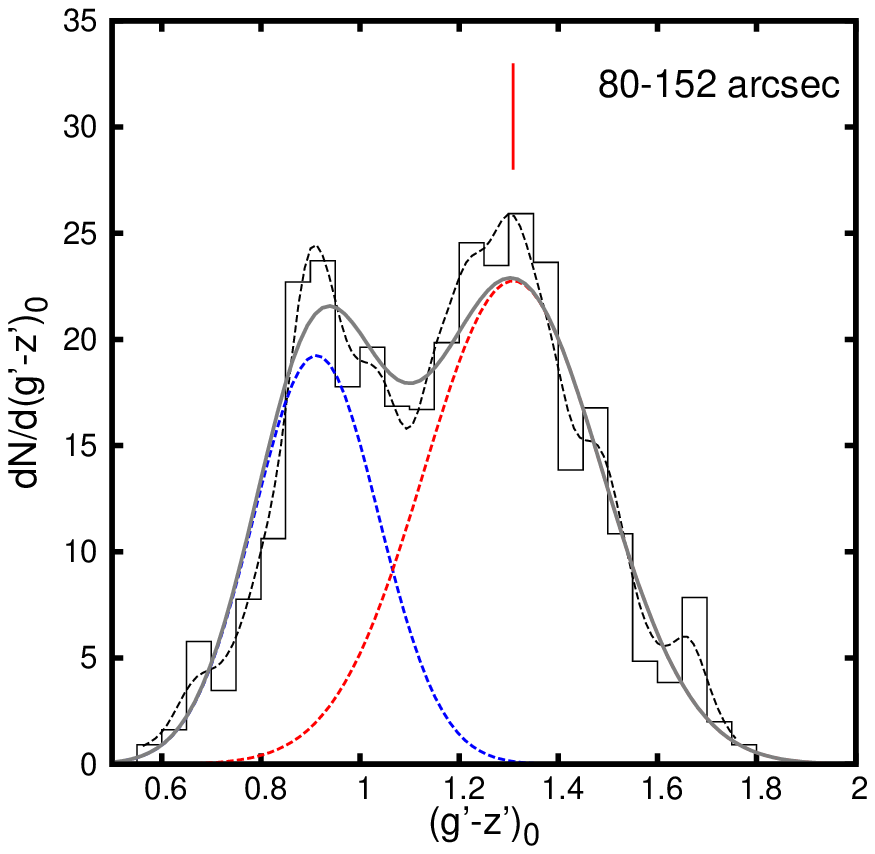}}
\resizebox{0.49\hsize}{!}{\includegraphics{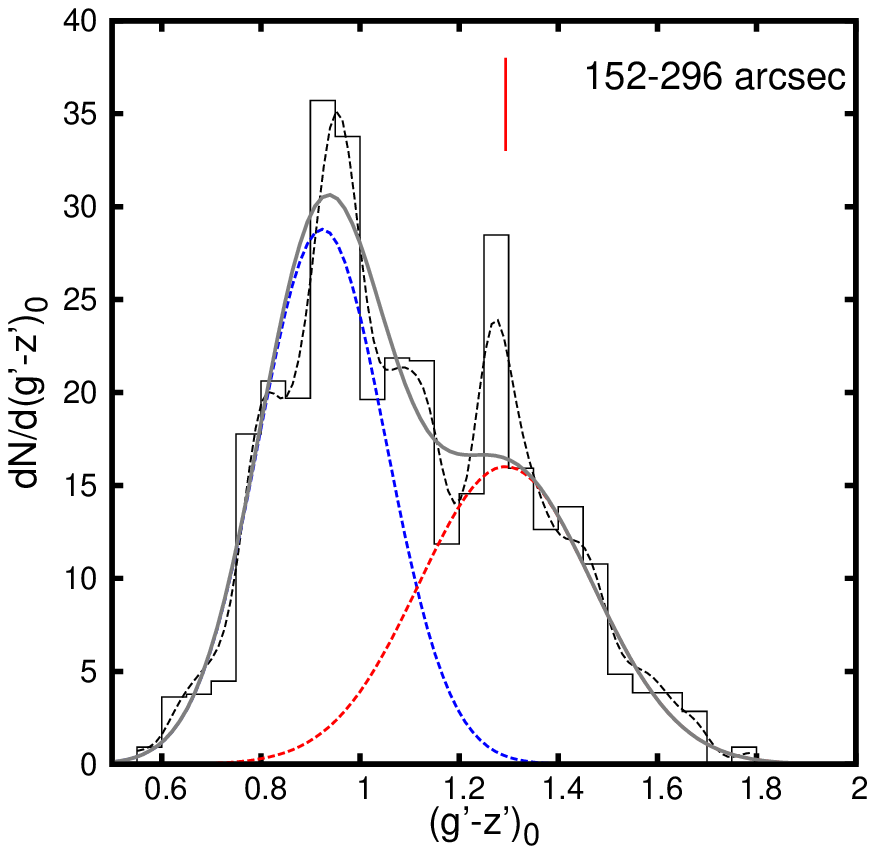}}
\resizebox{0.49\hsize}{!}{\includegraphics{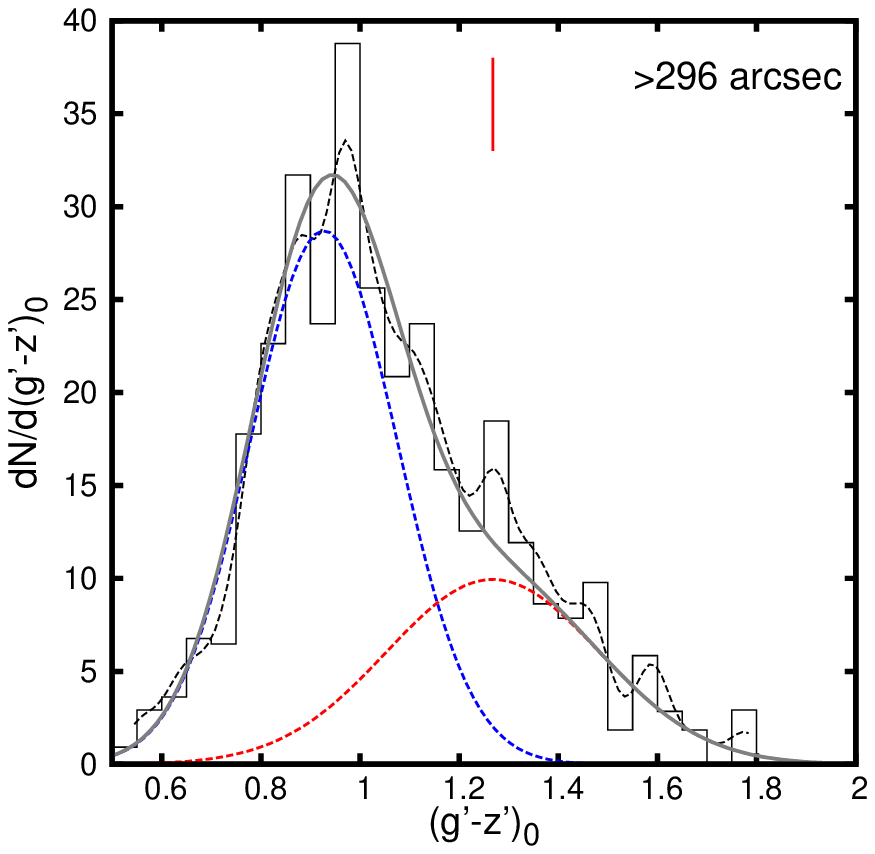}}
\caption{$(g'-z')_0$ colour distribution of the GC candidates at different galactocentric radii ($<$80, 80-152, 152-296 and $>$296 arcsec). The dashed black line represents the smoothed colour distribution. Blue and red dashed lines show the best-fit Gaussian components obtained by GMM for blue and red GC subpopulations. Their sum is depicted by the grey solid line. Vertical red lines indicate the mean position of the red peak obtained by GMM.}
\label{figure10}
\end{figure}

\begin{table*}
\centering
\caption{Values obtained by GMM for the $(g'-z')_0$ colour distributions at different galactocentric radii (0-80, 80-152, 152-296 and $>$296 arcsec).}
\label{gmm_2comp_rgal}
\begin{tabular}{ccccccccc}
\toprule
\toprule
\multicolumn{1}{c}{\textbf{$r_\mathrm{gal}$}} &
\multicolumn{1}{c}{\textbf{$\mu_b$}} &
\multicolumn{1}{c}{\textbf{$\sigma_b$}} &
\multicolumn{1}{c}{\textbf{$\mu_r$}} &
\multicolumn{1}{c}{\textbf{$\sigma_r$}} &
\multicolumn{1}{c}{\textbf{$f_b$}}  &
\multicolumn{1}{c}{\textbf{$P(\chi^2)$}} &
\multicolumn{1}{c}{\textbf{$D$}} &
\multicolumn{1}{c}{\textbf{$k$}} \\
\multicolumn{1}{c}{} &
\multicolumn{1}{c}{(mag)} &
\multicolumn{1}{c}{(mag)} &
\multicolumn{1}{c}{(mag)} &
\multicolumn{1}{c}{(mag)} &
\multicolumn{4}{c}{} \\
\midrule
 $<80$ arcsec     & 0.94$\pm$0.05 & 0.16$\pm$0.02 & 1.38$\pm$0.03 & 0.17$\pm$0.01 & 0.29$\pm$0.08 & 0.001 & 2.63$\pm$0.33 & -0.52  \\
 $80-152$ arcsec  & 0.91$\pm$0.03 & 0.13$\pm$0.03 & 1.31$\pm$0.03 & 0.18$\pm$0.02 & 0.35$\pm$0.07 & 0.001 & 2.55$\pm$0.25 & -0.74  \\
 $152-296$ arcsec & 0.92$\pm$0.02 & 0.13$\pm$0.01 & 1.29$\pm$0.04 & 0.17$\pm$0.02 & 0.53$\pm$0.08 & 0.001 & 2.40$\pm$0.33 & -0.56  \\
 $>296$ arcsec    & 0.93$\pm$0.03 & 0.15$\pm$0.02 & 1.27$\pm$0.08 & 0.21$\pm$0.03 & 0.61$\pm$0.14 & 0.001 & 1.84$\pm$0.51 & -0.03  \\
\bottomrule
\end{tabular}
\end{table*}

Figure\,\ref{figure10} shows that in all the radial histograms (corrected by background) the presence of typical blue and red GCs is clearly observed at the modal values $(g'-z')_0\sim0.95$ and $(g'-z')_0\sim1.35$ mag, respectively. As seen in other galaxies, at large galactocentric radii ($>$152 arcsec) the blue GCs begin to dominate the sample, while the red subpopulation dominates towards the inner region showing a broader colour distribution ($\sigma \sim 0.17-0.21$ mag) compared with the blue one.

Another feature observed in Figure\,\ref{figure10} is the shift in the mean position of the red peak towards blue colours as we move away from the galactic center. In order to clearly visualize and quantify this colour (metallicity) gradient, we obtained the mean colour of each subpopulation as a function of the normalized galactocentric radius ($r_\mathrm{gal}/R_\mathrm{eff}$), considering the $R_\mathrm{eff}$ value obtained in Section\,\ref{sec:elipse}. 

\begin{figure}
\centering
\resizebox{0.99\hsize}{!}{\includegraphics{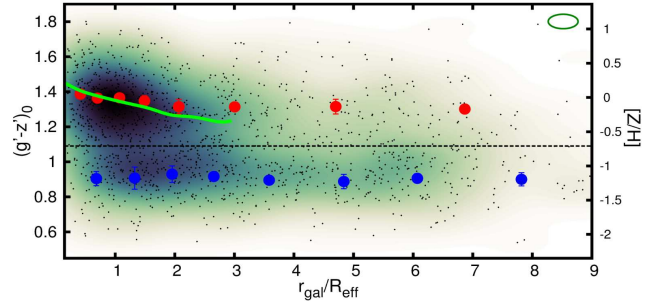}}
\caption{Smoothed colour $(g'-z')_0$ versus normalized projected galactocentric distance. Black dots show the GC candidates, and red and blue filled circles show the mean value of both subpopulations at different radial bins. The green solid line shows the inner colour profile of the halo of the galaxy. The elliptical green symbol depicts the kernel used for smoothing. The right axis shows the colour transformation into metallicity using the relation of \citet{usher12}.}
\label{figure11}
\end{figure}

We considered the separation between both GC families using the colour cut $(g'-z')_0=1.09$ mag, corresponding to the GMM value where an object has the same probability of belonging to the blue or red subpopulation (see Tables\,\ref{gmm_2comp} and \ref{gmm_2comp_rgal}). Then, we separated this two subsamples in several galactocentric bins with the same number of objects, and we obtained the mean colour for each bin. In this case, we consider 80 and 95 GCs for blue and red subsamples, respectively. The Figure\,\ref{figure11} shows the smoothed isocountor map of the GC radial colour distribution, the mean colours for each subpopulation, and the colour profile of the galaxy (green solid line) obtained in Section\,\ref{sec:elipse}. The smoothed colour distribution shows additional evidence of the presence of two subpopulations of GCs (blue and red already mentioned). The red subpopulation is more concentrated towards the center of the galaxy while the blue one display a more extended distribution. The red mean colours in each radial bin also reveal a colour gradient. In order to quantify these colour gradients, we fit the expression $(g'-z')_0=a+b\,log(r_\mathrm{gal}/R_\mathrm{eff})$ \citep{harris09} to the blue and red candidates in the ranges $r_\mathrm{gal}/R_\mathrm{eff}<8$ and $r_\mathrm{gal}/R_\mathrm{eff}<7$. The values obtained were $a=0.859\pm0.031$, $b=-0.038\pm0.026$ and $a=1.218\pm0.020$, $b=-0.084\pm0.012$, respectively. Using the colour-metallicity relation of \citet{usher12}: 
\begin{equation}\label{eq:met}
\begin{aligned}
 {[Z/H]} &= (2.56\pm0.09)\times(g-z) + (-3.50\pm0.11) \\
         &                            for \,(g-z)>0.84,
\end{aligned}
\end{equation}
we transform the colour slopes into metallicity gradients, obtaining $\Delta[Z/H]/log(r_\mathrm{gal}/R_\mathrm{eff})=-0.09\pm0.07$ dex per dex for the blue subpopulation, and $\Delta[Z/H]/log(r_\mathrm{gal}/R_\mathrm{eff})=-0.21\pm0.03$ dex per dex for the red one. These values are shallower when compared to those obtained by \citet{forbes11} for the E galaxy NGC\,1407 ($-0.38\pm0.06$ for the blue subpopulation and $-0.43\pm0.07$ for the red one) belonging to another subgroup of the Eridanus supergroup. However, they are comparable to those found in other massive galaxies located in different environments. For example, \citet{harris09,harris09b} found the following values for the GC subpopulations belonging to the galaxy NGC\,7626, located in the Pegasus group, and M87, located in the Virgo cluster: $-0.12\pm0.02$ (blue), $-0.17\pm0.03$ (red) and $-0.16\pm0.06$ (blue), $-0.18\pm0.07$ (red), respectively. \citet{faifer11} derived the values $-0.18\pm0.07$ (blue) and $-0.17\pm0.08$ (red) for the shell galaxy NGC\,3923.

The presence of these colour (metallicity) gradients, mainly evidenced by the red GC subpopulation within $\sim2\,R_\mathrm{eff}$ ($\sim$14\,kpc), and the subsequently flattening towards larger radii, suggest a two-phase assembly for both clusters families and, therefore, for NGC\,1395 \citep{forte09,oser10,forbes11}. In addition, Figure\,\ref{figure11} also shows that the inner colour of the halo of the galaxy is in good agreement with the average colour of the red clusters. This similarity suggests a strong link between them, indicating a likely joint stellar formation \citep{forte09,forte14}. It should be noticed that the absence of a significant metallicity gradient in the blue GC subpopulation suggests that this family would be comprised of a strong mixing of blue clusters formed {\it in situ} in the galaxy and accreted from different satellite galaxies \citep{harris16}.

\subsection{Spatial distribution}
\label{sec:dens}
\begin{figure}
\centering
\resizebox{0.95\hsize}{!}{\includegraphics{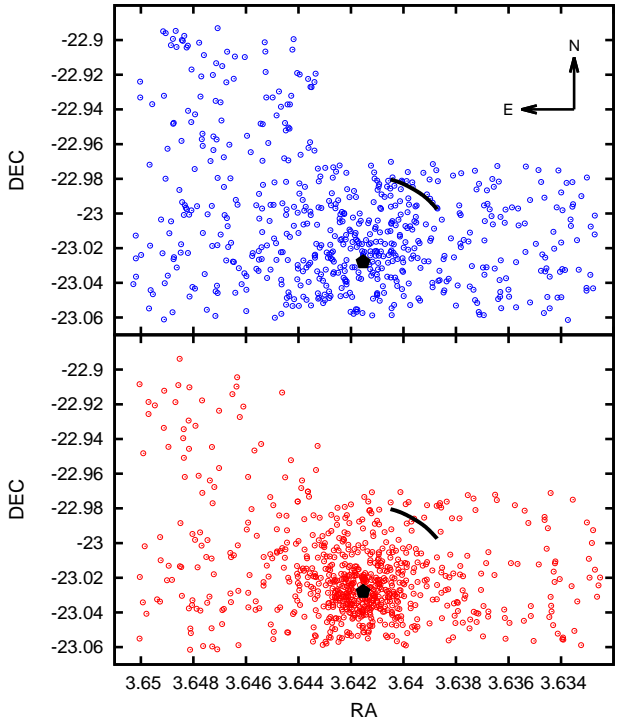}}
\caption{Spatial distribution of the blue {\it (top panel)} and red {\it (bottom panel)} GC candidates around NGC\,1395. 
Black pentagons indicate the position of the galactic center. Black strip indicates the location of the shell (see Section\,\ref{sec:tidal}).}
\label{figure12}
\end{figure}

As we mentioned in the previous section, we considered the colour value $(g'-z')_0=1.09$ mag to separate between the blue and red subpopulations. Figure\,\ref{figure12} shows the projected spatial distribution of the blue and red GC candidates of NGC\,1395 (top and bottom panels, respectively). Both families fill the entire GMOS mosaic, showing the typical spatial distribution found in early-type galaxies \citep{faifer11,escudero15}, with the red clusters more concentrated toward the galactic center in comparison with the blue ones.

\begin{figure}
\centering
\resizebox{0.99\hsize}{!}{\includegraphics{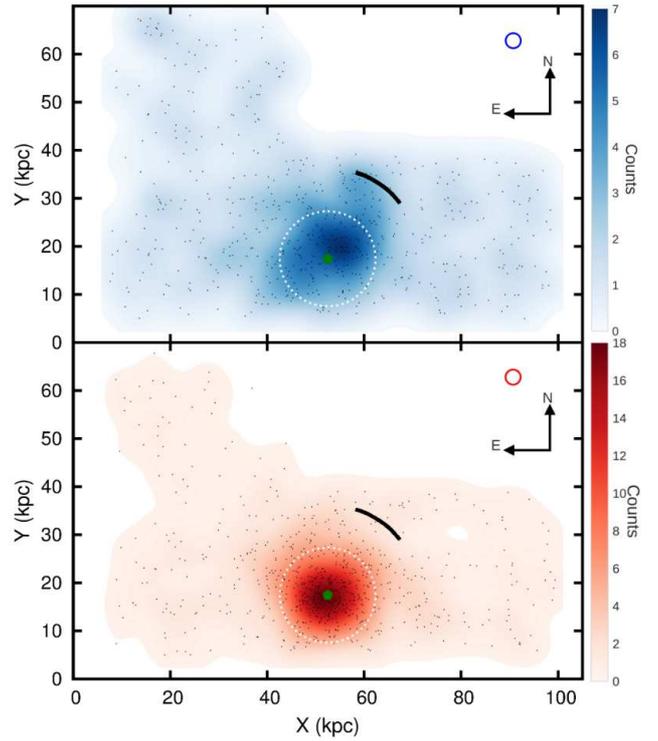}}
\caption{Smoothed spatial distribution of the blue {\it (top panel)} and red {\it (bottom panel)} GC candidates around NGC\,1395. The white dashed circle shows the region of 10 kpc of galactocentric radius where the asymmetry of GCs is observed. The green pentagons indicate the position of the galactic center. The circular blue and red symbols depict the kernel used for smoothing. Black strip indicates the location of the shell (see Section\,\ref{sec:tidal}). The colour bars indicate the number of objects within the kernel area used, in this case 0.19 arcmin$^{2}$.}
\label{figure13}
\end{figure}

In order to identify possible substructures or anisotropies in the spatial distribution of GCs, we built the smoothed density maps of the blue and red candidates (Figure\,\ref{figure13}). In addition to the different degree of concentration shown by both families, it is observed what seems to be a mild overdensity of objects ($r_\mathrm{gal}<1.6$ arcmin; $\sim10$ kpc) slightly offset from the galactic center, in a northwest direction towards the shell location (see Section\,\ref{sec:tidal}). 
\citet{dabrusco15} suggest that this type of overdensities may be driven by accretion of satellite galaxies, major dissipationless mergers, or wet dissipation mergers. Although the first case seems to fit better with the GC population results found for NGC\,1395, to confirm or discard this scenario it will be necessary to obtain spectroscopic data to study the kinematics and the stellar population of the GCs in this region.


\begin{figure}
\centering
\resizebox{0.99\hsize}{!}{\includegraphics{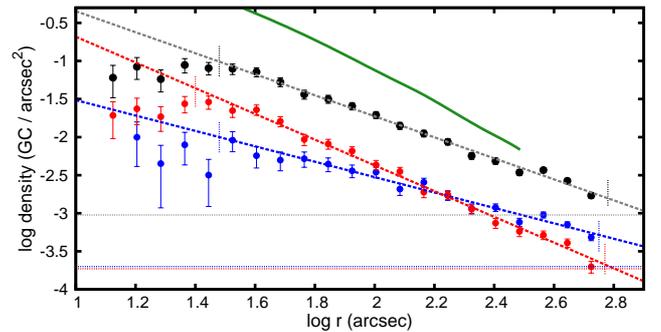}}
\caption{Projected density profiles for all GC candidates (black filled circles shifted $+0.4$ on the vertical axis to avoid overlapping) and for the blue and red subpopulations (blue and red filled circles). The blue, red and gray dashed lines show the fits obtained considering a power law function. Vertical dotted lines indicate the ranges used for the fits. The horizontal dotted grey (shifted $+0.4$ on the vertical axis), blue and red lines correspond to the background levels for the entire sample, blue and red GCs, respectively. The green solid line depicts the galaxy light profile.}
\label{figure14}
\end{figure}

To quantify the spatial distribution of the GC candidates of the galaxy, we obtained the one-dimensional radial distribution of the entire sample, as well as those for each subpopulation (Figure\,\ref{figure14}). In order to obtain a good sampling, the construction of the density profiles was carried out by counting objects in concentric circular annuli with $\Delta\,log(r)=0.08$. Each bin was corrected for contamination and per effective area. Figure\,\ref{figure14} shows that the GC system of NGC\,1395 is very extended, outpacing the boundary of our GMOS mosaic ($>$8.5 arcmin; $>$56 kpc at the distance assumed for NGC\,1395). In addition, we also show the surface brightness profile of the galaxy (see Section\,\ref{sec:brightness_prof}) with a green line. As observed in other early type galaxies \citep{lee08,harris09,escudero15}, the light profile of the halo of NGC\,1395 is steeper compared to the distribution of the entire GC system. However, it shows a close match with the projected density profile of the red candidates. 

Subsequently, two scaling laws, a de Vaucouleurs ($r^{1/4}$) and a power law, were fitted to the density profiles of the whole GC sample, as well as to each GC subpopulation. These scaling laws, widely used in the literature \citep[e.g.,][]{faifer11,escudero15,hudson17}, show good fits on the GC density profiles. We perform least squares fits in order to estimate which of these functions provide a better aproximation to our profiles. Since the profiles flatten towards the galactic center, we excluded the most internal points from the fits ($r<0.34$ arcmin). The lack of GCs towards the inner region of the galaxy may be due to the probable incompleteness of our photometric sample, or/and to dynamical effects, such as tidal disruption and/or erosion of GCs \citep{kruijssen15,brockamp14}. 

Table\,\ref{ley_pot} lists the values obtained from the fits of the different scaling laws and the background contamination levels expected in each case. Both functions give similar results in terms of the residual errors. In addition, a fit considering the data for $r>1$ arcmin ($\sim$100 per cent completeness) was made, obtaining similar results within the errors to the values listed in Table \ref{ley_pot}. Figure\,\ref{figure14} shows that both subpopulations do not seem to have reached a background level within our GMOS mosaic. The slope found for the red GCs is similar to that reported for other massive early-type galaxies \citep{bassino06,faifer11,sesto16}. However, the blue GC candidates present a shallower slope, similar to that found for NGC\,3923 \citep{faifer11,miller17}. Interestingly, this is a massive elliptical shell galaxy as well, located in a poor group of galaxies. 

\begin{table}
\centering
\caption{Fitted parameters for the surface density profiles considering a power law and a de Vaucouleurs law, for the whole sample and for the blue and red GC subpopulations. Fifth and sixth columns indicate the rms of the fits and the background level estimations, respectively.}
\label{ley_pot}
\begin{tabular}{cccccc}
\multicolumn{1}{c}{ } \\
\toprule
\toprule
\multicolumn{1}{c}{\textbf{Popul.}} &
\multicolumn{1}{c}{\textbf{Scal. law}} &
\multicolumn{1}{c}{\textbf{Slope}} &
\multicolumn{1}{c}{\textbf{Zero point}} &
\multicolumn{1}{c}{\textbf{rms}} &
\multicolumn{1}{c}{\textbf{bg}} \\
\multicolumn{1}{c}{} &
\multicolumn{1}{c}{} &
\multicolumn{1}{c}{} &
\multicolumn{1}{c}{} &
\multicolumn{1}{c}{} &
\multicolumn{1}{c}{(log($''^{-2}$))} \\
\midrule
All  &  Power    &  -1.36$\pm$0.03  &  0.60$\pm$0.07  &  0.05 & -3.42  \\
     &  de Vauc. &  -0.67$\pm$0.02  &  0.02$\pm$0.09  &  0.06 & ---    \\
\midrule
Blue &  Power    &  -1.01$\pm$0.05  & -0.50$\pm$0.13  &  0.07 & -3.72  \\
     &  de Vauc. &  -0.48$\pm$0.02  & -1.00$\pm$0.10  &  0.06 & ---    \\
\midrule
Red  &  Power    &  -1.69$\pm$0.04  &  1.01$\pm$0.09  &  0.06 & -3.73  \\
     &  de Vauc. &  -0.85$\pm$0.03  &  0.39$\pm$0.09  &  0.07 & ---    \\
\bottomrule
\end{tabular}
\end{table}

We studied the azimuthal distribution for the whole system and for both subpopulations in order to determine the ellipticities and position angles of the GCs proyected spatial distribution. To do this, we used the expresion of \citet{mcLaughlin94}:
\begin{equation}\label{eq:azim}
\sigma(R,\theta)=kR^{-\alpha}[cos^2(\theta-PA)+(1-\epsilon)^{-2}sin^2(\theta-PA)]^{-\alpha/2}.
\end{equation}
In this expression, $\sigma(R,\theta)$ represents the number of GC candidates, $k$, the normalization constant, $PA$, the position angle measured counterclockwise from the north, $\epsilon$, the ellipticity and $\alpha$, the slope of the power law obtained in the surface density fit. 

\begin{figure}
\centering
\resizebox{0.99\hsize}{!}{\includegraphics{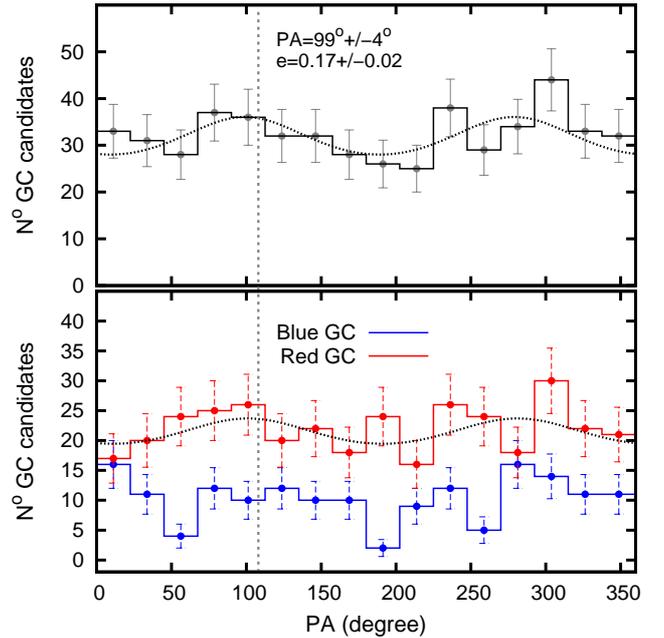}}
\caption{Azimuthal distribution of the all GC candidates within the annulus $0.35<r_\mathrm{gal}<1.84$ arcmin {\it (top panel)}, and for the blue and red GC subpopulations (blue and red histograms, respectively; {\it bottom panel}). The dashed vertical line represents the average $PA$ value of the galaxy light obtained in this work ($\langle PA \rangle=108$ degrees). Dotted black curves in both panels show the fit obtained for the GC sample and for the red GC subpopulation.}
\label{figure15}
\end{figure}

We counted GC candidates in wedges of 22.5 degrees within a circular ring of $0.35<r_\mathrm{gal}<1.84$ arcmin centered on the galaxy. The inner radius was considered to avoid the low completeness of objects in the region near the galactic center. The outer radius corresponds to the maximum radius that can be reached at the edge of the mosaic, which allow us to avoid to perform corrections for areal incompleteness. Figure\,\ref{figure15} shows the histograms of the azimuthal analysis of the whole GC system as well as those corresponding to each subpopulation. Using Equation\,\ref{eq:azim}, allowing $k$, $PA$ and $\epsilon$ to vary freely, the following values were obtained for the whole GC system: $\epsilon=0.17\pm0.02$ and $PA=99^\circ\pm4^\circ$. This indicates that the system exhibits an elongation and orientation similar to the stellar component of the galaxy ($\langle \epsilon \rangle=0.17$, $\langle PA \rangle=108^\circ$; see Section\,\ref{sec:tidal}). In particular, the ellipticity of the projected spatial distribution of the red candidates ($\epsilon=0.11\pm0.03$, $PA=101^\circ\pm8^\circ$) is slightly lower than the ellipticity of the galaxy. In addition, we did not find any significant elongation for the projected spatial distribution of the blue subpopulation, and therefore, in this case, the $PA$ is unconstrained. The asymmetry shown by the blue candidates in the central region could be responsible for the greater elongation obtained for the GC system (blue+red).

\subsection{Diffuse hot gas in the group} 
\label{sec:rayosx}
Eridanus supergroup presents diffuse X-ray emission, mainly around the galaxies NGC\,1407 and NGC\,1395. This weak emission extends $\sim$30 arcmin ($\sim$200 kpc) and $\sim$20 arcmin ($\sim$135 kpc) around both galaxies, respectively. Contours of the extended X-ray emission around NGC\,1395, based on ROSAT observations, have been presented in \citet{omar05}. 

Different studies such as \citet{forte05}, \citet{forbes12} and \citet{escudero15} have shown that, in some massive early-type galaxies, the slope of the projected density profile of the blue GCs in the external region of the galaxy, and the X-ray brightness profile of the host galaxy halo, exhibit a similar behavior. This close relationship between both tracers of the galaxy halo, would indicate that they share the same gravitational potential in equilibrium \citep{forbes12}. However, this phenomenom is not well understood yet. Therefore, to identify more galaxies of this kind will help to clarify this issue.

In order to disentangle if a similar relation occurs in NGC\,1395, we compared the blue density profile obtained in Section\,\ref{sec:dens} with those built from archival images of XMM-Newton (ObsID:\,0305930101) and Chandra (ObsID:\,799) X-ray space observatories. Finally, we only considered the profile obtained from XMM-Newton images, since they display a continuous radial coverage and an exposure time three times greater than those of Chandra.

The XMM data reduction was carried out following \citet{nagino09}. From the reduced final image, the X-ray surface brightness profile of the galaxy in the range $0.3-2.5$ keV, corrected by background, was extracted considering concentric annuli, reaching a distance of $\sim$10 arcmin ($\sim$66 kpc) from the galactic center. Subsequently, in order to compare both profiles (blue GCs and hot gas), we decided to fit a $\beta$-model to them \citep{cavaliere76}:
\begin{equation}\label{xprof}
S(r) = S_0[1+(r/r_c)^2]^{-3\beta_x+0.5},
\end{equation}
with $r_c$ the core radius and $\beta_x$ the slope of the profile.

\begin{figure}
\centering
\resizebox{0.99\hsize}{!}{\includegraphics{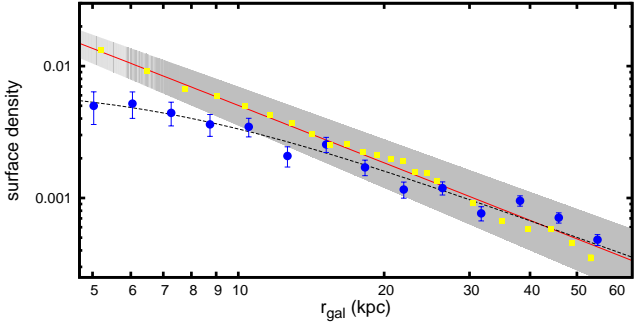}}
\caption{Globular cluster density profile (blue filled circles) and X-ray density profiles (yellow filled squares) for NGC\,1395. The red solid line shows the fitted $\beta$-model for the X-ray surface brightness with a $\pm5$\% slope uncertainty (shaded region), while the black dashed line shows the $\beta$-model fit obtained over the density profile of the blue subpopulation.}
\label{figure16}
\end{figure}

Figure\,\ref{figure16} shows the comparison between the density profile of the blue GCs and the X-ray profile of NGC\,1395. The latter was shifted on the y axis to obtain a better comparison and visualization of them. In addition, a $\pm5$\% variation on the $\beta$ value (shaded region in Figure\,\ref{figure16}) was considered according to \citet{forbes12}. The fit of Equation\,\ref{xprof} was carried out by discarding the most internal points ($<0.4$ arcmin; $\sim$2.5 kpc), mainly due to the incompleteness of our blue GC profile in that region.
The obtained values are $\beta=0.39\pm0.04$ for the blue GCs and $\beta=0.41\pm0.01$ for the X-ray profile. As in other few well studied massive early-type galaxies, Figure\,\ref{figure16} shows that the blue GCs profiles display a good agreement with the X-ray emission profile for $r_\mathrm{gal}\gtrsim\,15$ kpc. When comparing these values with those obtained from the literature, it is observed that NGC\,1395 presents similar slope values to other massive galaxies, such as the central galaxy of the Fornax cluster NGC\,1399 ($\beta_X=0.35$, $\beta_\mathrm{blue}=0.42\pm0.05$), the central group galaxy NGC\,5846 ($\beta_X=0.45$, $\beta_\mathrm{blue}=0.51\pm0.10$) \citep{forbes12}, and the lenticular galaxy NGC\,6861 in the Telescopium group ($\beta_X=0.38$, $\beta_\mathrm{blue}=0.42\pm0.01$) \citep{escudero15}.

\subsection{Luminosity function and total GC population} 
\label{sec:lumin}
The distance to NGC\,1395 have been estimated in the literature using different methods, including Cepheids \citep{ferrarese00}, surface brightness fluctuation \citep[SBF;][]{tully13} and Tully-Fisher relation \citep{freedman01}. In this paper we use the luminosity function of the GC system (GCLF) as an independent method to those previously mentioned. The advantage of this method is that it allows us to estimate the total population of GCs of the galaxy, using the joint information provided by the density profile analysis of the system.

\begin{figure}
\centering
\resizebox{0.99\hsize}{!}{\includegraphics{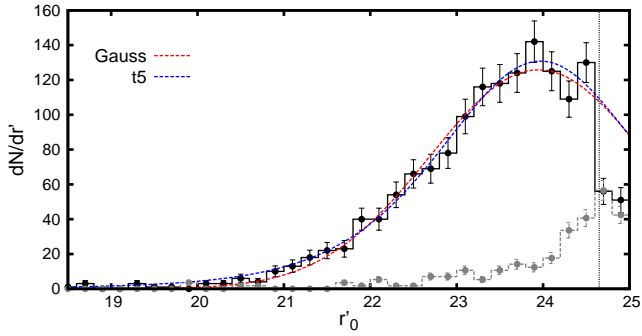}}
\caption{Globular cluster luminosity function of NGC\,1395. Solid line shows the corrected LF (background and completeness) for the GC system. The histogram corresponding to the comparison fields is shown by dashed grey line. Blue and red dashed lines represent the Gaussian and t5 fit to the corrected GCLF. Vertical dotted line indicates the magnitude range not included in the fits ($r'_0>24.6$ mag).}
\label{figure17}
\end{figure}

The construction of the GCLF was performed as follows. From our initial photometric catalogue, which contains all the unresolved objects detected in the GMOS mosaic, we performed the colour cuts mentioned in Section\,\ref{sec:color} but without limiting the magnitude range. Using this sample, we counted GC candidates in bins of 0.2 mag, correcting the distribution for completeness and background (Section\,\ref{sec:comp} and Section\,\ref{sec:test}). Figure\,\ref{figure17} shows the corrected luminosity distribution (solid line) of the GC system of NGC\,1395 in the $r'$ band. In addition, the Figure shows the histogram corresponding to the comparison fields corrected by an areal factor (grey dashed line).

The GCLF of NGC\,1395 seems to show the turnover (TO) magnitude at $r'_0\sim24$ mag. To quantify its position, both Gaussian and t5 functions were used on the histogram. These functions are widely used in the literature \citep{jacoby92,villegas10,faifer11,escudero15} since they show acceptable fits on the GC luminosity distributions. The obtained fits for the GC system are displayed in Figure\,\ref{figure17}, while Table\,\ref{tab_gclf} lists the TO magnitudes and dispersions ($\sigma$) of both functions. 

Taking advantage of the ``universality'' shown by the GCLF, we used the value $M_R=-8.06\pm0.15$ mag of \citet{gomez04} and our estimated value of the magnitude of the TO to obtain the distance modulus of NGC\,1395. To do this, we transform $M_R$ using the expression of \citet{lupton05}\footnote{http://www.sdss.org/dr12/algorithms/sdssubvritransform/}:
\begin{equation}\label{lupton}
R = r'-0.2936(r'-i')-0.1439.
\end{equation}
Using the mean value $\langle r'-i' \rangle_0=0.28$ mag for the GC system, we obtain $M_r=-7.83$ mag.

Subsequently, considering the averaged value of the magnitude of the TO corresponding to the whole GC population (Table\,\ref{tab_gclf}), we get $(m-M)=31.79\pm0.16$ mag, which translates into a distance of 22.8$\pm$1.7 Mpc. This value is in good agreement with the initially used in this work by \citet{tully13} (see Table\,\ref{Tab0}). 

As shown by \citet{larsen01} and \citet{escudero15}, a significant difference between the TO magnitudes of blue and red GC subpopulations is observed. This difference, possibly originated by metallicity effects \citep{ashman95}, results in a widening of the GCLF when considering the entire GC sample, and therefore in a shift in the value of TO, especially if photometric bands are centered towards the blue of the optical spectrum. To minimize this metallicity effect, one option is to use the TO magnitude of the blue subpopulation as a distance indicator \citep{kissler00}. Therefore, we repeated the same procedure followed for the whole GC sample, and we obtained the GCLFs of both subpopulations. 

\begin{figure}
\centering
\resizebox{0.99\hsize}{!}{\includegraphics{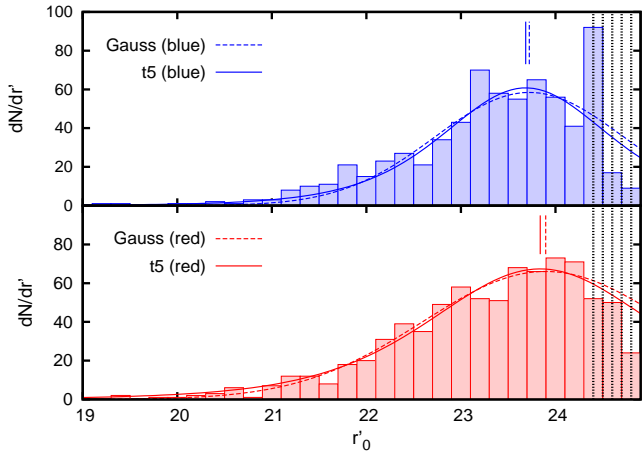}}
\caption{Corrected luminosity function for blue {\it (top panel)} and red {\it (bottom panel)} subpopulation. Dashed and solid lines show the Gaussian and t5 fits, respectively. The dashed and solid vertical lines indicate the position of the magnitudes of the TO. Vertical black dotted lines indicate the magnitude range not included in the fits ($r'_0>24.5$ mag).}
\label{figure18}
\end{figure}

Figure\,\ref{figure18} shows the corrected histograms for the blue and red clusters, while Table\,\ref{tab_gclf}, the values obtained from the fitted functions. The difference obtained between the positions of the TO magnitudes of both GC subpopulations (being the TO of the blue subpopularion slightly brighter than the TO of the red one), results significant within 1$\sigma$.

In order to determine the distance modulus of NGC\,1395 using its blue subpopulation, we used the 2010 version of the McMaster globular cluster database \citep{harris96} to re-fit the GCLF of the Milky Way in the $R$ filter. According to the analysis performed by \citet{harris16} on the metallicity distribution function for the GCs in the Galaxy, we are left with those clusters with $[\mathrm{Fe/H}]<-1$ dex (metal-poor GCs) and reddenings $E_{B-V}<1.6$. Then, we corrected by reddening the magnitudes of the GCs using the values of \citep{schlafly11}, and we built up the luminosity function using the averaged shifted histogram method \citep{scott83} with a bin of 0.3 mag. Then, we fit a Gaussian and a t5 function to determine the TO magnitudes, $M_R^0$, and the corresponding dispersions ($M_R^0=-8.15\pm0.05$, $\sigma=0.92\pm0.05$ mag and $M_R^0=-8.13\pm0.05$, $\sigma=0.89\pm0.06$ mag, respectively). 

\begin{figure}
\centering
\resizebox{0.99\hsize}{!}{\includegraphics{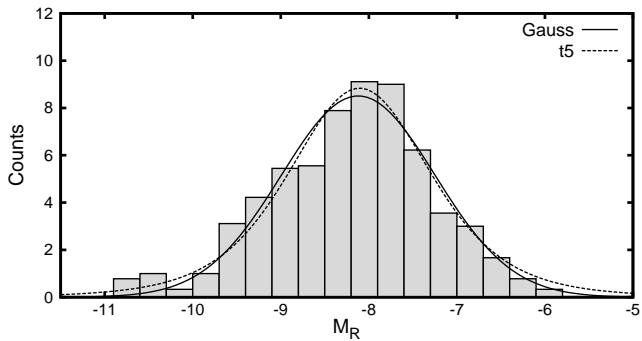}}
\caption{Globular cluster luminosity function of the Milky Way in the $R$ band considering only the metal-poor subpopulation. The solid and dashed black lines show the Gaussian and t5 fits, respectively.}
\label{figure19}
\end{figure}

Figure\,\ref{figure19} shows the resulting GCLF together with the obtained fits. Considering the averaged value of $M_R^0$, the final result for the Milky Way is $M_R^0=-8.14\pm0.07$ mag. We transform this value using Equation\,\ref{lupton} and the value $\langle r'-i' \rangle_0=0.27$ mag (see Table\,\ref{gmm_2comp}) for the blue subpopulation of NGC\,1395, obtaining $M_r=-7.92$ mag. Finally, using this $M_r$ value and the averaged TO magnitude for the blue candidates of NGC\,1395, we got the distance modulus $(m-M)=31.62\pm0.11$ mag ($\sim21.1\pm1.1$ Mpc). This value is slightly lower than the previously obtained for the whole sample of GCs and than that of \citet{tully13}, but it is still in agreement within 2$\sigma$ with the value obtained by these authors. In addition, the distance modulus obtained here for NGC\,1395 results comparable to that obtained by \citet{forbes06},  using the GCLF, for NGC\,1407 in the Eridanus supergroup.

\begin{table}
\centering
\caption{Values of the TO magnitudes and dispersions for the different GC samples, obtained from the fit of Gaussian and t5 functions.}
\label{tab_gclf}
\begin{tabular}{ccccc}
\toprule
\toprule
\multicolumn{1}{c}{\textbf{Population}} &
\multicolumn{2}{c}{\textbf{Gaussian}} &
\multicolumn{2}{c}{\textbf{t5}} \\
\multicolumn{1}{c}{} &
\multicolumn{1}{c}{$TO$} &
\multicolumn{1}{c}{$\sigma$} &
\multicolumn{1}{c}{$TO$} &
\multicolumn{1}{c}{$\sigma$} \\
\midrule
All  &  23.94$\pm$0.09 & 1.25$\pm$0.07  &  23.97$\pm$0.06 & 1.22$\pm$0.06  \\
Blue &  23.72$\pm$0.11 & 0.99$\pm$0.10  &  23.68$\pm$0.07 & 0.91$\pm$0.08  \\
Red  &  23.89$\pm$0.11 & 1.30$\pm$0.09  &  23.84$\pm$0.09 & 1.22$\pm$0.09  \\
\bottomrule
\end{tabular}
\end{table}

As mentioned at the beginning of this section, using the information provided by the GCLF and the density profile analysis of the system, we estimated the total GC population of NGC\,1395. Initially, we integrated the fitted power law density profile obtained in Section\,\ref{sec:dens} in the range 0.4 to 15.2 arcmin. The internal value corresponds to the limit at which the profile begins to flatten, while the external value corresponds to a distance of 100 kpc, since there are still a considerable number of GC candidates up to the edge of our GMOS mosaic ($\sim8.8$ arcmin; $\sim$58\,kpc) (Figure\,\ref{figure12}). In addition, this value has been used in the study of other GC systems of massive galaxies \citep{bassino06,harris09,escudero15}.

The number of GCs obtained from the integration of the profile is 2680. Then, we assume a constant density value for the inner region of the profile ($<0.4$ arcmin), obtaining an additional 50 GCs. According to the adopted value of the TO magnitude and $\sigma$ for the whole sample ($\langle TO \rangle=23.95$, $\langle \sigma \rangle=1.24$ mag), these 2730 GCs represent a 64 per cent of the total GC population, which are the GCs brighter than $r'_0=24.4$ mag. Taking into account the uncertainties in the estimation of different parameters obtained in the analysis of the spatial distribution and the luminosity function, we estimate a total of $4270\pm800$ GCs for the system of NGC\,1395. This value corresponds to a specific frequency of $S_N=N_{GC}10^{0.4(M_V+15)}=5.3\pm1.2$ considering $M_V=-22.27$ mag \citep{harris81}. Additionally, we estimate the total subpopulations of blue and red GCs to be $2300\pm650$ and $1860\pm380$, respectively.

As shown in Section\,\ref{sec:dens}, the GC system of NGC\,1395, in particular the blue subpopulation, shows a shallower slope of the density profile compared to other GC systems belonging to massive E galaxies \citep[see e.g.,][]{faifer11}. 
In recent years, using wide-field imaging, different authors have found that some GC systems belonging to massive galaxies extend to large galactocentric radii \citep[$\gtrsim$100\,kpc;][]{kartha16,miller17,taylor17}.
In this context, we decided to extend the integration of the density profile to a maximum galactocentric radius of 165\,kpc. This value is obtained by extending the density profile of the blue GC subpopulation until it reaches the background value estimated in Table\,\ref{ley_pot}. On the other hand, the extended X-ray emission around the galaxy and its similarity with the blue profile, provides additional support to the election of the radius mentioned above. Considering this new extension for the system, we re-estimated the total GC population and $S_N$ in $6000\pm1100$ and $7.40\pm1.40$, respectively. These values result more similar to those found for other large elliptical galaxies, such as NGC\,4365 with $S_N=7.75\pm0.13$ \citep{blom12}, NGC\,1399 $S_N=6.72\pm0.81$ \citep{spitler08}. For the following analysis, we will adopt this value for the total population corresponding to 165\,kpc.


\section{Global properties of NGC\,1395 from its GC system}
\label{sec:correlacion}
Using the results obtained in previous sections, we estimated some structural parameters of NGC\,1395 considering different correlations and models obtained from the literature. The results obtained here were compared with early-type massive galaxies located in the Fornax+Eridanus complex. Table\,\ref{galaxies} lists some global parameters of the considered galaxies obtained from published data. 

Following \citet{harris17}, we estimate the total halo mass ($M_h$) of NGC\,1395 using the number of GCs obtained in the previous section. To do this, we use the simple empirical expression $\eta_M=M_{GCS}/M_h$ \citep{blakeslee97,hudson14,harris15}, where $M_{GCS}$ is the total mass in the galaxy GC system and $M_h$ is the total mass of the galaxy that includes the baryonic mass and the dark matter halo mass. \citet{harris17} obtained a value of $\eta_M=2.9\pm0.2\times10^{-5}$ including the variation of the mean mass of the GCs as the mass of the galaxy increases. This near-constant mass ratio across a range of $\sim10^5$ in galaxy mass (from ultra-diffuse galaxies up to entire clusters of galaxies), suggests that the amount of gas available for GC formation at high redshift ($8\gtrsim z \gtrsim2$) was in almost direct proportion to the dark matter halo potential \citep{harris15}. 

Initially, we estimate the GC mean mass ($\langle M_{GC} \rangle$) using the expression of \citet{harris17}:
\begin{equation}\label{eq:masa_GC}
\begin{aligned}
 {log\langle M_{GC} \rangle} &= 5.698 + 0.1294\,M_V^{T} + 0.0054\,(M_V^{T})^2. \\
\end{aligned}
\end{equation}
Adopting $M_V=-22.27$ mag for NGC\,1395, the total mass of its GC system is $M_{GCS}=N_{GC}\langle M_{GC} \rangle=1.87\times10^{9}$ M$\odot$, whereby the total mass for the galaxy results in $M_h=6.46\times10^{13}$ M$\odot$ (log($M_h$)=13.81 M$\odot$).

Using this same procedure for the galaxies mentioned in Table \ref{galaxies}, and according to their values of $M_V$ and $N_{GC}$, we obtain that each subgroup of the Eridanus supergroup have similar masses. It is necessary to mention that the $N_{GC}$ of NGC\,1332 was estimated using the expression of \citet{aragon06} given by $S_N^\mathrm{local}/S_N^\mathrm{global}=5.71+0.25\,M_V$ with the value $S_N^\mathrm{local}=2.2\pm0.7$ estimated by \citet{kundu01}. In this context, the values obtained here are of the order to those estimated by \citet{makarov11} based on the virial mass estimates of each group. Although NGC\,1399 presents a similar value to those of the Eridanus subgroups, it should be noticed that $N_{GC}$ is probably underestimated \citep[see][]{harris17}. Regarding to NGC\,1316, the situation is completely different since the mass assembly of this galaxy is still ongoing, showing a dominant GC component with intermediate-age stellar populations \citep{goudfrooij12,sesto16}.

Subsequently, the obtained value $M_h$ allowed us to estimate the stellar mass ($M_\mathrm{stellar}$) of the galaxy using the expression of \citet{yang08}: 
\begin{equation}\label{ms_mh}
M_\mathrm{stellar} = M_s \, {(M_\mathrm{halo}/M_n)^{\alpha+\beta}\over(1+M_\mathrm{halo}/M_n)^{\beta}}~
\end{equation}
with scaling parameters $log\,(M_s)=9.98$, $log\,(M_n)=10.7$, $\alpha=0.64$ and $\beta=2.88$ from \citet{harris13}.
We obtained for NGC\,1395 the value $M_\mathrm{stellar}=9.32\times10^{11}$ M$\odot$ (log($M_\mathrm{stellar}$)=11.97 M$\odot$), which is $\sim$9 times greater than the value estimated by \citet{colbert04} ($M_\mathrm{stellar}=1\times10^{11}$ M$\odot$), using the luminosity of the galaxy in the $K$ filter. 

We estimated the stellar mass for each galaxy in Table\,\ref{galaxies} by Equation\,\ref{ms_mh}. In addition, we added the dynamical mass values estimated by \citet{harris13} using the velocity dispersion and the effective radius of the galaxy. According to these authors, \textit{``since the luminosity-weighted velocity dispersion is dominated by light from within $R_\mathrm{eff}$, and the dark-matter halo contributes a small fraction of the mass within $R_\mathrm{eff}$, $M_\mathrm{dyn}$ is close to being the baryonic mass of the galactic bulge''.}
As can be seen in Table \ref{galaxies}, in this case both values are in a very good agreement.

Another important parameter that has a strong correlation with the number of GCs, is the mass of the central supermassive black hole \citep[SMBH,][]{burkert10,harrisg14} in the galaxy. \citet{harrisg14} suggest that if there is a causal link between them, would be through AGN feedback and its influence on star and GC formation, and not due to a statistical origin. In this context, \citet{forte17} recently found $(g-z)$ colour modulation patterns in the GC systems associated with galaxies included in the Virgo and Fornax HST-Advanced Camera Surveys. These features suggest that the GC formation process would be associated with large scale feedback effects connected with violent star forming events and/or with SMBHs. 

We use two different large-scale galaxy properties to estimate the mass of the SMBH of NGC\,1395: $N_{GC}$ and the velocity dispersion ($\sigma_e$). Using the expressions of \citet{harrisg14}:
\begin{equation}\label{bh_ncg}
log\,N_{GC} = 2.952\pm0.059 + (0.840\pm0.080)(log\,M_{SMBH}-8.5)
\end{equation}
\begin{equation}\label{bh_sigma}
log\,M_{SMBH} = 8.413\pm0.080 + (4.730\pm0.539)(log\,\sigma_e-2.30)
\end{equation}
we obtain $log(M_{SMBH})=9.48$ and $log(M_{SMBH})=8.78$ M$\odot$, respectively. We consider for NGC\,1395 the mean value of both quanitities, $\langle log(M_{SMBH}) \rangle_0=9.13$ M$\odot$.

\begin{table*}
\centering
\caption{Global properties of early-type galaxies. Columns (1)-(7) list the galaxies, morphological type, environment in which they are located, absolute $V$ magnitude, total number of GCs, velocity dispersion and effective radius. Column (8) shows the stellar mass obtained using Equation\,\ref{ms_mh}; column (9), the dynamical mass from \citet{harris13}; column (10)-(11), the SMBH mass and the total halo mass. The last column lists the data references: (a) \citet{pota13}, (b) \citet{spitler08}, (c) \citet{harris13}, (d) \citet{barth16}, (e) \citet{harrisg14}, (f) \citet{gebhardt07}, (g) \citet{nowak08}, (h) \citet{richtler12}.}
\label{galaxies}
\begin{tabular}{cccccccccccc}
\toprule
\toprule
\multicolumn{1}{c}{\textbf{Galaxy}} &
\multicolumn{1}{c}{\textbf{Type}} &
\multicolumn{1}{c}{\textbf{Env.}} &
\multicolumn{1}{c}{\textbf{$M_V^T$}} &
\multicolumn{1}{c}{\textbf{$N_{GC}$}} &
\multicolumn{1}{c}{\textbf{$\sigma$}} &
\multicolumn{1}{c}{\textbf{$R_\mathrm{eff}$}} &
\multicolumn{1}{c}{\textbf{log($M_\mathrm{stellar}$)}} &
\multicolumn{1}{c}{\textbf{log($M_\mathrm{dyn}$)}} &
\multicolumn{1}{c}{\textbf{log($M_{SMBH}$)}} &
\multicolumn{1}{c}{\textbf{log($M_h$)}} &
\multicolumn{1}{c}{\textbf{Ref.}} \\
\multicolumn{1}{c}{} &
\multicolumn{1}{c}{} &
\multicolumn{1}{c}{} &
\multicolumn{1}{c}{(mag)} &
\multicolumn{1}{c}{} &
\multicolumn{1}{c}{(km\,s$^{-1}$)} &
\multicolumn{1}{c}{(kpc)} &
\multicolumn{1}{c}{(M$\odot$)} &
\multicolumn{1}{c}{(M$\odot$)} &
\multicolumn{1}{c}{(M$\odot$)} &
\multicolumn{1}{c}{(M$\odot$)} &
\multicolumn{1}{c}{} \\ 
\multicolumn{1}{c}{(1)} &
\multicolumn{1}{c}{(2)} &
\multicolumn{1}{c}{(3)} &
\multicolumn{1}{c}{(4)} &
\multicolumn{1}{c}{(5)} &
\multicolumn{1}{c}{(6)} &
\multicolumn{1}{c}{(7)} &
\multicolumn{1}{c}{(8)} &
\multicolumn{1}{c}{(9)} &
\multicolumn{1}{c}{(10)} &
\multicolumn{1}{c}{(11)} &
\multicolumn{1}{c}{(12)} \\
\midrule
NGC\,1395 &  E2      & Group    & -22.27 & 6000$\pm$1100 & 238 &  7.10 & 11.97 & 11.60 & 9.13 & 13.81  & This work \\
NGC\,1407 &  E0      & Group    & -22.26 & 6400$\pm$700  & 270 & 10.38 & 11.98 & 11.85 & 9.67 & 13.84  &  a,b,c    \\
NGC\,1332 &  S0      & Group    & -21.62 & 3200$\pm$1000 & 320 &  2.75 & 11.75 & 11.42 & 8.77 & 13.47  &  c,d      \\
NGC\,1399 &  E1      & Cluster  & -22.34 & 6625$\pm$1180 & 337 &  9.70 & 12.01 & 12.01 & 8.70 & 13.86  &  c,e,f    \\
NGC\,1316 &  S0\,pec & Group    & -23.16 & 1500$\pm$500  & 227 & 10.19 & 11.65 & 11.69 & 8.17 & 13.31  &  c,g,h    \\
\bottomrule
\end{tabular}
\end{table*}

\section{Stellar population analysis} 
\label{sec:espec}
As presented in Section\,\ref{sec:elipse}, the detection of shells and boxy isophotes in NGC\,1395, would indicate that the galaxy have experienced, at least, one recent merger event. Therefore, using public spectroscopic data, we decided to study the stellar population in the central region of NGC\,1395 in order to inspect if it has undergone recent star formation. Some works in the literature have presented an analysis of the stellar population of NGC\,1395. In the following we give a short review of these papers, and their results will be used to test our analysis.

{\bf{\citet{beuing02}}} present the analysis of several line-strength indices (H$\beta$, Mg1 , Mg2 , Mgb, Fe5015, Fe5270, Fe5335, Fe5406, Fe5709, Fe5782, NaD, TiO1 and TiO2) of 148 early-type galaxies, including NGC\,1395, using longslit spectra. These indices were measured within an aperture of 7.4 arcsec, and were subsequently corrected by velocity dispersion effects and calibrated to the Lick/IDS system. From the measurement of the equivalent width (EW) of the [OIII] line, these authors classified NGC\,1395 as a class ``0'' galaxy, since it does not show significant emission lines (EW[OIII]$<$0.3\,\AA). As the corrections for emission are typically smaller than the errors in the estimation of indices, they were not corrected for this effect. 

{\bf{\citet{thomas05}}} estimated the stellar parameters (age, [Z/H] and [$\alpha$/Fe] ratio) of 124 early-type galaxies, using recalibrated Lick indices H$\beta$, Mgb, and $\langle$Fe$\rangle$ ($\langle$Fe$\rangle$=(Fe5270+Fe5335)/2) of \citet{beuing02} and stellar population models of \citet{thomas03}. In this work, the values obtained for NGC\,1395 are: age=7.6$\pm$1.4 Gyr, [Z/H]=0.439$\pm$0.033 dex and [$\alpha$/Fe]=0.353$\pm$0.013 dex. 

{\bf{\citet{serra10}}} present an analysis of the stellar population of several elliptical galaxies using HI observations \citep{meyer04} and the same Lick indices published by \citet{thomas05}, but only considering the solar value of [$\alpha$/Fe] to compare with SSP models. The age obtained by these authors for NGC\,1395 is slightly younger than that of Thomas et al. ($6.0_{-0.9}^{+1.6}$ Gyr).

In this context, we decided to study the stellar population of NGC\,1395 using a different method to the traditional line-strengths analysis in order to verify if the same results are obtained. To this aim we used the 6dFGS spectrum (see Section \ref{sec:obs}) and the full spectra fitting technique which has the advantage of using all the information available in the spectrum, making it possible even to perform an analysis at lower S/N. The spectrum covers the central 6.7 arcsec ($\sim$\,$R_\mathrm{eff}/10$) of NGC\,1395, being comparable to the aperture value used by \citet{beuing02}.
In this particular case, we determined the stellar kinematics and stellar population of the galaxy using the ULySS \citep[University of Lyon Spectroscopic Analysis Software;][]{koleva09} code and the SSP models of the MILES library \citep{vazdekis10} which cover a wide range of age ($0.03-14$ Gyr) and metallicity ($-2.27<[Z/H]<0.4$ dex). 
We adopt an $\alpha$-element ratio of [$\alpha$/Fe]=0.4 dex according to the value previously obtained by \citet{thomas05}. These models, with a resolution of 2.51\,\AA\, were degraded to the resolution of the science spectrum. In addition, we have considered the parameter {\tt \textbackslash CLEAN} to exclude outliers during the fit, and also a multiplicative polynomial of high order (order 20) to correct for inaccuracies in the model prediction.

In order to obtain reliable results, we performed several runs considering different spectral ranges. The best fit was obtained in the wavelength range $3900-6800$\,\AA. To determine the first two moments of the line-of-sight velocity distribution (velocity and velocity dispersion), and the integrated SSP equivalent age and metallicity of the galaxy, we considered 100 Monte Carlo simulations to estimate the errors and the coupling between the different parameters, adopting the mean value of the simulations as final values. 

\begin{figure*}
\centering
\includegraphics[width=0.9\textwidth]{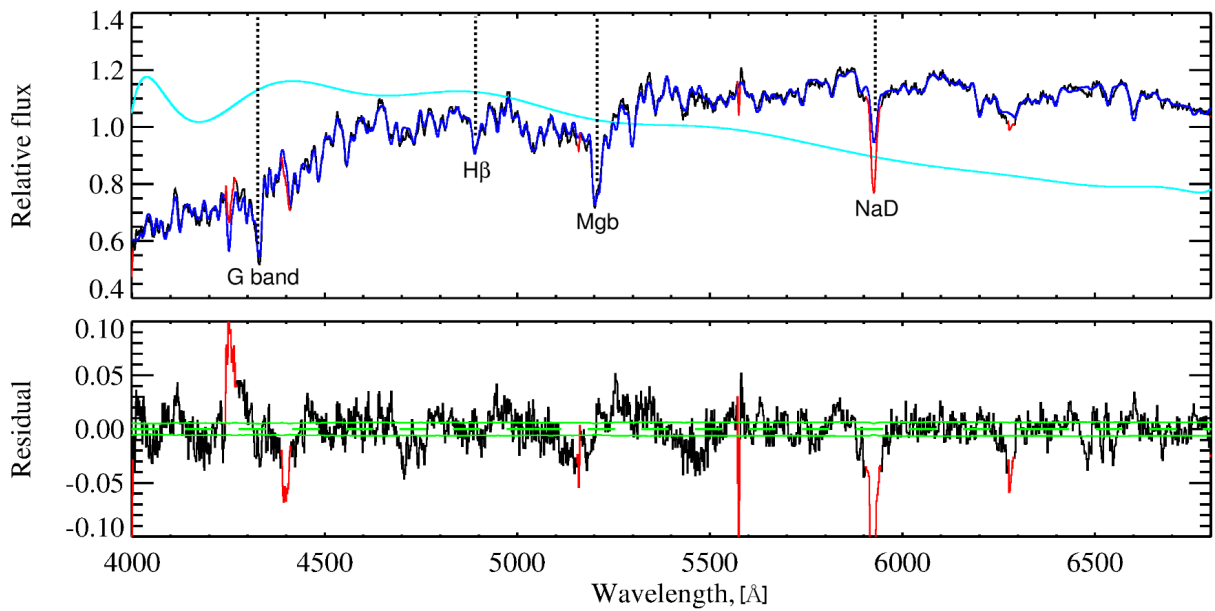}
\caption{Observed spectra of NGC\,1395 (black line) with the best fit obtained by ULySS (blue line). The light blue line is the considered multiplicative polynomial. Red lines indicate the discarded lines during the fit, due to the presence of telluric lines in this regions and automatic rejection of outliers. The residual spectrum obtained from the fit is displayed below NGC\,1395's spectrum, where the green lines indicate the 1-$\sigma$ deviation.}
\label{figure20}
\end{figure*}

The values obtained by the code for the radial velocity and velocity dispersion of the stellar component of the galaxy were $V_r=1726\pm3$ km\,s$^{-1}$ and $\sigma=255\pm4$ km\,s$^{-1}$, respectively, which are in agreement with those obtained by \citet{smith00}. In addition, the calculated values of age and metallicity for NGC\,1395 are $12.1\pm0.2$ Gyr and [Z/H]=$0.34\pm0.01$ dex, respectively.
Figure\,\ref{figure20} shows the integrated spectrum of NGC\,1395 with the best fit and residual spectra obtained by ULySS.

According to our results, the central region of the galaxy is dominated by an old stellar population showing a greater age compared to that found by \citet{thomas05} using absorption line strength indices. In addition, we obtained a super-solar metallicity slightly lower than that estimated by those authors. \citet{spolaor08b} obtained a similar difference in the age of NGC\,1407 when compared with the published value of \citet{thomas05}.
Although the age-metalicity degeneracy is reduced using spectral fitting techniques \citep{sanchez11,carlsten17}, it will be necessary to obtain higher-quality longslit data to confirm our results and to study in detail the stellar population of NGC\,1395 at different galactocentric radii.

\section{SUMMARY AND CONCLUSIONS}
\label{sec:disc_concl}
Using excellent Gemini/GMOS photometric data, we studied and characterized the GC system of the elliptical galaxy NGC\,1395, located in one of the subgroups that make up the Eridanus supergroup. In addition, we used public spectroscopic data to determine the stellar population parameters in its central region.

The analysis of the images allowed us to obtain global photometric parameters in the filters $g'$, $r'$, $i'$ and $z'$. We detected the presence of a faint inner shell located at a galactocentric radius of $\sim$3 arcmin ($\sim$20 kpc) in the northwest direction, connected with a ``radial feature'' forming an umbrella-like structure. In addition to these substructures, the presence of boxy isophotes provides additional piece of evidence that NGC\,1395 has experienced, at least, one recent interaction/merger event with a low-mass satellite. 
 
We estimated that the total GC population of NGC\,1395 includes $6000\pm1100$ members, considering a radial extension of the system of 165\,kpc. This value translates into a specific frequency of $S_N=7.4\pm1.4$, and imply a halo mass of $M_h=6.46\times10^{13}$ M$\odot$. Similar masses have been obtained for the remaining massive subgroups belonging to the Eridanus supergroup and the Fornax cluster (NGC\,1407, NGC\,1332, NGC\,1399 and NGC\,1316), based on the collection of data and models from the literature. Considering the sum of the halo mass of the galaxies mentioned above, an overall mass of $M=2.56\times10^{14}$ M$\odot$ is obtained for the Fornax-Eridanus complex, being in agreement within the mass range estimated by \citet{nasonova11} ($M_\mathrm{tot}=[1.30-3.93]\times10^{14}$ M$\odot$).

We used Gaussian mixture models to identify the different GC subpopulations in NGC\,1395. The analysis of the GC colour distribution exhibits a bimodal appearance in the colours $(g'-z')_0$, $(g'-i')_0$ and $(r'-z')_0$, indicating the presence of at least two GC subpopulations: the typical blue (metal-poor) and red (metal-rich) subpopulations. In particular, the red subpopulation presents a steep colour (metallicity) gradient ($<2\,R_\mathrm{eff}$; $\sim$14 kpc) as a function of galactocentric radius compared to the blue one, with the latter exhibiting relatively shallow metallicity gradients throughout the system. The estimated values for both GC families ($\Delta[Z/H]/log(r_\mathrm{gal}/R_\mathrm{eff})=-0.21\pm0.03$ dex\,per\,dex for the blue family and $\Delta[Z/H]/log(r_\mathrm{gal}/R_\mathrm{eff})=-0.09\pm0.07$ dex\,per\,dex for the red one), are similar to other systems associated to massive galaxies located in different environments.
These radial metallicity trends within $\sim$2$R_\mathrm{eff}$ and the subsequently flattening towards larger radii, would indicate that the inner populations of the galaxy (i.e. stars and GCs) have been formed by {\it in situ} dissipative processes. In addition, the observed similarity between the inner colour of the halo of the galaxy and the average colour of the red subpopulation would indicate a likely joint stellar formation, reinforcing the scenario previously mentioned. However, as shown in the spectroscopic analysis of NGC\,1395, there is uncertainty about the stellar population age in the central region of the galaxy, with which dissipative accretion processes can not be ruled out.
On the other hand, the absence of a significant colour (metallicity) gradient in the blue candidates suggests a growth of the outer halo due to continuous accretion of low mass systems with its GCs. Particularly, in this last point, the presence of a mild GC overdensity, mainly of blue candidates, is observed in the same direction as the shell structure. Unfortunately, it is difficult to confirm if both features have the same origin. It will be necessary to obtain spectroscopic data to analyze the kinematics and stellar population of these objects, as well as the use of numerical simulations, in order to constraint the different scenarios that would originate these structures. 

Another feature observed in the blue subpopulation is its extended and shallow radial spatial distribution. By comparing the density profile of the blue clusters with the X-ray surface brightness profile of the galaxy, there is a good agreement between them outside $\sim$15 kpc ($\sim$2$R_\mathrm{eff}$). This coincidence is probably due to the fact that both, blue GCs and the hot gas, would be in equilibrium within the gravitational potential of NGC\,1395 \citep{forbes11}. In this context, given the similarity between these two halo tracers, it is reasonable to assume that the GC system of the galaxy could be extended up to $\sim$160\,kpc. 

Regarding the information obtained from the analysis of the blue GCLF, we estimated the distance modulus of the galaxy in $(m-M)=31.62\pm0.11$ mag, being in good agreement with the values reported in the literature. 

Considering the full spectral fitting technique implemented by the publicly software ULySS, we determined the stellar population parameters of the central region ($\sim$\,$R_\mathrm{eff}/10$) of NGC\,1395 using 6dFGS data. We derived an age of $12.1\pm0.2$ Gyr and a metallicity [Z/H]=$0.34\pm0.01$ dex in that region by fitting the integrated spectrum of the galaxy with SSP models of the MILES library. We obtained a significant discrepancy between our derived population parameters (mainly in the age value) and those estimated by \citet{thomas05} using Lick indices. Similar differences have been found in other elliptical shell galaxies studied by \citet{carlsten17} using both methods. The presence of weak emission lines in the spectrum without an adequate correction and the use of few indices in the fit would explain this difference. However, it will be necessary to obtain new higher-quality spectroscopic data to provide more precise stellar population parameters. 

Combining all the results obtained in this work, we have found that the assembly of the elliptical galaxy NGC\,1395, and consequently the formation of its GC system, would fit in the scenario of growth in ``two-phases'' or ``inside-out'' \citep{oser10,naab14,hirschmann15}. In this picture, the old age, steep metallicity gradients, and [$\alpha$/Fe]=0.35\,dex \citep{thomas05} displayed by the galaxy in its central region, would indicate that the core of NGC\,1395 had a rapid formation timescale, possibly dominated by {\it in situ} dissipative processes that formed the vast bulk of star and GCs. Subsequently, the continuos accretion of stars and mainly blue GCs belonging to low-mass satellite galaxies, have been increasing the mass growth of the galaxy, contributing to the metallicity gradient flattening. In addition, the presence of tidal structures indicates that the galaxy is still accreting group members.

\section*{Acknowledgments}
We thank the anonymous referee for his/her constructive comments.
This work was funded with grants from Consejo Nacional de Investigaciones
Cientificas y Tecnicas de la Republica Argentina, and Universidad Nacional
de La Plata (Argentina). Based on observations obtained at the Gemini Observatory, 
which is operated by the Association of Universities for Research in Astronomy, Inc., 
under a cooperative agreement with the NSF on behalf of the Gemini partnership: the National 
Science Foundation (United States), the National Research Council (Canada), 
CONICYT (Chile), Minist\'{e}rio da Ci\^{e}ncia, Tecnologia e Inova\c{c}\~{a}o (Brazil) and 
Ministerio de Ciencia, Tecnolog\'{i}a e Innovaci\'{o}n Productiva (Argentina). 
The Gemini program ID are GS-2012B-Q-44 and GS-2014B-Q-28. This research has made use of the 
NED, which is operated by the Jet Propulsion Laboratory, Caltech, under contract with the 
National Aeronautics and Space Administration. 

\bibliographystyle{mnras}
\bibliography{biblio_1395}

\begin{thebibliography}{}
\makeatletter
\relax
\def\mn@urlcharsother{\let\do\@makeother \do\$\do\&\do\#\do\^\do\_\do\%\do\~}
\def\mn@doi{\begingroup\mn@urlcharsother \@ifnextchar [ {\mn@doi@}
  {\mn@doi@[]}}
\def\mn@doi@[#1]#2{\def\@tempa{#1}\ifx\@tempa\@empty \href
  {http://dx.doi.org/#2} {doi:#2}\else \href {http://dx.doi.org/#2} {#1}\fi
  \endgroup}
\def\mn@eprint#1#2{\mn@eprint@#1:#2::\@nil}
\def\mn@eprint@arXiv#1{\href {http://arxiv.org/abs/#1} {{\tt arXiv:#1}}}
\def\mn@eprint@dblp#1{\href {http://dblp.uni-trier.de/rec/bibtex/#1.xml}
  {dblp:#1}}
\def\mn@eprint@#1:#2:#3:#4\@nil{\def\@tempa {#1}\def\@tempb {#2}\def\@tempc
  {#3}\ifx \@tempc \@empty \let \@tempc \@tempb \let \@tempb \@tempa \fi \ifx
  \@tempb \@empty \def\@tempb {arXiv}\fi \@ifundefined
  {mn@eprint@\@tempb}{\@tempb:\@tempc}{\expandafter \expandafter \csname
  mn@eprint@\@tempb\endcsname \expandafter{\@tempc}}}

\bibitem[\protect\citeauthoryear{{Annunziatella}, {Mercurio}, {Brescia},
  {Cavuoti}  \& {Longo}}{{Annunziatella} et~al.}{2013}]{annunziatella13}
{Annunziatella} M.,  {Mercurio} A.,  {Brescia} M.,  {Cavuoti} S.,   {Longo} G.,
   2013, \mn@doi [PASP] {10.1086/669333}, \href
  {http://adsabs.harvard.edu/abs/2013PASP..125...68A} {125, 68}

\bibitem[\protect\citeauthoryear{{Arag{\'o}n-Salamanca}, {Bedregal}  \&
  {Merrifield}}{{Arag{\'o}n-Salamanca} et~al.}{2006}]{aragon06}
{Arag{\'o}n-Salamanca} A.,  {Bedregal} A.~G.,   {Merrifield} M.~R.,  2006,
  \mn@doi [\aap] {10.1051/0004-6361:20065948}, \href
  {http://adsabs.harvard.edu/abs/2006A%26A...458..101A} {458, 101}

\bibitem[\protect\citeauthoryear{{Ashman}, {Conti}  \& {Zepf}}{{Ashman}
  et~al.}{1995}]{ashman95}
{Ashman} K.~M.,  {Conti} A.,   {Zepf} S.~E.,  1995, \mn@doi [\aj]
  {10.1086/117595}, \href {http://adsabs.harvard.edu/abs/1995AJ....110.1164A}
  {110, 1164}

\bibitem[\protect\citeauthoryear{{Atkinson}, {Abraham}  \&
  {Ferguson}}{{Atkinson} et~al.}{2013}]{atkinson13}
{Atkinson} A.~M.,  {Abraham} R.~G.,   {Ferguson} A.~M.~N.,  2013, \mn@doi
  [\apj] {10.1088/0004-637X/765/1/28}, \href
  {http://adsabs.harvard.edu/abs/2013ApJ...765...28A} {765, 28}

\bibitem[\protect\citeauthoryear{{Balogh}, {Navarro}  \& {Morris}}{{Balogh}
  et~al.}{2000}]{balogh00}
{Balogh} M.~L.,  {Navarro} J.~F.,   {Morris} S.~L.,  2000, \mn@doi [\apj]
  {10.1086/309323}, \href {http://adsabs.harvard.edu/abs/2000ApJ...540..113B}
  {540, 113}

\bibitem[\protect\citeauthoryear{{Barth}, {Darling}, {Baker}, {Boizelle},
  {Buote}, {Ho}  \& {Walsh}}{{Barth} et~al.}{2016}]{barth16}
{Barth} A.~J.,  {Darling} J.,  {Baker} A.~J.,  {Boizelle} B.~D.,  {Buote}
  D.~A.,  {Ho} L.~C.,   {Walsh} J.~L.,  2016, \mn@doi [\apj]
  {10.3847/0004-637X/823/1/51}, \href
  {http://adsabs.harvard.edu/abs/2016ApJ...823...51B} {823, 51}

\bibitem[\protect\citeauthoryear{{Bassino} \& {Caso}}{{Bassino} \&
  {Caso}}{2017}]{bassino17}
{Bassino} L.~P.,  {Caso} J.~P.,  2017, \mn@doi [\mnras]
  {10.1093/mnras/stw3390}, \href
  {http://adsabs.harvard.edu/abs/2017MNRAS.466.4259B} {466, 4259}

\bibitem[\protect\citeauthoryear{{Bassino}, {Faifer}, {Forte}, {Dirsch},
  {Richtler}, {Geisler}  \& {Schuberth}}{{Bassino} et~al.}{2006}]{bassino06}
{Bassino} L.~P.,  {Faifer} F.~R.,  {Forte} J.~C.,  {Dirsch} B.,  {Richtler} T.,
   {Geisler} D.,   {Schuberth} Y.,  2006, \mn@doi [\aap]
  {10.1051/0004-6361:20054563}, \href
  {http://adsabs.harvard.edu/abs/2006A%26A...451..789B} {451, 789}

\bibitem[\protect\citeauthoryear{{Bekki}}{{Bekki}}{2014}]{bekki14}
{Bekki} K.,  2014, \mn@doi [\mnras] {10.1093/mnras/stt2216}, \href
  {http://adsabs.harvard.edu/abs/2014MNRAS.438..444B} {438, 444}

\bibitem[\protect\citeauthoryear{{Bertin}}{{Bertin}}{2011}]{bertin11}
{Bertin} E.,  2011, in {Evans} I.~N.,  {Accomazzi} A.,  {Mink} D.~J.,   {Rots}
  A.~H.,  eds,  Astronomical Society of the Pacific Conference Series Vol. 442,
  Astronomical Data Analysis Software and Systems XX. p.~435

\bibitem[\protect\citeauthoryear{{Bertin} \& {Arnouts}}{{Bertin} \&
  {Arnouts}}{1996}]{bertin96}
{Bertin} E.,  {Arnouts} S.,  1996, \mn@doi [\aaps] {10.1051/aas:1996164}, \href
  {http://adsabs.harvard.edu/abs/1996A%26AS..117..393B} {117, 393}

\bibitem[\protect\citeauthoryear{{Bertin}, {Mellier}, {Radovich}, {Missonnier},
  {Didelon}  \& {Morin}}{{Bertin} et~al.}{2002}]{bertin02}
{Bertin} E.,  {Mellier} Y.,  {Radovich} M.,  {Missonnier} G.,  {Didelon} P.,
  {Morin} B.,  2002, in {Bohlender} D.~A.,  {Durand} D.,   {Handley} T.~H.,
  eds,  Astronomical Society of the Pacific Conference Series Vol. 281,
  Astronomical Data Analysis Software and Systems XI. p.~228

\bibitem[\protect\citeauthoryear{{Beuing}, {Bender}, {Mendes de Oliveira},
  {Thomas}  \& {Maraston}}{{Beuing} et~al.}{2002}]{beuing02}
{Beuing} J.,  {Bender} R.,  {Mendes de Oliveira} C.,  {Thomas} D.,   {Maraston}
  C.,  2002, \mn@doi [\aap] {10.1051/0004-6361:20021321}, \href
  {http://adsabs.harvard.edu/abs/2002A%26A...395..431B} {395, 431}

\bibitem[\protect\citeauthoryear{{B{\'{\i}}lek}, {Cuillandre}, {Gwyn},
  {Ebrov{\'a}}, {Barto{\v s}kov{\'a}}, {Jungwiert}  \&
  {J{\'{\i}}lkov{\'a}}}{{B{\'{\i}}lek} et~al.}{2016}]{bilek16}
{B{\'{\i}}lek} M.,  {Cuillandre} J.-C.,  {Gwyn} S.,  {Ebrov{\'a}} I.,
  {Barto{\v s}kov{\'a}} K.,  {Jungwiert} B.,   {J{\'{\i}}lkov{\'a}} L.,  2016,
  \mn@doi [\aap] {10.1051/0004-6361/201526608}, \href
  {http://adsabs.harvard.edu/abs/2016A%26A...588A..77B} {588, A77}

\bibitem[\protect\citeauthoryear{{Blakeslee}, {Tonry}  \&
  {Metzger}}{{Blakeslee} et~al.}{1997}]{blakeslee97}
{Blakeslee} J.~P.,  {Tonry} J.~L.,   {Metzger} M.~R.,  1997, \mn@doi [\aj]
  {10.1086/118488}, \href {http://adsabs.harvard.edu/abs/1997AJ....114..482B}
  {114, 482}

\bibitem[\protect\citeauthoryear{{Blom}, {Spitler}  \& {Forbes}}{{Blom}
  et~al.}{2012}]{blom12}
{Blom} C.,  {Spitler} L.~R.,   {Forbes} D.~A.,  2012, \mn@doi [\mnras]
  {10.1111/j.1365-2966.2011.19963.x}, \href
  {http://adsabs.harvard.edu/abs/2012MNRAS.420...37B} {420, 37}

\bibitem[\protect\citeauthoryear{{Blom}, {Forbes}, {Foster}, {Romanowsky}  \&
  {Brodie}}{{Blom} et~al.}{2014}]{blom14}
{Blom} C.,  {Forbes} D.~A.,  {Foster} C.,  {Romanowsky} A.~J.,   {Brodie}
  J.~P.,  2014, \mn@doi [\mnras] {10.1093/mnras/stu095}, \href
  {http://adsabs.harvard.edu/abs/2014MNRAS.439.2420B} {439, 2420}

\bibitem[\protect\citeauthoryear{{Brockamp}, {K{\"u}pper}, {Thies}, {Baumgardt}
   \& {Kroupa}}{{Brockamp} et~al.}{2014}]{brockamp14}
{Brockamp} M.,  {K{\"u}pper} A.~H.~W.,  {Thies} I.,  {Baumgardt} H.,   {Kroupa}
  P.,  2014, \mn@doi [\mnras] {10.1093/mnras/stu562}, \href
  {http://adsabs.harvard.edu/abs/2014MNRAS.441..150B} {441, 150}

\bibitem[\protect\citeauthoryear{{Brough}, {Forbes}, {Kilborn}, {Couch}  \&
  {Colless}}{{Brough} et~al.}{2006}]{brough06}
{Brough} S.,  {Forbes} D.~A.,  {Kilborn} V.~A.,  {Couch} W.,   {Colless} M.,
  2006, \mn@doi [\mnras] {10.1111/j.1365-2966.2006.10387.x}, \href
  {http://adsabs.harvard.edu/abs/2006MNRAS.369.1351B} {369, 1351}

\bibitem[\protect\citeauthoryear{{Burkert} \& {Tremaine}}{{Burkert} \&
  {Tremaine}}{2010}]{burkert10}
{Burkert} A.,  {Tremaine} S.,  2010, \mn@doi [\apj]
  {10.1088/0004-637X/720/1/516}, \href
  {http://adsabs.harvard.edu/abs/2010ApJ...720..516B} {720, 516}

\bibitem[\protect\citeauthoryear{{Capaccioli} et~al.,}{{Capaccioli}
  et~al.}{2015}]{capaccioli15}
{Capaccioli} M.,  et~al., 2015, \mn@doi [\aap] {10.1051/0004-6361/201526252},
  \href {http://adsabs.harvard.edu/abs/2015A%26A...581A..10C} {581, A10}

\bibitem[\protect\citeauthoryear{{Carlsten}, {Hau}  \& {Zenteno}}{{Carlsten}
  et~al.}{2017}]{carlsten17}
{Carlsten} S.~G.,  {Hau} G.~K.~T.,   {Zenteno} A.,  2017, \mn@doi [\mnras]
  {10.1093/mnras/stx2182}, \href
  {http://adsabs.harvard.edu/abs/2017MNRAS.472.2889C} {472, 2889}

\bibitem[\protect\citeauthoryear{{Caso}, {Bassino}  \& {G{\'o}mez}}{{Caso}
  et~al.}{2015}]{caso15}
{Caso} J.~P.,  {Bassino} L.~P.,   {G{\'o}mez} M.,  2015, \mn@doi [\mnras]
  {10.1093/mnras/stv2015}, \href
  {http://adsabs.harvard.edu/abs/2015MNRAS.453.4421C} {453, 4421}

\bibitem[\protect\citeauthoryear{{Cavaliere} \& {Fusco-Femiano}}{{Cavaliere} \&
  {Fusco-Femiano}}{1976}]{cavaliere76}
{Cavaliere} A.,  {Fusco-Femiano} R.,  1976, \aap, \href
  {http://adsabs.harvard.edu/abs/1976A%26A....49..137C} {49, 137}

\bibitem[\protect\citeauthoryear{{Cho}, {Sharples}, {Blakeslee}, {Zepf},
  {Kundu}, {Kim}  \& {Yoon}}{{Cho} et~al.}{2012}]{cho12}
{Cho} J.,  {Sharples} R.~M.,  {Blakeslee} J.~P.,  {Zepf} S.~E.,  {Kundu} A.,
  {Kim} H.-S.,   {Yoon} S.-J.,  2012, \mn@doi [\mnras]
  {10.1111/j.1365-2966.2012.20873.x}, \href
  {http://adsabs.harvard.edu/abs/2012MNRAS.422.3591C} {422, 3591}

\bibitem[\protect\citeauthoryear{{Colbert}, {Heckman}, {Ptak}, {Strickland}  \&
  {Weaver}}{{Colbert} et~al.}{2004}]{colbert04}
{Colbert} E.~J.~M.,  {Heckman} T.~M.,  {Ptak} A.~F.,  {Strickland} D.~K.,
  {Weaver} K.~A.,  2004, \mn@doi [\apj] {10.1086/380899}, \href
  {http://adsabs.harvard.edu/abs/2004ApJ...602..231C} {602, 231}

\bibitem[\protect\citeauthoryear{{D'Abrusco}, {Fabbiano}  \&
  {Zezas}}{{D'Abrusco} et~al.}{2015}]{dabrusco15}
{D'Abrusco} R.,  {Fabbiano} G.,   {Zezas} A.,  2015, \mn@doi [\apj]
  {10.1088/0004-637X/805/1/26}, \href
  {http://adsabs.harvard.edu/abs/2015ApJ...805...26D} {805, 26}

\bibitem[\protect\citeauthoryear{{D'Abrusco} et~al.,}{{D'Abrusco}
  et~al.}{2016}]{dabrusco16}
{D'Abrusco} R.,  et~al., 2016, \mn@doi [\apj] {10.3847/2041-8205/819/2/L31},
  \href {http://adsabs.harvard.edu/abs/2016ApJ...819L..31D} {819, L31}

\bibitem[\protect\citeauthoryear{{Dekel}, {Sari}  \& {Ceverino}}{{Dekel}
  et~al.}{2009}]{dekel09}
{Dekel} A.,  {Sari} R.,   {Ceverino} D.,  2009, \mn@doi [\apj]
  {10.1088/0004-637X/703/1/785}, \href
  {http://adsabs.harvard.edu/abs/2009ApJ...703..785D} {703, 785}

\bibitem[\protect\citeauthoryear{{Desai} et~al.,}{{Desai}
  et~al.}{2012}]{desai12}
{Desai} S.,  et~al., 2012, \mn@doi [\apj] {10.1088/0004-637X/757/1/83}, \href
  {http://adsabs.harvard.edu/abs/2012ApJ...757...83D} {757, 83}

\bibitem[\protect\citeauthoryear{{Duc} et~al.,}{{Duc} et~al.}{2015}]{duc15}
{Duc} P.-A.,  et~al., 2015, \mn@doi [\mnras] {10.1093/mnras/stu2019}, \href
  {http://adsabs.harvard.edu/abs/2015MNRAS.446..120D} {446, 120}

\bibitem[\protect\citeauthoryear{{Erwin}, {Pohlen}  \& {Beckman}}{{Erwin}
  et~al.}{2008}]{erwin08}
{Erwin} P.,  {Pohlen} M.,   {Beckman} J.~E.,  2008, \mn@doi [\aj]
  {10.1088/0004-6256/135/1/20}, \href
  {http://adsabs.harvard.edu/abs/2008AJ....135...20E} {135, 20}

\bibitem[\protect\citeauthoryear{{Escudero}, {Faifer}, {Bassino},
  {Calder{\'o}n}  \& {Caso}}{{Escudero} et~al.}{2015}]{escudero15}
{Escudero} C.~G.,  {Faifer} F.~R.,  {Bassino} L.~P.,  {Calder{\'o}n} J.~P.,
  {Caso} J.~P.,  2015, \mn@doi [\mnras] {10.1093/mnras/stv283}, \href
  {http://adsabs.harvard.edu/abs/2015MNRAS.449..612E} {449, 612}

\bibitem[\protect\citeauthoryear{{Faifer} et~al.,}{{Faifer}
  et~al.}{2011}]{faifer11}
{Faifer} F.~R.,  et~al., 2011, \mn@doi [\mnras]
  {10.1111/j.1365-2966.2011.19018.x}, \href
  {http://adsabs.harvard.edu/abs/2011MNRAS.416..155F} {416, 155}

\bibitem[\protect\citeauthoryear{{Ferrarese} et~al.,}{{Ferrarese}
  et~al.}{2000}]{ferrarese00}
{Ferrarese} L.,  et~al., 2000, \mn@doi [\apj] {10.1086/308309}, \href
  {http://adsabs.harvard.edu/abs/2000ApJ...529..745F} {529, 745}

\bibitem[\protect\citeauthoryear{{Forbes}, {S{\'a}nchez-Bl{\'a}zquez}, {Phan},
  {Brodie}, {Strader}  \& {Spitler}}{{Forbes} et~al.}{2006}]{forbes06}
{Forbes} D.~A.,  {S{\'a}nchez-Bl{\'a}zquez} P.,  {Phan} A.~T.~T.,  {Brodie}
  J.~P.,  {Strader} J.,   {Spitler} L.,  2006, \mn@doi [\mnras]
  {10.1111/j.1365-2966.2006.09763.x}, \href
  {http://adsabs.harvard.edu/abs/2006MNRAS.366.1230F} {366, 1230}

\bibitem[\protect\citeauthoryear{{Forbes}, {Spitler}, {Strader}, {Romanowsky},
  {Brodie}  \& {Foster}}{{Forbes} et~al.}{2011}]{forbes11}
{Forbes} D.~A.,  {Spitler} L.~R.,  {Strader} J.,  {Romanowsky} A.~J.,  {Brodie}
  J.~P.,   {Foster} C.,  2011, \mn@doi [\mnras]
  {10.1111/j.1365-2966.2011.18373.x}, \href
  {http://adsabs.harvard.edu/abs/2011MNRAS.413.2943F} {413, 2943}

\bibitem[\protect\citeauthoryear{{Forbes}, {Ponman}  \& {O'Sullivan}}{{Forbes}
  et~al.}{2012}]{forbes12}
{Forbes} D.~A.,  {Ponman} T.,   {O'Sullivan} E.,  2012, \mn@doi [\mnras]
  {10.1111/j.1365-2966.2012.21368.x}, \href
  {http://adsabs.harvard.edu/abs/2012MNRAS.425...66F} {425, 66}

\bibitem[\protect\citeauthoryear{{Forte}}{{Forte}}{2017}]{forte17}
{Forte} J.~C.,  2017, \mn@doi [\mnras] {10.1093/mnras/stx643}, \href
  {http://adsabs.harvard.edu/abs/2017MNRAS.468.3917F} {468, 3917}

\bibitem[\protect\citeauthoryear{{Forte}, {Faifer}  \& {Geisler}}{{Forte}
  et~al.}{2005}]{forte05}
{Forte} J.~C.,  {Faifer} F.,   {Geisler} D.,  2005, \mn@doi [\mnras]
  {10.1111/j.1365-2966.2004.08572.x}, \href
  {http://adsabs.harvard.edu/abs/2005MNRAS.357...56F} {357, 56}

\bibitem[\protect\citeauthoryear{{Forte}, {Faifer}  \& {Geisler}}{{Forte}
  et~al.}{2007}]{forte07}
{Forte} J.~C.,  {Faifer} F.,   {Geisler} D.,  2007, \mn@doi [\mnras]
  {10.1111/j.1365-2966.2007.12515.x}, \href
  {http://adsabs.harvard.edu/abs/2007MNRAS.382.1947F} {382, 1947}

\bibitem[\protect\citeauthoryear{{Forte}, {Vega}  \& {Faifer}}{{Forte}
  et~al.}{2009}]{forte09}
{Forte} J.~C.,  {Vega} E.~I.,   {Faifer} F.,  2009, \mn@doi [\mnras]
  {10.1111/j.1365-2966.2009.15023.x}, \href
  {http://adsabs.harvard.edu/abs/2009MNRAS.397.1003F} {397, 1003}

\bibitem[\protect\citeauthoryear{{Forte}, {Vega}  \& {Faifer}}{{Forte}
  et~al.}{2012}]{forte12}
{Forte} J.~C.,  {Vega} E.~I.,   {Faifer} F.,  2012, \mn@doi [\mnras]
  {10.1111/j.1365-2966.2011.20341.x}, \href
  {http://adsabs.harvard.edu/abs/2012MNRAS.421..635F} {421, 635}

\bibitem[\protect\citeauthoryear{{Forte}, {Vega}, {Faifer}, {Smith Castelli},
  {Escudero}, {Gonz{\'a}lez}  \& {Sesto}}{{Forte} et~al.}{2014}]{forte14}
{Forte} J.~C.,  {Vega} E.~I.,  {Faifer} F.~R.,  {Smith Castelli} A.~V.,
  {Escudero} C.,  {Gonz{\'a}lez} N.~M.,   {Sesto} L.,  2014, \mn@doi [\mnras]
  {10.1093/mnras/stu658}, \href
  {http://adsabs.harvard.edu/abs/2014MNRAS.441.1391F} {441, 1391}

\bibitem[\protect\citeauthoryear{{Freedman} et~al.,}{{Freedman}
  et~al.}{2001}]{freedman01}
{Freedman} W.~L.,  et~al., 2001, \mn@doi [\apj] {10.1086/320638}, \href
  {http://adsabs.harvard.edu/abs/2001ApJ...553...47F} {553, 47}

\bibitem[\protect\citeauthoryear{{Fukugita}, {Shimasaku}  \&
  {Ichikawa}}{{Fukugita} et~al.}{1995}]{fukugita95}
{Fukugita} M.,  {Shimasaku} K.,   {Ichikawa} T.,  1995, \mn@doi [PASP]
  {10.1086/133643}, \href {http://adsabs.harvard.edu/abs/1995PASP..107..945F}
  {107, 945}

\bibitem[\protect\citeauthoryear{{Fukugita}, {Ichikawa}, {Gunn}, {Doi},
  {Shimasaku}  \& {Schneider}}{{Fukugita} et~al.}{1996}]{fukugita96}
{Fukugita} M.,  {Ichikawa} T.,  {Gunn} J.~E.,  {Doi} M.,  {Shimasaku} K.,
  {Schneider} D.~P.,  1996, \mn@doi [\aj] {10.1086/117915}, \href
  {http://adsabs.harvard.edu/abs/1996AJ....111.1748F} {111, 1748}

\bibitem[\protect\citeauthoryear{{Gebhardt} et~al.,}{{Gebhardt}
  et~al.}{2007}]{gebhardt07}
{Gebhardt} K.,  et~al., 2007, \mn@doi [\apj] {10.1086/522938}, \href
  {http://adsabs.harvard.edu/abs/2007ApJ...671.1321G} {671, 1321}

\bibitem[\protect\citeauthoryear{{G{\'o}mez} \& {Richtler}}{{G{\'o}mez} \&
  {Richtler}}{2004}]{gomez04}
{G{\'o}mez} M.,  {Richtler} T.,  2004, \mn@doi [\aap]
  {10.1051/0004-6361:20034610}, \href
  {http://adsabs.harvard.edu/abs/2004A%26A...415..499G} {415, 499}

\bibitem[\protect\citeauthoryear{{Gonz{\'a}lez-L{\'o}pezlira},
  {Lomel{\'{\i}}-N{\'u}{\~n}ez}, {{\'A}lamo-Mart{\'{\i}}nez},
  {{\'O}rdenes-Brice{\~n}o}, {Loinard}, {Georgiev}  \& {et
  al.}}{{Gonz{\'a}lez-L{\'o}pezlira} et~al.}{2017}]{gonzalez17}
{Gonz{\'a}lez-L{\'o}pezlira} R.~A.,  {Lomel{\'{\i}}-N{\'u}{\~n}ez} L.,
  {{\'A}lamo-Mart{\'{\i}}nez} K.,  {{\'O}rdenes-Brice{\~n}o} Y.,  {Loinard} L.,
   {Georgiev} I.~Y.,   {et al.} 2017, \mn@doi [\apj]
  {10.3847/1538-4357/835/2/184}, \href
  {http://adsabs.harvard.edu/abs/2017ApJ...835..184G} {835, 184}

\bibitem[\protect\citeauthoryear{{Goudfrooij}}{{Goudfrooij}}{2012}]{goudfrooij12}
{Goudfrooij} P.,  2012, \mn@doi [\apj] {10.1088/0004-637X/750/2/140}, \href
  {http://adsabs.harvard.edu/abs/2012ApJ...750..140G} {750, 140}

\bibitem[\protect\citeauthoryear{{Hargis} \& {Rhode}}{{Hargis} \&
  {Rhode}}{2012}]{hargis12}
{Hargis} J.~R.,  {Rhode} K.~L.,  2012, \mn@doi [\aj]
  {10.1088/0004-6256/144/6/164}, \href
  {http://adsabs.harvard.edu/abs/2012AJ....144..164H} {144, 164}

\bibitem[\protect\citeauthoryear{{Harris}}{{Harris}}{1996}]{harris96}
{Harris} W.~E.,  1996, \mn@doi [\aj] {10.1086/118116}, \href
  {http://adsabs.harvard.edu/abs/1996AJ....112.1487H} {112, 1487}

\bibitem[\protect\citeauthoryear{{Harris}}{{Harris}}{2009a}]{harris09b}
{Harris} W.~E.,  2009a, \mn@doi [\apj] {10.1088/0004-637X/699/1/254}, \href
  {http://adsabs.harvard.edu/abs/2009ApJ...699..254H} {699, 254}

\bibitem[\protect\citeauthoryear{{Harris}}{{Harris}}{2009b}]{harris09}
{Harris} W.~E.,  2009b, \mn@doi [\apj] {10.1088/0004-637X/703/1/939}, \href
  {http://adsabs.harvard.edu/abs/2009ApJ...703..939H} {703, 939}

\bibitem[\protect\citeauthoryear{{Harris} \& {van den Bergh}}{{Harris} \& {van
  den Bergh}}{1981}]{harris81}
{Harris} W.~E.,  {van den Bergh} S.,  1981, \mn@doi [\aj] {10.1086/113047},
  \href {http://adsabs.harvard.edu/abs/1981AJ.....86.1627H} {86, 1627}

\bibitem[\protect\citeauthoryear{{Harris}, {Harris}  \& {Alessi}}{{Harris}
  et~al.}{2013}]{harris13}
{Harris} W.~E.,  {Harris} G.~L.~H.,   {Alessi} M.,  2013, \mn@doi [\apj]
  {10.1088/0004-637X/772/2/82}, \href
  {http://adsabs.harvard.edu/abs/2013ApJ...772...82H} {772, 82}

\bibitem[\protect\citeauthoryear{{Harris}, {Poole}  \& {Harris}}{{Harris}
  et~al.}{2014}]{harrisg14}
{Harris} G.~L.~H.,  {Poole} G.~B.,   {Harris} W.~E.,  2014, \mn@doi [\mnras]
  {10.1093/mnras/stt2337}, \href
  {http://adsabs.harvard.edu/abs/2014MNRAS.438.2117H} {438, 2117}

\bibitem[\protect\citeauthoryear{{Harris}, {Harris}  \& {Hudson}}{{Harris}
  et~al.}{2015}]{harris15}
{Harris} W.~E.,  {Harris} G.~L.,   {Hudson} M.~J.,  2015, \mn@doi [\apj]
  {10.1088/0004-637X/806/1/36}, \href
  {http://adsabs.harvard.edu/abs/2015ApJ...806...36H} {806, 36}

\bibitem[\protect\citeauthoryear{{Harris}, {Blakeslee}, {Whitmore}, {Gnedin},
  {Geisler}  \& {Rothberg}}{{Harris} et~al.}{2016}]{harris16}
{Harris} W.~E.,  {Blakeslee} J.~P.,  {Whitmore} B.~C.,  {Gnedin} O.~Y.,
  {Geisler} D.,   {Rothberg} B.,  2016, \mn@doi [\apj]
  {10.3847/0004-637X/817/1/58}, \href
  {http://adsabs.harvard.edu/abs/2016ApJ...817...58H} {817, 58}

\bibitem[\protect\citeauthoryear{{Harris}, {Ciccone}, {Eadie}, {Gnedin},
  {Geisler}, {Rothberg}  \& {Bailin}}{{Harris} et~al.}{2017}]{harris17}
{Harris} W.~E.,  {Ciccone} S.~M.,  {Eadie} G.~M.,  {Gnedin} O.~Y.,  {Geisler}
  D.,  {Rothberg} B.,   {Bailin} J.,  2017, \mn@doi [\apj]
  {10.3847/1538-4357/835/1/101}, \href
  {http://adsabs.harvard.edu/abs/2017ApJ...835..101H} {835, 101}

\bibitem[\protect\citeauthoryear{{Hirschmann}, {Naab}, {Ostriker}, {Forbes},
  {Duc}, {Dav{\'e}}, {Oser}  \& {Karabal}}{{Hirschmann}
  et~al.}{2015}]{hirschmann15}
{Hirschmann} M.,  {Naab} T.,  {Ostriker} J.~P.,  {Forbes} D.~A.,  {Duc} P.-A.,
  {Dav{\'e}} R.,  {Oser} L.,   {Karabal} E.,  2015, \mn@doi [\mnras]
  {10.1093/mnras/stv274}, \href
  {http://adsabs.harvard.edu/abs/2015MNRAS.449..528H} {449, 528}

\bibitem[\protect\citeauthoryear{{Hudson} \& {Robison}}{{Hudson} \&
  {Robison}}{2017}]{hudson17}
{Hudson} M.~J.,  {Robison} B.,  2017, preprint, \href
  {http://adsabs.harvard.edu/abs/2017arXiv170702609H} {} (\mn@eprint {arXiv}
  {1707.02609})

\bibitem[\protect\citeauthoryear{{Hudson}, {Harris}  \& {Harris}}{{Hudson}
  et~al.}{2014}]{hudson14}
{Hudson} M.~J.,  {Harris} G.~L.,   {Harris} W.~E.,  2014, \mn@doi [apjl]
  {10.1088/2041-8205/787/1/L5}, \href
  {http://adsabs.harvard.edu/abs/2014ApJ...787L...5H} {787, L5}

\bibitem[\protect\citeauthoryear{{Jacoby} et~al.,}{{Jacoby}
  et~al.}{1992}]{jacoby92}
{Jacoby} G.~H.,  et~al., 1992, \mn@doi [PASP] {10.1086/133035}, \href
  {http://adsabs.harvard.edu/abs/1992PASP..104..599J} {104, 599}

\bibitem[\protect\citeauthoryear{{Jones} et~al.,}{{Jones}
  et~al.}{2004}]{jones04}
{Jones} D.~H.,  et~al., 2004, \mn@doi [\mnras]
  {10.1111/j.1365-2966.2004.08353.x}, \href
  {http://adsabs.harvard.edu/abs/2004MNRAS.355..747J} {355, 747}

\bibitem[\protect\citeauthoryear{{Jones} et~al.,}{{Jones}
  et~al.}{2009}]{jones09}
{Jones} D.~H.,  et~al., 2009, \mn@doi [\mnras]
  {10.1111/j.1365-2966.2009.15338.x}, \href
  {http://adsabs.harvard.edu/abs/2009MNRAS.399..683J} {399, 683}

\bibitem[\protect\citeauthoryear{{Kartha} et~al.,}{{Kartha}
  et~al.}{2016}]{kartha16}
{Kartha} S.~S.,  et~al., 2016, \mn@doi [\mnras] {10.1093/mnras/stw185}, \href
  {http://adsabs.harvard.edu/abs/2016MNRAS.458..105K} {458, 105}

\bibitem[\protect\citeauthoryear{{Kissler-Patig}}{{Kissler-Patig}}{2000}]{kissler00}
{Kissler-Patig} M.,  2000, in {Schielicke} R.~E.,  ed.,  Reviews in Modern
  Astronomy Vol. 13, Reviews in Modern Astronomy. p.~13 (\mn@eprint {}
  {astro-ph/0002070})

\bibitem[\protect\citeauthoryear{{Koleva}, {Prugniel}, {Bouchard}  \&
  {Wu}}{{Koleva} et~al.}{2009}]{koleva09}
{Koleva} M.,  {Prugniel} P.,  {Bouchard} A.,   {Wu} Y.,  2009, \mn@doi [\aap]
  {10.1051/0004-6361/200811467}, \href
  {http://adsabs.harvard.edu/abs/2009A%26A...501.1269K} {501, 1269}

\bibitem[\protect\citeauthoryear{{Kormendy} \& {Bender}}{{Kormendy} \&
  {Bender}}{1996}]{kormendy96}
{Kormendy} J.,  {Bender} R.,  1996, \mn@doi [\apj] {10.1086/310095}, \href
  {http://adsabs.harvard.edu/abs/1996ApJ...464L.119K} {464, L119}

\bibitem[\protect\citeauthoryear{{Kruijssen}}{{Kruijssen}}{2015}]{kruijssen15}
{Kruijssen} J.~M.~D.,  2015, \mn@doi [\mnras] {10.1093/mnras/stv2026}, \href
  {http://adsabs.harvard.edu/abs/2015MNRAS.454.1658K} {454, 1658}

\bibitem[\protect\citeauthoryear{{Kundu} \& {Whitmore}}{{Kundu} \&
  {Whitmore}}{2001}]{kundu01}
{Kundu} A.,  {Whitmore} B.~C.,  2001, \mn@doi [\aj] {10.1086/322095}, \href
  {http://adsabs.harvard.edu/abs/2001AJ....122.1251K} {122, 1251}

\bibitem[\protect\citeauthoryear{{Lane}, {Salinas}  \& {Richtler}}{{Lane}
  et~al.}{2013}]{lane13}
{Lane} R.~R.,  {Salinas} R.,   {Richtler} T.,  2013, \mn@doi [\aap]
  {10.1051/0004-6361/201220231}, \href
  {http://adsabs.harvard.edu/abs/2013A%26A...549A.148L} {549, A148}

\bibitem[\protect\citeauthoryear{{Larsen}, {Brodie}, {Huchra}, {Forbes}  \&
  {Grillmair}}{{Larsen} et~al.}{2001}]{larsen01}
{Larsen} S.~S.,  {Brodie} J.~P.,  {Huchra} J.~P.,  {Forbes} D.~A.,
  {Grillmair} C.~J.,  2001, \mn@doi [\aj] {10.1086/321081}, \href
  {http://adsabs.harvard.edu/abs/2001AJ....121.2974L} {121, 2974}

\bibitem[\protect\citeauthoryear{{Lee}, {Park}, {Kim}, {Hwang}, {Kim}  \&
  {Geisler}}{{Lee} et~al.}{2008}]{lee08}
{Lee} M.~G.,  {Park} H.~S.,  {Kim} E.,  {Hwang} H.~S.,  {Kim} S.~C.,
  {Geisler} D.,  2008, \mn@doi [\apj] {10.1086/587469}, \href
  {http://adsabs.harvard.edu/abs/2008ApJ...682..135L} {682, 135}

\bibitem[\protect\citeauthoryear{{Li}, {Ho}, {Barth}  \& {Peng}}{{Li}
  et~al.}{2011}]{li11}
{Li} Z.-Y.,  {Ho} L.~C.,  {Barth} A.~J.,   {Peng} C.~Y.,  2011, \mn@doi [ApJs]
  {10.1088/0067-0049/197/2/22}, \href
  {http://adsabs.harvard.edu/abs/2011ApJS..197...22L} {197, 22}

\bibitem[\protect\citeauthoryear{{Lupton}}{{Lupton}}{2005}]{lupton05}
{Lupton} R.,  2005,
  {\url{http://www.sdss.org/dr12/algorithms/sdssubvritransform/}}

\bibitem[\protect\citeauthoryear{{Makarov} \& {Karachentsev}}{{Makarov} \&
  {Karachentsev}}{2011}]{makarov11}
{Makarov} D.,  {Karachentsev} I.,  2011, \mn@doi [\mnras]
  {10.1111/j.1365-2966.2010.18071.x}, \href
  {http://adsabs.harvard.edu/abs/2011MNRAS.412.2498M} {412, 2498}

\bibitem[\protect\citeauthoryear{{Malin} \& {Carter}}{{Malin} \&
  {Carter}}{1983}]{malin83}
{Malin} D.~F.,  {Carter} D.,  1983, \mn@doi [\apj] {10.1086/161467}, \href
  {http://adsabs.harvard.edu/abs/1983ApJ...274..534M} {274, 534}

\bibitem[\protect\citeauthoryear{{Mazzei}, {Marino}  \& {Rampazzo}}{{Mazzei}
  et~al.}{2014}]{mazzei14}
{Mazzei} P.,  {Marino} A.,   {Rampazzo} R.,  2014, \mn@doi [\apj]
  {10.1088/0004-637X/782/1/53}, \href
  {http://adsabs.harvard.edu/abs/2014ApJ...782...53M} {782, 53}

\bibitem[\protect\citeauthoryear{{McLaughlin}, {Harris}  \&
  {Hanes}}{{McLaughlin} et~al.}{1994}]{mcLaughlin94}
{McLaughlin} D.~E.,  {Harris} W.~E.,   {Hanes} D.~A.,  1994, \mn@doi [\apj]
  {10.1086/173744}, \href {http://adsabs.harvard.edu/abs/1994ApJ...422..486M}
  {422, 486}

\bibitem[\protect\citeauthoryear{{Meyer} et~al.,}{{Meyer}
  et~al.}{2004}]{meyer04}
{Meyer} M.~J.,  et~al., 2004, \mn@doi [\mnras]
  {10.1111/j.1365-2966.2004.07710.x}, \href
  {http://adsabs.harvard.edu/abs/2004MNRAS.350.1195M} {350, 1195}

\bibitem[\protect\citeauthoryear{{Miller} et~al.,}{{Miller}
  et~al.}{2017}]{miller17}
{Miller} B.,  et~al., 2017, \mn@doi [Galaxies] {10.3390/galaxies5030029}, \href
  {http://adsabs.harvard.edu/abs/2017Galax...5...29M} {5, 29}

\bibitem[\protect\citeauthoryear{{Muratov} \& {Gnedin}}{{Muratov} \&
  {Gnedin}}{2010}]{muratov10}
{Muratov} A.~L.,  {Gnedin} O.~Y.,  2010, \mn@doi [\apj]
  {10.1088/0004-637X/718/2/1266}, \href
  {http://adsabs.harvard.edu/abs/2010ApJ...718.1266M} {718, 1266}

\bibitem[\protect\citeauthoryear{{Naab}, {Johansson}  \& {Ostriker}}{{Naab}
  et~al.}{2009}]{naab09}
{Naab} T.,  {Johansson} P.~H.,   {Ostriker} J.~P.,  2009, \mn@doi [\apj]
  {10.1088/0004-637X/699/2/L178}, \href
  {http://adsabs.harvard.edu/abs/2009ApJ...699L.178N} {699, L178}

\bibitem[\protect\citeauthoryear{{Naab} et~al.,}{{Naab} et~al.}{2014}]{naab14}
{Naab} T.,  et~al., 2014, \mn@doi [\mnras] {10.1093/mnras/stt1919}, \href
  {http://adsabs.harvard.edu/abs/2014MNRAS.444.3357N} {444, 3357}

\bibitem[\protect\citeauthoryear{{Nagino} \& {Matsushita}}{{Nagino} \&
  {Matsushita}}{2009}]{nagino09}
{Nagino} R.,  {Matsushita} K.,  2009, \mn@doi [\aap]
  {10.1051/0004-6361/200810978}, \href
  {http://adsabs.harvard.edu/abs/2009A%26A...501..157N} {501, 157}

\bibitem[\protect\citeauthoryear{{Nasonova}, {de Freitas Pacheco}  \&
  {Karachentsev}}{{Nasonova} et~al.}{2011}]{nasonova11}
{Nasonova} O.~G.,  {de Freitas Pacheco} J.~A.,   {Karachentsev} I.~D.,  2011,
  \mn@doi [\aap] {10.1051/0004-6361/201016004}, \href
  {http://adsabs.harvard.edu/abs/2011A%26A...532A.104N} {532, A104}

\bibitem[\protect\citeauthoryear{{Niemi}, {Hein{\"a}m{\"a}ki}, {Nurmi}  \&
  {Saar}}{{Niemi} et~al.}{2010}]{niemi10}
{Niemi} S.-M.,  {Hein{\"a}m{\"a}ki} P.,  {Nurmi} P.,   {Saar} E.,  2010,
  \mn@doi [\mnras] {10.1111/j.1365-2966.2010.16457.x}, \href
  {http://adsabs.harvard.edu/abs/2010MNRAS.405..477N} {405, 477}

\bibitem[\protect\citeauthoryear{{Norris} et~al.,}{{Norris}
  et~al.}{2008}]{norris08}
{Norris} M.~A.,  et~al., 2008, \mn@doi [\mnras]
  {10.1111/j.1365-2966.2008.12826.x}, \href
  {http://adsabs.harvard.edu/abs/2008MNRAS.385...40N} {385, 40}

\bibitem[\protect\citeauthoryear{{Nowak}, {Saglia}, {Thomas}, {Bender},
  {Davies}  \& {Gebhardt}}{{Nowak} et~al.}{2008}]{nowak08}
{Nowak} N.,  {Saglia} R.~P.,  {Thomas} J.,  {Bender} R.,  {Davies} R.~I.,
  {Gebhardt} K.,  2008, \mn@doi [\mnras] {10.1111/j.1365-2966.2008.13960.x},
  \href {http://adsabs.harvard.edu/abs/2008MNRAS.391.1629N} {391, 1629}

\bibitem[\protect\citeauthoryear{{Nulsen}}{{Nulsen}}{1989}]{nulsen89}
{Nulsen} P.~E.~J.,  1989, \mn@doi [\apj] {10.1086/168051}, \href
  {http://adsabs.harvard.edu/abs/1989ApJ...346..690N} {346, 690}

\bibitem[\protect\citeauthoryear{{O'Sullivan}, {Forbes}  \&
  {Ponman}}{{O'Sullivan} et~al.}{2001}]{osullivan01}
{O'Sullivan} E.,  {Forbes} D.~A.,   {Ponman} T.~J.,  2001, \mn@doi [\mnras]
  {10.1046/j.1365-8711.2001.04890.x}, \href
  {http://adsabs.harvard.edu/abs/2001MNRAS.328..461O} {328, 461}

\bibitem[\protect\citeauthoryear{{Omar} \& {Dwarakanath}}{{Omar} \&
  {Dwarakanath}}{2005}]{omar05}
{Omar} A.,  {Dwarakanath} K.~S.,  2005, \mn@doi [Journal of Astrophysics and
  Astronomy] {10.1007/BF02702451}, \href
  {http://adsabs.harvard.edu/abs/2005JApA...26....1O} {26, 1}

\bibitem[\protect\citeauthoryear{{Oser}, {Ostriker}, {Naab}, {Johansson}  \&
  {Burkert}}{{Oser} et~al.}{2010}]{oser10}
{Oser} L.,  {Ostriker} J.~P.,  {Naab} T.,  {Johansson} P.~H.,   {Burkert} A.,
  2010, \mn@doi [\apj] {10.1088/0004-637X/725/2/2312}, \href
  {http://adsabs.harvard.edu/abs/2010ApJ...725.2312O} {725, 2312}

\bibitem[\protect\citeauthoryear{{Pastorello}, {Forbes}, {Foster}, {Brodie},
  {Usher}, {Romanowsky}, {Strader}  \& {Arnold}}{{Pastorello}
  et~al.}{2014}]{pastorello14}
{Pastorello} N.,  {Forbes} D.~A.,  {Foster} C.,  {Brodie} J.~P.,  {Usher} C.,
  {Romanowsky} A.~J.,  {Strader} J.,   {Arnold} J.~A.,  2014, \mn@doi [\mnras]
  {10.1093/mnras/stu937}, \href
  {http://adsabs.harvard.edu/abs/2014MNRAS.442.1003P} {442, 1003}

\bibitem[\protect\citeauthoryear{{Pellegrini}}{{Pellegrini}}{2010}]{pellegrini10}
{Pellegrini} S.,  2010, \mn@doi [\apj] {10.1088/0004-637X/717/2/640}, \href
  {http://adsabs.harvard.edu/abs/2010ApJ...717..640P} {717, 640}

\bibitem[\protect\citeauthoryear{{Peng} et~al.,}{{Peng} et~al.}{2006}]{peng06}
{Peng} E.~W.,  et~al., 2006, \mn@doi [\apj] {10.1086/498210}, \href
  {http://adsabs.harvard.edu/abs/2006ApJ...639...95P} {639, 95}

\bibitem[\protect\citeauthoryear{{Pop}, {Pillepich}, {Amorisco}  \&
  {Hernquist}}{{Pop} et~al.}{2017}]{pop17}
{Pop} A.-R.,  {Pillepich} A.,  {Amorisco} N.~C.,   {Hernquist} L.,  2017,
  preprint, \href {http://adsabs.harvard.edu/abs/2017arXiv170606102P} {}
  (\mn@eprint {arXiv} {1706.06102})

\bibitem[\protect\citeauthoryear{{Pota} et~al.,}{{Pota} et~al.}{2013}]{pota13}
{Pota} V.,  et~al., 2013, \mn@doi [\mnras] {10.1093/mnras/sts029}, \href
  {http://adsabs.harvard.edu/abs/2013MNRAS.428..389P} {428, 389}

\bibitem[\protect\citeauthoryear{{Pota} et~al.,}{{Pota} et~al.}{2015}]{pota15}
{Pota} V.,  et~al., 2015, \mn@doi [\mnras] {10.1093/mnras/stv831}, \href
  {http://adsabs.harvard.edu/abs/2015MNRAS.450.3345P} {450, 3345}

\bibitem[\protect\citeauthoryear{{Radovich} et~al.,}{{Radovich}
  et~al.}{2015}]{radovich15}
{Radovich} M.,  et~al., 2015, \mn@doi [\aap] {10.1051/0004-6361/201425600},
  \href {http://adsabs.harvard.edu/abs/2015A%26A...579A...7R} {579, A7}

\bibitem[\protect\citeauthoryear{{Richtler}, {Bassino}, {Dirsch}  \&
  {Kumar}}{{Richtler} et~al.}{2012}]{richtler12}
{Richtler} T.,  {Bassino} L.~P.,  {Dirsch} B.,   {Kumar} B.,  2012, \mn@doi
  [\aap] {10.1051/0004-6361/201118589}, \href
  {http://adsabs.harvard.edu/abs/2012A%26A...543A.131R} {543, A131}

\bibitem[\protect\citeauthoryear{{Rodriguez-Gomez} et~al.,}{{Rodriguez-Gomez}
  et~al.}{2016}]{rodriguez16}
{Rodriguez-Gomez} V.,  et~al., 2016, \mn@doi [\mnras] {10.1093/mnras/stw456},
  \href {http://adsabs.harvard.edu/abs/2016MNRAS.458.2371R} {458, 2371}

\bibitem[\protect\citeauthoryear{{Romanowsky}, {Strader}, {Spitler}, {Johnson},
  {Brodie}, {Forbes}  \& {Ponman}}{{Romanowsky} et~al.}{2009}]{romanowsky09}
{Romanowsky} A.~J.,  {Strader} J.,  {Spitler} L.~R.,  {Johnson} R.,  {Brodie}
  J.~P.,  {Forbes} D.~A.,   {Ponman} T.,  2009, \mn@doi [\aj]
  {10.1088/0004-6256/137/6/4956}, \href
  {http://adsabs.harvard.edu/abs/2009AJ....137.4956R} {137, 4956}

\bibitem[\protect\citeauthoryear{{Salinas}, {Alabi}, {Richtler}  \&
  {Lane}}{{Salinas} et~al.}{2015}]{salinas15}
{Salinas} R.,  {Alabi} A.,  {Richtler} T.,   {Lane} R.~R.,  2015, \mn@doi
  [\aap] {10.1051/0004-6361/201425574}, \href
  {http://adsabs.harvard.edu/abs/2015A%26A...577A..59S} {577, A59}

\bibitem[\protect\citeauthoryear{{S{\'a}nchez-Bl{\'a}zquez}, {Ocvirk},
  {Gibson}, {P{\'e}rez}  \& {Peletier}}{{S{\'a}nchez-Bl{\'a}zquez}
  et~al.}{2011}]{sanchez11}
{S{\'a}nchez-Bl{\'a}zquez} P.,  {Ocvirk} P.,  {Gibson} B.~K.,  {P{\'e}rez} I.,
   {Peletier} R.~F.,  2011, \mn@doi [\mnras]
  {10.1111/j.1365-2966.2011.18749.x}, \href
  {http://adsabs.harvard.edu/abs/2011MNRAS.415..709S} {415, 709}

\bibitem[\protect\citeauthoryear{{Sanderson} \& {Helmi}}{{Sanderson} \&
  {Helmi}}{2013}]{sanderson13}
{Sanderson} R.~E.,  {Helmi} A.,  2013, \mn@doi [\mnras]
  {10.1093/mnras/stt1307}, \href
  {http://adsabs.harvard.edu/abs/2013MNRAS.435..378S} {435, 378}

\bibitem[\protect\citeauthoryear{{Schlafly} \& {Finkbeiner}}{{Schlafly} \&
  {Finkbeiner}}{2011}]{schlafly11}
{Schlafly} E.~F.,  {Finkbeiner} D.~P.,  2011, \mn@doi [\apj]
  {10.1088/0004-637X/737/2/103}, \href
  {http://adsabs.harvard.edu/abs/2011ApJ...737..103S} {737, 103}

\bibitem[\protect\citeauthoryear{{Scott} \& {Thompson}}{{Scott} \&
  {Thompson}}{1983}]{scott83}
{Scott} D.~W.,  {Thompson} J.~R.,  1983, Proceedings of the Fifteenth Symposium
  on the Interface, pp 173--179

\bibitem[\protect\citeauthoryear{{Serra} \& {Oosterloo}}{{Serra} \&
  {Oosterloo}}{2010}]{serra10}
{Serra} P.,  {Oosterloo} T.~A.,  2010, \mn@doi [\mnras]
  {10.1111/j.1745-3933.2009.00779.x}, \href
  {http://adsabs.harvard.edu/abs/2010MNRAS.401L..29S} {401, L29}

\bibitem[\protect\citeauthoryear{{S\'ersic}}{{S\'ersic}}{1968}]{sersic68}
{S\'ersic} J.~L.,  1968, {Atlas de galaxias australes}

\bibitem[\protect\citeauthoryear{{Sesto}, {Faifer}  \& {Forte}}{{Sesto}
  et~al.}{2016}]{sesto16}
{Sesto} L.~A.,  {Faifer} F.~R.,   {Forte} J.~C.,  2016, \mn@doi [\mnras]
  {10.1093/mnras/stw1627}, \href
  {http://adsabs.harvard.edu/abs/2016MNRAS.461.4260S} {461, 4260}

\bibitem[\protect\citeauthoryear{{Smith}, {Lucey}, {Hudson}, {Schlegel}  \&
  {Davies}}{{Smith} et~al.}{2000}]{smith00}
{Smith} R.~J.,  {Lucey} J.~R.,  {Hudson} M.~J.,  {Schlegel} D.~J.,   {Davies}
  R.~L.,  2000, \mn@doi [\mnras] {10.1046/j.1365-8711.2000.03251.x}, \href
  {http://adsabs.harvard.edu/abs/2000MNRAS.313..469S} {313, 469}

\bibitem[\protect\citeauthoryear{{Spitler}, {Forbes}, {Strader}, {Brodie}  \&
  {Gallagher}}{{Spitler} et~al.}{2008}]{spitler08}
{Spitler} L.~R.,  {Forbes} D.~A.,  {Strader} J.,  {Brodie} J.~P.,   {Gallagher}
  J.~S.,  2008, \mn@doi [\mnras] {10.1111/j.1365-2966.2007.12823.x}, \href
  {http://adsabs.harvard.edu/abs/2008MNRAS.385..361S} {385, 361}

\bibitem[\protect\citeauthoryear{{Spolaor}, {Forbes}, {Hau}, {Proctor}  \&
  {Brough}}{{Spolaor} et~al.}{2008a}]{spolaor08a}
{Spolaor} M.,  {Forbes} D.~A.,  {Hau} G.~K.~T.,  {Proctor} R.~N.,   {Brough}
  S.,  2008a, \mn@doi [\mnras] {10.1111/j.1365-2966.2008.12891.x}, \href
  {http://adsabs.harvard.edu/abs/2008MNRAS.385..667S} {385, 667}

\bibitem[\protect\citeauthoryear{{Spolaor}, {Forbes}, {Proctor}, {Hau}  \&
  {Brough}}{{Spolaor} et~al.}{2008b}]{spolaor08b}
{Spolaor} M.,  {Forbes} D.~A.,  {Proctor} R.~N.,  {Hau} G.~K.~T.,   {Brough}
  S.,  2008b, \mn@doi [\mnras] {10.1111/j.1365-2966.2008.12892.x}, \href
  {http://adsabs.harvard.edu/abs/2008MNRAS.385..675S} {385, 675}

\bibitem[\protect\citeauthoryear{{Tal}, {van Dokkum}, {Nelan}  \&
  {Bezanson}}{{Tal} et~al.}{2009}]{tal09}
{Tal} T.,  {van Dokkum} P.~G.,  {Nelan} J.,   {Bezanson} R.,  2009, \mn@doi
  [\aj] {10.1088/0004-6256/138/5/1417}, \href
  {http://adsabs.harvard.edu/abs/2009AJ....138.1417T} {138, 1417}

\bibitem[\protect\citeauthoryear{{Taylor}, {Puzia}, {Mu{\~n}oz}, {Mieske},
  {Lan{\c c}on}, {Zhang}, {Eigenthaler}  \& {Bovill}}{{Taylor}
  et~al.}{2017}]{taylor17}
{Taylor} M.~A.,  {Puzia} T.~H.,  {Mu{\~n}oz} R.~P.,  {Mieske} S.,  {Lan{\c
  c}on} A.,  {Zhang} H.,  {Eigenthaler} P.,   {Bovill} M.~S.,  2017, \mn@doi
  [\mnras] {10.1093/mnras/stx1021}, \href
  {http://adsabs.harvard.edu/abs/2017MNRAS.469.3444T} {469, 3444}

\bibitem[\protect\citeauthoryear{{Thomas}, {Maraston}  \& {Bender}}{{Thomas}
  et~al.}{2003}]{thomas03}
{Thomas} D.,  {Maraston} C.,   {Bender} R.,  2003, \mn@doi [\mnras]
  {10.1046/j.1365-8711.2003.06248.x}, \href
  {http://adsabs.harvard.edu/abs/2003MNRAS.339..897T} {339, 897}

\bibitem[\protect\citeauthoryear{{Thomas}, {Maraston}, {Bender}  \& {Mendes de
  Oliveira}}{{Thomas} et~al.}{2005}]{thomas05}
{Thomas} D.,  {Maraston} C.,  {Bender} R.,   {Mendes de Oliveira} C.,  2005,
  \mn@doi [\apj] {10.1086/426932}, \href
  {http://adsabs.harvard.edu/abs/2005ApJ...621..673T} {621, 673}

\bibitem[\protect\citeauthoryear{{Trentham}, {Tully}  \& {Mahdavi}}{{Trentham}
  et~al.}{2006}]{trentham06}
{Trentham} N.,  {Tully} R.~B.,   {Mahdavi} A.,  2006, \mn@doi [\mnras]
  {10.1111/j.1365-2966.2006.10378.x}, \href
  {http://adsabs.harvard.edu/abs/2006MNRAS.369.1375T} {369, 1375}

\bibitem[\protect\citeauthoryear{{Tully} et~al.,}{{Tully}
  et~al.}{2013}]{tully13}
{Tully} R.~B.,  et~al., 2013, \mn@doi [\aj] {10.1088/0004-6256/146/4/86}, \href
  {http://adsabs.harvard.edu/abs/2013AJ....146...86T} {146, 86}

\bibitem[\protect\citeauthoryear{{Usher} et~al.,}{{Usher}
  et~al.}{2012}]{usher12}
{Usher} C.,  et~al., 2012, \mn@doi [\mnras] {10.1111/j.1365-2966.2012.21801.x},
  \href {http://adsabs.harvard.edu/abs/2012MNRAS.426.1475U} {426, 1475}

\bibitem[\protect\citeauthoryear{{Vazdekis}, {S{\'a}nchez-Bl{\'a}zquez},
  {Falc{\'o}n-Barroso}, {Cenarro}, {Beasley}, {Cardiel}, {Gorgas}  \&
  {Peletier}}{{Vazdekis} et~al.}{2010}]{vazdekis10}
{Vazdekis} A.,  {S{\'a}nchez-Bl{\'a}zquez} P.,  {Falc{\'o}n-Barroso} J.,
  {Cenarro} A.~J.,  {Beasley} M.~A.,  {Cardiel} N.,  {Gorgas} J.,   {Peletier}
  R.~F.,  2010, \mn@doi [\mnras] {10.1111/j.1365-2966.2010.16407.x}, \href
  {http://adsabs.harvard.edu/abs/2010MNRAS.404.1639V} {404, 1639}

\bibitem[\protect\citeauthoryear{{Villegas} et~al.,}{{Villegas}
  et~al.}{2010}]{villegas10}
{Villegas} D.,  et~al., 2010, \mn@doi [\apj] {10.1088/0004-637X/717/2/603},
  \href {http://adsabs.harvard.edu/abs/2010ApJ...717..603V} {717, 603}

\bibitem[\protect\citeauthoryear{{Wellons} et~al.,}{{Wellons}
  et~al.}{2016}]{wellons16}
{Wellons} S.,  et~al., 2016, \mn@doi [\mnras] {10.1093/mnras/stv2738}, \href
  {http://adsabs.harvard.edu/abs/2016MNRAS.456.1030W} {456, 1030}

\bibitem[\protect\citeauthoryear{{Yang}, {Mo}  \& {van den Bosch}}{{Yang}
  et~al.}{2008}]{yang08}
{Yang} X.,  {Mo} H.~J.,   {van den Bosch} F.~C.,  2008, \mn@doi [\apj]
  {10.1086/528954}, \href {http://adsabs.harvard.edu/abs/2008ApJ...676..248Y}
  {676, 248}

\bibitem[\protect\citeauthoryear{{Zolotov}, {Willman}, {Brooks}, {Governato},
  {Hogg}, {Shen}  \& {Wadsley}}{{Zolotov} et~al.}{2010}]{zolotov10}
{Zolotov} A.,  {Willman} B.,  {Brooks} A.~M.,  {Governato} F.,  {Hogg} D.~W.,
  {Shen} S.,   {Wadsley} J.,  2010, \mn@doi [\apj]
  {10.1088/0004-637X/721/1/738}, \href
  {http://adsabs.harvard.edu/abs/2010ApJ...721..738Z} {721, 738}

\makeatother
\end{thebibliography}

\begin{appendix}
\section{Classification parameters: Class Star and Spread Model}
\label{DAO_PSF}
One of the most reliable software for performing stellar photometry using the PSF fitting technique, is the well known {\sc{DAOPHOT}}. However, in recent years the software SE{\sc{xtractor}} along with PSFEx, has begun to be implemented in several works to accomplish the PSF modeling \citep{desai12,radovich15,gonzalez17}. A detailed analysis of the different capacities provided by PSFEx was carried out by \citet{annunziatella13}.
Here, we evaluate the reliability of that software, by analyzing the new classifier parameter {\sc{spread model}} in comparison with the classical {\sc{class star}} parameter of SE{\sc{xtractor}}.

The range of values given by the {\sc{class star}} parameter varies between 0 and 1, assigning the value 0 to resolved objects and 1 for unresolved ones. This procedure is performed using a neural network. On the other hand, {\sc{spread model}} is a normalized simplified linear discriminant between the best fitting local PSF model and a slightly more extended model, made by the same PSF convolved with a circular exponential disk model with scalelength$=$FWHM/16, where FWHM is the full-width at half maximum of the PSF model \citep{desai12}. This parameter classify resolved objects with values around zero, and extended objects with positive values. A good separation compromise for point sources is {\sc{spread model}}$<$0.0035.

\begin{figure}
\centering
\resizebox{0.98\hsize}{!}{\includegraphics{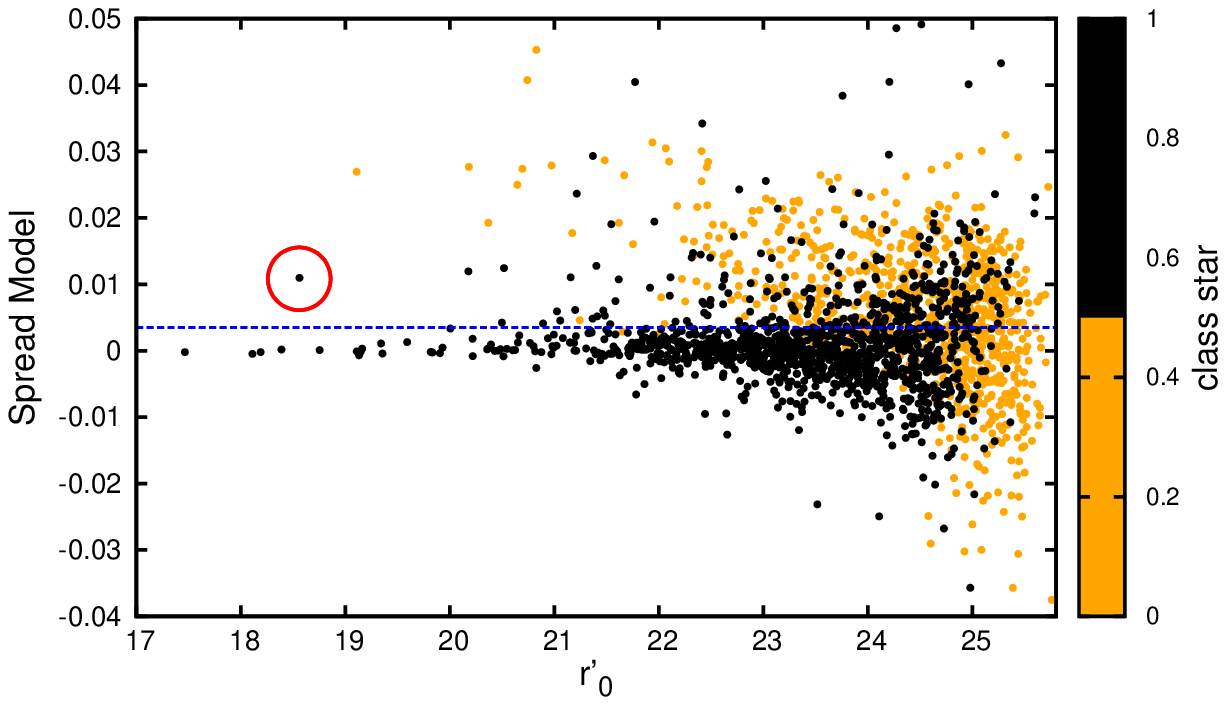}}
\resizebox{0.98\hsize}{!}{\includegraphics{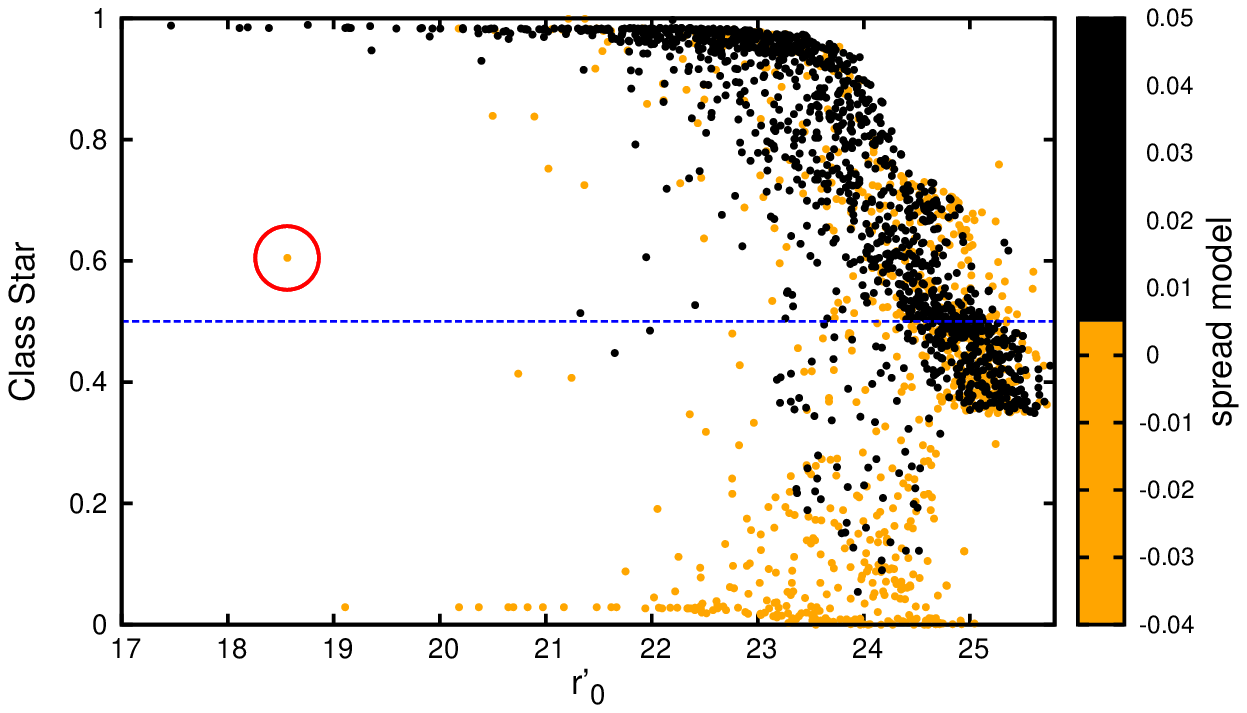}}
\caption{Object classification diagrams using {\sc{spread model}} {\it(top panel)} and {\sc{class star}} {\it (bottom panel)} parameters as a function of the $r'$ magnitude. Blue horizontal dashed lines show the boundary used in this work to separate between resolved (orange dots) and unresolved (black dots) sources. The red circle indicates the resolved object classified by {\sc{spread model}}.}
\label{figure_a1}
\end{figure}

Figure\,\ref{figure_a1} shows {\sc{spread model}} (top panel) and {\sc{class star}} (bottom panel) values as a function of $r'$ magnitude, for all the detected objects in the field that contains NGC\,1395. As shown in the figure, the separation value between resolved from unresolved sources considered by {\sc{spread model}} is comparable to {\sc{class star}}=0.5. It is clear from the Figure that both parameters become uncertain towards the faint luminosity end ($r'_0>24.5$ mag). 

\begin{figure*}
\centering
\resizebox{0.34\hsize}{!}{\includegraphics{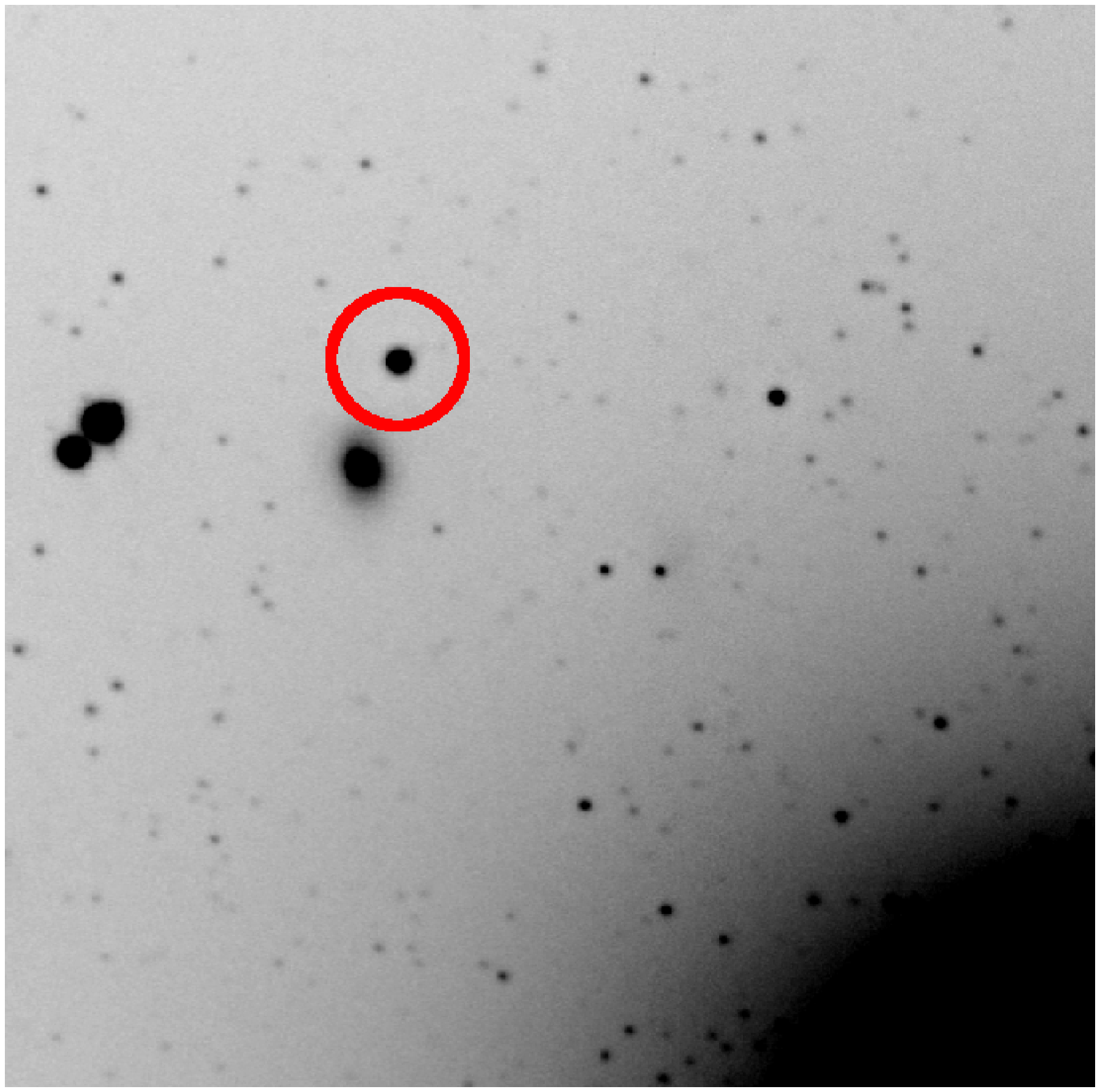}}
\resizebox{0.56\hsize}{!}{\includegraphics{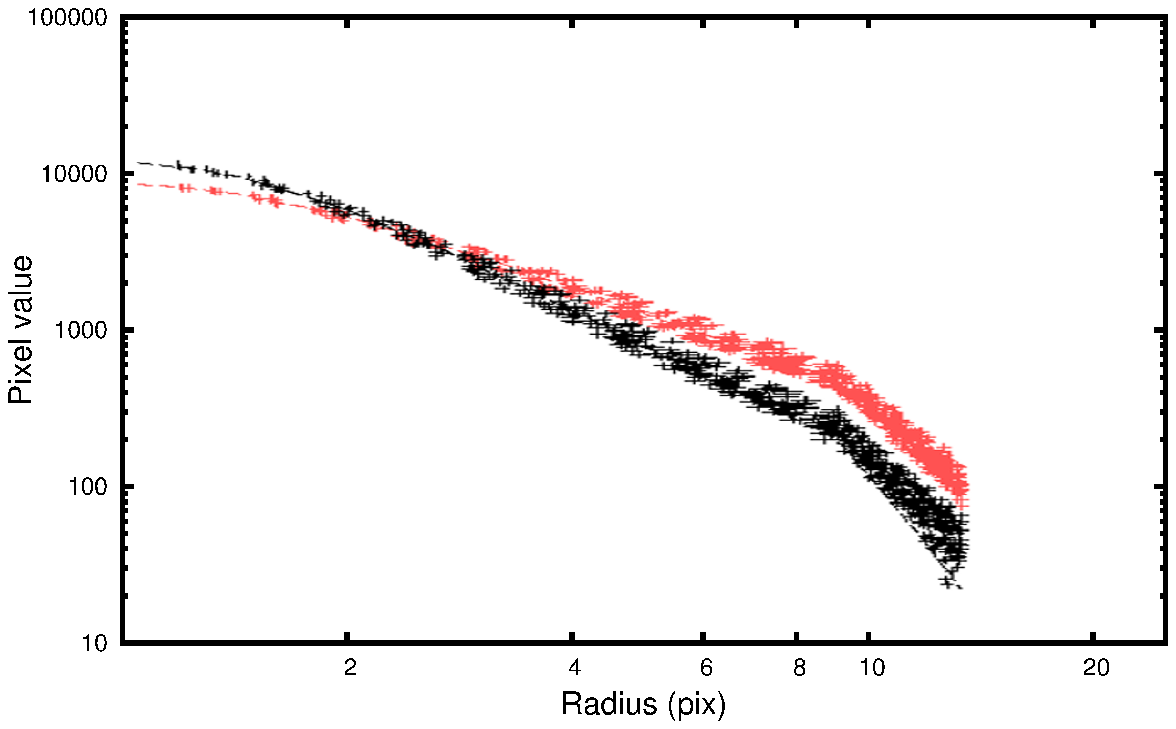}}
\caption{{\it Left panel:} Gemini-GMOS image of 2$\times$2 arcmin of the field that contains the galaxy NGC\,1395. The red circle shows the location of the resolved (according to the classification made by {\sc{spread model}}) object mentioned in the text. {\it Right panel:} Comparison between the radial profiles of the aformentioned extended source (red pluses) and an unresolved source (black pluses) in the GMOS field.}
\label{figure_a2}
\end{figure*}

Although both parameters are similar, the correct classification of relatively bright, compact, although marginally resolved objects becomes difficult using only the parameter {\sc{class star}}, which is less efficient for this aim. A clear example is observed when analyzing the object located at a projected radius of 2.1 arcmin ($\sim$14 kpc) from the center of NGC\,1395 (red circle in the left panel of Figure\,\ref{figure_a2}). According to the classification criterion adopted in this work, this object is catalogued by {\sc{class star}} as an unresolved source assigning it a value of 0.55. On the other hand, the value 0.011 is obtained by the parameter {\sc{spread model}}, which classifies it securely as an extended object. The red circles in Figure\,\ref{figure_a1} depict the location of this object in the considered classification diagrams.
Using the task {\sc{imexa}} of IRAF, we compare the radial profile of this object with that of an unresolved source in the image. As shown in the right panel of Figure \ref{figure_a2}, this object results clearly extended in our images.  

In conclusion, the new classifier of SExtractor+PSFEx {\sc{spread model}} allows us to obtain a reliable separation between the resolved and unresolved objects present in GMOS images, being a good complement to be used together with the index {\sc{class star}}.

\end{appendix}


\end{document}